\documentclass[aps,pre,superscriptaddress,groupedaddress]{revtex4}
\pdfoutput=1
\usepackage{graphicx}  
\usepackage{bm}        
\usepackage{amssymb}   
\usepackage{amsmath}
\usepackage{amsthm}
\usepackage{color}
\usepackage{pgfplots}
\usepackage{cases}
\usepackage{enumerate}
\usepackage{todonotes}

\begin{document}

\title{ Information thermodynamics for interacting stochastic systems without bipartite structure}
\author{R. Ch\' etrite}
\affiliation{Laboratoire J.A. Dieudonn\'e, UMR CNRS 7351, Universit\' e de Nice Sophia-Antipolis, Parc Valrose, 06108 Nice Cedex 02, France}
\author{ M.L. Rosinberg}
\affiliation{Sorbonne Universit\'e, CNRS, Laboratoire de Physique Th\'eorique de la Mati\`ere Condens\'ee, LPTMC,\\  F-75005 Paris, France}
\author{T. Sagawa}
\affiliation{Department of Applied Physics, School of Engineering, The University of Tokyo, 7-3-1 Hongo, Bunkyo-ku, Tokyo 113-8656, Japan}
\author{ G. Tarjus}
\affiliation{Sorbonne Universit\'e, CNRS, Laboratoire de Physique Th\'eorique de la Mati\`ere Condens\'ee, LPTMC,\\  F-75005 Paris, France}

\begin{abstract}

Fluctuations in biochemical networks, {\it e.g.}, in a living cell, have a complex origin that precludes a description of such systems in terms of bipartite or multipartite processes, as is usually done in the framework of stochastic and/or information thermodynamics. This means that fluctuations in each subsystem are not independent:  subsystems jump simultaneously if the dynamics is modeled  as a Markov jump process, or noises are correlated for diffusion processes. In this paper,  we consider  information and thermodynamic exchanges between a pair of coupled systems that do not satisfy the bipartite property. The generalization of information-theoretic measures, such as learning rates and transfer entropy rates,  to this situation is non-trivial and  also involves introducing several additional rates. We describe how this can be achieved in the framework of general continuous-time Markov processes, without restricting the study to the steady-state regime.  We illustrate our general formalism on the case of  diffusion processes and derive an extension of the second law of information thermodynamics in which the difference of transfer entropy rates in the forward and backward time directions replaces the learning rate. As a side result, we also generalize an important relation linking information theory and estimation theory. To further obtain analytical expressions we treat in detail the case of Ornstein-Uhlenbeck processes, and 
discuss the ability of the various information measures to detect a directional coupling in the presence of correlated noises.  Finally, we apply our formalism to the analysis of the directional influence between cellular processes in a concrete example, which also requires considering the case of a non-bipartite and non-Markovian process.

 \end{abstract} 


\maketitle

\tableofcontents

\section{Introduction}

\label{sec:Intro}

One recent and important field of application of information theory is biological systems, in particular gene regulation and signal transduction systems. Cells must sense, process and adapt to their environment or their own physiological state, which are noisy processes  subjected to fluctuations (see {\it e.g.} \cite{BS2014}). In addition, information transfers consume energy, so that there is  a competition between the information gain and the energy cost. This fundamental issue is the realm of information thermodynamics, a recent and active field of research, as reviewed in  \cite{PHS2015}. In this framework, the present work is motivated by the observation that there exist, broadly speaking, two different sources of fluctuations contributing to the stochasticity of biochemical processes, for instance in cell metabolic networks. The first one -- commonly called ``intrinsic"-- is due the small numbers of biomolecules involved in a given reaction. The second one -- the ``extrinsic" source-- arises from the heterogeneity in the physical environment of the cell and the occurrence of (many) other biochemical reactions  (see, {\it  e.g.},  ~\cite{E2002,TWW2006,DCLME2008,UW2011,GW2014,K2014,HT2016,LEM2017,LNTRL2017}).  This implies that the noise in the input biochemical signal -- to be detected-- and the noise of the reactions that form the network are correlated. In short,  stochastic noises have a nontrivial structure and fluctuations in each subsystem are  not independent.

In contrast, in the context of stochastic and information thermodynamics,  the so-called {\it bipartite} assumption is usually made ({\it e.g.}, for modeling Maxwell's demons): one  assumes that subsystems cannot jump simultaneously if the dynamics is modeled  as a Markov jump process or that noises are uncorrelated  if the dynamics is modeled  as diffusion processes.This simplifies the theoretical analysis and allows the contribution of each components of the  system to the entropy production to be clearly identified~\cite{AJM2009,BHS2013,HE2014,IS2013,DE2014,BHS2014,HBS2014,HS2014,IS2015,SS2015,H2015,HBS2016,SLP2016,I2016,RH2016,MS2018,I2018}.  Although the abandon of the bipartite (or multipartite~\cite{H2015}) structure seriously complicates the interpretation of information and thermodynamics exchanges, our objective in the present work is to show that a detailed description is still available.  The price to pay is that  several  information-theoretic measures must be added to those already introduced in the literature (information flow, aka learning rate, and transfer entropy),  which characterize how  information is exchanged between two interacting systems in the course of their dynamical evolution. 

In this paper,  we will mostly consider non-equilibrium systems that can be modeled by continuous-time Markov processes  (diffusions, jump processes, or both).  It turns out however that many definitions or relations are also valid beyond the Markovian description and we will therefore provide a general framework.  Moreover, in order to offer a sufficiently general perspective, we assume the presence of multiplicative noises (additive noises being regarded as a only special case) and  we do not restrict the study to steady-state situations, as is often done. On the other hand,  as far as stochastic thermodynamics is concerned, we only consider averaged quantities and do not derive fluctuation relations. We leave this important issue to future investigations. 

The main results of the paper can be summarized as follows:

1) We introduce a set of information-theoretic measures to consistently characterize information exchanges in a generic non-bipartite stochastic system composed of two interacting subsystems. This includes learning rates and transfer entropy rates. In particular, we define a learning rate that contains the footprints of time-irreversibility.  To make the reading of the following sections easier, the definitions of all these information measures are listed in Table I.

2) We derive a set of inequalities among these quantities and show under which circumstances some inequalities become equalities, which may signal the presence of a so-called ``sufficient statistic"~\cite{CT2006,MS2018}. For Markov processes, we then  propose a generalization of the notion of sensory capacity introduced in \cite{HBS2016}.

3) We give  explicit expressions of the information measures for Markov diffusion processes in terms of  probability currents, diffusion tensor, and propagators. In passing, we obtain a generalization to non-bipartite systems of a classical result linking information theory and estimation theory~\cite{D1970,D1971}.

5) We derive a generalized version of the so-called ``second law of information thermodynamics"~\cite{PHS2015} that applies to non-bipartite diffusion processes, showing that the entropy production rate in one of the subsystems is lower bounded by the difference in  transfer entropy rates in the forward and backward time directions.

6) We illustrate the behavior of the various information-theoretic measures in the case of a stationary bidimensional Ornstein-Uhlenbeck process and show their evolution as one systematically varies the parameter quantifying the correlation between the noises.

7) We apply our formalism to the study of the directional influence between cellular processes in the metabolism of  {\it E. coli}~\cite{K2014,LNTRL2017}  and exhibit an intriguing feature that may indicate some optimality in the transmission of information.


\vspace{0.3cm}

\begin{table*}[!h]
  \centering
  \renewcommand{\arraystretch}{2.5}
  \begin{tabular}{|c|c|c|c}
    \hline
    \mbox{{\bf Information measures}} & \mbox{{\bf Name}} & \mbox{{\bf Definition}} \\
     \hline
    $ l_X^+(t)$ & $\mbox{Forward learning rate (a.k.a. information flow)}$& $\frac{1}{h} \Bigl\langle \ln\frac{P(Y_t\vert X_{t+h})}{P(Y_t\vert X_t)} \Bigr\rangle\vert_{0^+}$\\
    $l_X^-(t)$ & $\mbox{Backward learning rate}$& $\frac{1}{h}  \Bigl\langle \ln\frac{P(Y_{t+h}\vert X_{t+h})}{P(Y_{t+h}\vert X_t)} \Bigr\rangle\vert_{0^+}$\\
     $l^s_X(t)$ & $\mbox{Symmetric learning rate}$& $l_X^s(t)=\frac{1}{2}[l_X^+(t)+l_X^-(t)]$ \\
     \hline
     $ {\cal T}_{X\to Y}(t)$ & $\mbox{Transfer entropy (TE) rate}$ & $\frac{1}{h} \Bigl\langle \ln\frac{P(Y_{t+h}\vert X_0^t,Y_0^t)}{P(Y_{t+h}\vert Y_0^t)} \Bigr\rangle\vert_{0^+}$ \\
    $ \overline{{\cal T}}_{X\to Y}(t)$ & $\mbox{Single-time-step TE rate}$ & $\frac{1}{h} \Bigl\langle \ln \frac{P(Y_{t+h}\vert X_t,Y_t)}{P(Y_{t+h}\vert Y_t)}\Bigr\rangle\vert_{0^+}$ \\
    \hline
    ${\cal T}^\dag_{X \to Y}(t)$ &$ \mbox{Backward TE rate}$ & $\frac{1}{h} \Bigl\langle \ln \frac{P(Y_t\vert X_{t+h}^T,Y_{t+h}^T)}{P(Y_t\vert Y_{t+h}^T)}\Bigr\rangle\vert_{0^+}$\\
     $\overline{{\cal T}}^\dag_{X \to Y}(t)$ &$ \mbox{Single-time-step backward TE rate}$ & $\frac{1}{h} \Bigl\langle \ln \frac{P(Y_t\vert X_{t+h},Y_{t+h})}{P(Y_t\vert Y_{t+h})}\Bigr\rangle\vert_{0^+}$\\
    \hline
    $\widehat{{\cal T}}_{X\to Y}(t)$ & $\mbox{Filtered TE rate}$ & $\frac{1}{h} \Bigl\langle \ln \frac{P(X_{t+h},Y_{t+h}\vert Y_0^t)}{P(X_{t+h}\vert Y_0^t)P(Y_{t+h}\vert Y_0^t)}\Bigr\rangle\vert_{0^+}$ \\   
    $\widehat{\overline{{\cal T}}}_{X\to Y}(t)$ & $\mbox{Single-time-step filtered TE rate}$ & $\frac{1}{h} \Bigl\langle \ln \frac{P(X_{t+h},Y_{t+h}\vert Y_t)}{P(X_{t+h}\vert Y_t)P(Y_{t+h}\vert Y_t)}\Bigr\rangle\vert_{0^+}$ \\ 
     \hline
    \end{tabular}
  \caption{Definitions of information measures for a  non-bipartite process ${\bf Z}_t=(X_t,Y_t)$. Similar quantities are defined by exchanging $X$ and $Y$. In the third column, it is implicit that the limit $h\to 0^+$ is taken.} 
\label{Table0}
\end{table*}

More specifically, the paper is organized as follows. In Sec. \ref{sec:Remind},  we describe our general setup and briefly review the existing results for bipartite processes. To this aim, we first present the  formal tools that will be used throughout the paper, in particular those related to continuous-time Markov processes. We then define the information-theoretic measures that are commonly considered in the framework of information thermodynamics and that satisfy some useful inequalities. We then recall the corresponding  formulations of the second law. In Sec. \ref{sec:Info},  the bipartite assumption is lifted and we introduce the relevant information measures.  The usual inequalities are then generalized. In Sec. \ref{sec:DIF}, to make all of the introduced definitions and relations more explicit, we focus on Markov diffusion processes.  In Sec. \ref{Sec:OU}  as a  special case,  we  consider a stationary bidimensional Ornstein-Uhlenbeck process with additive noises for which a full analytical study can be carried out. This allows us to  illustrate on a simple example the ability of the various information measures to infer a directional coupling in the underlying dynamics. Finally, in Sec. \ref{sec:Non-Markov}, we generalize the formalism to a class of non-Markovian processes and apply it to  cellular processes in the metabolism of  {\it E. coli}. This complements the previous experimental and theoretical investigations of Refs. \cite{K2014, LNTRL2017}.  A brief summary is given in Sec. VII and some demonstrations and technical details are presented in Appendices.
\section{Setup and brief reminder of the bipartite case}

\label{sec:Remind}

\subsection{Setup} 

\label{subsec:Def}

We are interested in the information and thermodynamic exchanges between two subsystems, denoted by $X$ and $Y$, of a stochastic system ${\bf Z}$ whose microscopic states at time $t$ are denoted by ${\bf Z}_t=(X_t,Y_t)$.  The random variables $X$ and $Y$ may be multivariate ($X$ and $Y$ are then vectors), continuous or discrete, and live in arbitrary, and not necessarily identical, spaces. In the following, the full process  ${\bf Z}_t$ can be Markovian or non-Markovian, but even in the former case the individual dynamics of $X_{t}$ and $Y_{t}$, viewed as coarse-grained descriptions of ${\bf Z}_t$, are in general non-Markovian.
 
When  $ {\bf Z}_t$ is a continuous-time Markovian process ~\cite{R1989,G2004,C2005,EK2005,RW2005}, which may involve a combination of drift, diffusion, and jump, the building block of its description is the transition probability $P(\mathbf{Z}_{t}=\mathbf{z}\vert \mathbf{Z}_{t'}=\mathbf{z'})$, for $t'\leq t$ and all $\bf z,\bf{z'}$.  Such an object is generated by a kernel $L_t({\bf z},{\bf z}')$, called the  Markovian generator, according to the forward Kolmogorov equation, 
\begin{equation}
\frac{d}{dt}P(\mathbf{Z}_{t}=\mathbf{z}\vert \mathbf{Z}_{t'}=\mathbf{z'})=\int d\mathbf{z''}P(\mathbf{Z}_{t}=\mathbf{z}''\vert \mathbf{Z}_{t'}=\mathbf{z'})L_{t}\left(\mathbf{z''},\mathbf{z}\right),
\label{KFE}
\end{equation}  
where $d\mathbf{z''}$ is the appropriate measure for either continuous or discrete space. For pure jump processes in a discrete space the Markovian generator is a matrix  involving the transition rates $W_{t}\left(\bf{z},\bf{z'}\right)$ (where by convention the transition is  from $\bf{z}$ to $\bf{z'}$) 
\begin{equation}
L_{t}\left({\bf z},{\bf z'}\right)=\delta_{{\bf z}\neq {\bf z'}}W_{t}\left({\bf z},{\bf z'}\right)-\delta_{{\bf z}={\bf z'}}\sum_{{\bf z''}\ne {\bf z}}W_{t}\left({\bf z},{\bf z''}\right)\ .
\label{eq:GPS}
\end{equation}
On the other hand, pure diffusion processes in continuous space are usually described by the stochastic differential equations 
\begin{align}
dX_{t}&=F_{X,t}\left(\mathbf{Z}_{t}\right)dt+\sum_j\sigma_{X,j,t}\left(\mathbf{Z}_{t}\right)dW_{j,t}\nonumber\\
dY_{t}&=F_{Y,t}\left(\mathbf{Z}_{t}\right)dt+\sum_j\sigma_{Y,j,t}\left(\mathbf{Z}_{t}\right)dW_{j,t}\ ,
\label{eq:BDP}
\end{align}
where $F_{X,t}$, $F_{Y,t}$, $\sigma_{X,j,t}$, and $\sigma_{Y,j,t}$  are time-dependent vector fields, and the $W_{j,t}$'s are independent Brownian motions. The non-negative covariance (diffusion) matrix $D_{t}\left(\mathbf{z}\right)$ is the $2\times 2$ block matrix with components 
\begin{align}
D_{XX,t}\left(\mathbf{z}\right)&\equiv\frac{1}{2}\sum_{j}\sigma_{X,j,t}\otimes\sigma_{X,j,t}(\mathbf{z})\,, \: D_{YY,t}\left(\mathbf{z}\right)\equiv\frac{1}{2}\sum_{j}\sigma_{Y,j,t}\otimes\sigma_{Y,j,t}\left(\mathbf{z}\right)\,,\nonumber\\
D_{XY,t}\left(\mathbf{z}\right)&=D_{YX,t}({\bf z})^{T}\equiv \frac{1}{2}\sum_{j}\sigma_{X,j,t}\otimes\sigma_{Y,j,t}\left(\mathbf{z}\right)\, ,
\label{CM'}
\end{align}
where the symbol  $\otimes$ applied to two vectors $U$ and $V$ means the matrix construction $(U\otimes V)^{ij}\equiv U^{i}V^{j}$. The associated Markovian generator is then obtained as  
\begin{equation}
L_{t}({\bf z},{\bf z'})=L^{FP}_{t}({\bf z'})\left[\delta({\bf z}-{\bf z'})\right]\ , 
\label{eq:GPD}
\end{equation}
where $L_{t}^{FP}$  is the (Fokker-Planck) second-order differential operator
\begin{equation}
L_{t}^{FP}\left(\mathbf{z}\right)=-\nabla_{x}\circ F_{X,t}\left(\mathbf{z}\right)-\nabla_{y}\circ F_{Y,t}\left(\mathbf{z}\right)+\nabla_{x}\circ \nabla_{x}\circ D_{XX,t}\left(\mathbf{z}\right)+\nabla_{y}\circ \nabla_{y}\circ D_{YY,t}\left(\mathbf{z}\right)+2\nabla_{x}\circ\nabla_{y}\circ D_{XY,t}\left(\mathbf{z}\right)
\label{eq:GFP}
\end{equation}
obtained by interpreting Eqs. (\ref{eq:BDP}) with  Ito convention. (In the above expression the last term is {\it a priori} ambiguous because $ D_{XY,t}$ is not necessarily symmetric. The notation  $\nabla_{x}\circ\nabla_{y}\circ D_{XY,t} $ should thus be interpreted as $\nabla_{x_{i}}\circ\nabla_{y_{j}}\circ (D_{XY,t})^{i,j} $, using  Einstein summation convention for repeated indices.) As usual, one can introduce  the probability currents
\begin{align}
J_{X,t}({\bf z}) &\equiv F_{X,t}({\bf z})P_t({\bf z})-\nabla_{x}[D_{XX,t}({\bf z})P_t({\bf z})]-\nabla_{y}[D_{XY,t}({\bf z})P_t({\bf z})] \nonumber\\
J_{Y,t}({\bf z}) &\equiv F_{Y,t}({\bf z})P_t({\bf z})-\nabla_{x}[D_{YX,t}({\bf z})P_t({\bf z})]-\nabla_{y}[D_{YY,t}({\bf z})P_t({\bf z})]\ ,
\label{eq:CourrantDiffT-1}
\end{align}
and recast the Fokker-Planck (FP) equation as the continuity equation
\begin{align}
\label{EqFP1}
\partial_tP_t({\bf z})+\nabla_{x} J_{X,t}({\bf z})+\nabla_{y} J_{Y,t}({\bf z})=0\ .
\end{align}
Note that the currents are defined up to a divergence-free vector. We will use the definitions in Eq. (\ref{eq:CourrantDiffT-1}) in the following, which means that correction terms must be added  in all expressions involving the currents if another decomposition is adopted.  

In continuous time, the Markov process ${\bf Z}$ is called {\it bipartite} if the transition probability $P({\bf Z}_{t+h}\vert {\bf Z}_t)$ satisfies the property
\begin{align}
\label{EqBip1}
P({\bf Z}_{t+h}\vert {\bf Z}_t)=P(X_{t+h}\vert {\bf Z}_t)P(Y_{t+h}\vert {\bf Z}_t)+{\rm O}(h^{2}), 
\end{align}
when $h \to 0^+$. From the forward Kolmogorov equation (\ref{KFE}), the above condition is equivalent to assuming that the Markovian generator $L_t({\bf z},{\bf z}')$ can be written as  
\begin{align}
L_t({\bf z},{\bf z}')=\delta(y-y')L_{t,y}(x,x')+\delta (x-x')L_{t,x}(y,y')\ ,
\label{BipG}
\end{align}
where $L_{t,y}(x,x')$ and $L_{t,x}(y,y')$ are called partial generators. The delta functions become  Kronecker matrices in the case of discrete space. The  partial generators must individually satisfy conservation of probability, {\it i.e.},
$\int dx' L_{t,y}(x,x')=\int dy' L_{t,x}(y,y')=0$.
In particular, a pure jump process is bipartite if the transition rates have the additive form 
$W_{t}[\mathbf z,\mathbf{z'}]=\delta_{y=y'}W_{t,y}(x,x')+\delta_{x=x'}W_{t,x}(y,y')$, 
which implies Eq. (\ref{BipG}), as can be readily checked. On the other hand, a pure diffusion process  is  bipartite if $D_{XY,t}=0$, {\it i.e.}, if the diffusion matrix is block diagonal. From  Eqs. (\ref{CM'}),  a sufficient condition is  that $\sigma_{X,j,t}\otimes\sigma_{Y,j,t}=0$ for all $j$, which means that the overall noises affecting $X_t$ and $Y_t$  are independent.

\subsection{Definition of information measures } 

We start our reminder of information thermodynamics by recalling the definitions of several information-theoretic measures that are  usually introduced in this framework.  As already stressed, a consequence of the abandon of the bipartite assumption will be a proliferation of information measures. It is thus desirable to use transparent notations as much as possible. (Already in the bipartite case, a given quantity may have different names or be defined with different signs, which is a source of confusion.)  It is also important to clearly state under which condition a relation is valid: in the following, the capital letter {\bf M} on the left of an equation indicates that the joint process  is Markovian, the capital letter {\bf B} indicates that the process is Markovian and bipartite, and the capital letter {\bf S} indicates that the process is stationary.
\\

\textit{1. Information flows, aka learning rates}
\\

Information flows quantify how the dynamical evolution of $X_t$ or $Y_t$ contributes to the change in the mutual information, $I(X_t:Y_t)\equiv \langle \ln (P({\bf Z}_t)/[P(X_t)P(Y_t)]\rangle$, where $P({\bf Z}_t)$ is the joint probability distribution and $P(X_t)$, $P(Y_t)$ its marginals. The latter characterizes the instantaneous correlation between $X$ and $Y$ at time $t$.  These information-theoretic measures were first considered in the context of interacting diffusion processes~\cite{AJM2009} and subsequently introduced in the analysis of the thermodynamics of continuously-coupled, discrete-space stochastic systems~\cite{HE2014, HBS2014,SS2015}.  Consider for instance the dynamical evolution of $X_t$. Introducing the time-shifted mutual information $I(X_{t+h}:Y_t)$ (with $h>0$) and taking the limit $h\to 0^+$, one  then defines~\cite{AJM2009} 
\begin{align}
\label{Iflowp}
l_X(t)&\equiv\lim_{h\rightarrow 0^{+}}\frac{1}{h} [I(X_{t+h}:Y_t)-I(X_t:Y_t)]
=\lim_{h\rightarrow 0^{+}}\frac{1}{h} \Bigl\langle \ln \frac{P(Y_t\vert X_{t+h})}{P(Y_t\vert X_t)} \Bigr\rangle,
\end{align}
where here and in the following we use the bracket symbol for an expectation. Similarly,  $l_Y(t)$ is defined from $I(Y_{t+h}:X_t)$. (For brevity, it will be  implicit in the following that similar quantities can be defined by exchanging $X$ and $Y$.) One could also introduce  Shannon entropies instead of  mutual informations, by using $I(X_t:Y_t)=H(X_t)-H(X_t\vert Y_t)=H(Y_t)-H(Y_t\vert X_t)$, with $H(X_t)\equiv -\langle \ln P(X_t)\rangle$ and $H(X_t\vert Y_t)\equiv -\langle \ln P(X_t\vert Y_t)\rangle$~\cite{CT2006}; however, we will try to avoid too many equivalent formulations throughout the paper.  Note that the definition (\ref{Iflowp}) is not restricted to a steady state. In a steady state the information flow identifies with the so-called {\it learning rate} $l_X$ defined in  \cite{BHS2014,HBS2016}. Hereafter, we will use the denomination learning rate for $l_X(t)$ whether or not the system is in a steady state~\cite{note00}. 
 
 As discussed in \cite{AJM2009,HE2014}, learning rates  have a clear meaning: For instance,  $l_X(t)>0$ reveals that the dynamical evolution of  $X$  increases the mutual information $I(X_t:Y_t)$. In other words,  the future of $X$ is more predictable than its present from the viewpoint of $Y$~\cite{AJM2009}, or $X$ is ``learning about" $Y$ through its dynamics~\cite{HE2014}. 

For a bipartite Markov process, one has the natural decomposition of the time derivative of $I(X_t:Y_t)$~\cite{AJM2009,HE2014,MS2018} 
\begin{align}
\label{EqderivI}
({\bf B})\:\:\:\: d_tI(X_t:Y_t)=l_X(t)+l_Y(t)\ ,
\end{align}
as will be explicitly illustrated below for Markov processes.
\\

\textit{2. Transfer entropy}
\\

Transfer entropy (TE) is an information-theoretic  measure that is used to assess directional dependencies between time series and possibly infer causal interactions~\cite{S2000,PKHS2001}.  Instead of $I(X_t:Y_t)$, one  considers the change in the mutual information between stochastic trajectories observed during some time interval, say from $0$ to $t$, and which are denoted by $X_0^t$ and $Y_0^t$ hereafter.  Specifically, we define the TE rate from $X$ to $Y$  in continuous time as
\begin{align}
\label{EqTErateXY}
 {\cal T}_{X\to Y}(t)&\equiv \lim_{h\rightarrow0^+}\frac 1{h}[I(X_0^t:Y_0^{t+h})-I(X_0^t:Y_0^t)] =\lim_{h\rightarrow0^+}\frac{1}{h}\:I(X_0^t:Y_{t+h}\vert Y_0^t)\nonumber\\
&=\lim_{h\rightarrow0^+}\frac{1}{h} \Bigl\langle \ln \frac{P(Y_{t+h}\vert X_0^t,Y_0^t)}{P(Y_{t+h}\vert Y_0^t)}\Bigr\rangle\ .
\end{align}
where we have assumed that $Y_0^{t+h}\sim(Y_0^t,Y_{t+h})$ for $h$ infinitesimal and used the chain rule for mutual information, $I(A:\{B,C\})=I(A:C)+I(A:B\vert C)$, where $I(A:B\vert C)$ is a conditional mutual information~\cite{CT2006}. Like the learning rate, ${\cal T}_{X\to Y}(t)$ has a clear interpretation in terms of information transfer: It quantifies how much the  knowledge of the trajectory $X_0^t$ reduces the uncertainty about $Y_{t+h}$ (for $h$ infinitesimal) when the trajectory $Y_0^t$ is already known.  As a conditional mutual information, ${\cal T}_{X\to Y}(t) $ is a non-negative quantity, whereas $l_X(t)$ has no definite sign.  Note that the present definition  is more general than the one adopted  in \cite{HBS2016} or \cite{MS2018} since we do not assume at this stage that the joint process is Markovian.  When the joint process is Markovian, one has $P(Y_{t+h}\vert X_0^t,Y_0^t)=P(Y_{t+h}\vert X_0^t,Y_t)=P(Y_{t+h}\vert X_t,Y_0^t)=P(Y_{t+h}\vert X_t,Y_t)$, and, after some manipulations, Eq. (\ref{EqTErateXY}) can be rewritten as
\begin{align}
\label{EqMTEnew}
({\bf M})\:\:\:\: {\cal T}_{X \to Y}(t)&=\lim_{h\rightarrow0^+}\frac{1}{h}\:I(X_t:Y_{t+h}\vert Y_0^t) =\lim_{h\rightarrow0^+}\frac{1}{h}\: [I(X_t:Y_0^{t+h})-I(X_t:Y_0^t)]\ ,
\end{align}
which clearly shows the difference with the learning rate  $l_Y(t)=\lim_{h\to 0^+} (1/h)[I(X_t:Y_{t+h})-I(X_t:Y_t)]$. Note that the original definition of transfer entropy in discrete time is even more general since  the number of time bins in the past of $X_t$ and $Y_t$ may be different~\cite{S2000}. This definition can also be extended to continuous time~\cite{SLP2016,BBHL2016,SL2018}. Finally, see Ref. \cite{WKP2013} for a  rigorous definition via a partition of the time interval. 

For a Markov bipartite process, in full analogy with Eq. (\ref{EqderivI}), one has the decomposition
\begin{align}
\label{EqdItraj}
 d_tI(X_0^t:Y_0^t)&\equiv \lim_{h\rightarrow0^+}\frac{1}{h}\: [I(X_0^{t+h}:Y_0^{t+h})-I(X_0^t:Y_0^t)]\nonumber\\
({\bf M})\:\:\:\:&= \lim_{h\rightarrow0^+}\frac{1}{h}\:\Bigl\langle \ln \frac{P(X_{t+h},Y_{t+h}\vert X_t,Y_t)}{P(X_{t+h}\vert X_0^t)P(Y_{t+h}\vert Y_0^t)}\Bigr\rangle\nonumber\\
({\bf B})\:\:\:\: &={\cal T}_{X \to Y}(t)+ {\cal T}_{Y \to X}(t)\ ,
\end{align}
where we have used Eq. (\ref{EqBip1}) and assumed ${\bf Z}_0^{t+h}\sim({\bf Z}_0^t,{\bf Z}_{t+h})$ for $h$ infinitesimal. Let us stress that the trajectory mutual information $I(X_0^t:Y_0^t)$ is a time-extensive quantity, in contrast with $I(X_t:Y_t)$. As a consequence, $d_tI(X_0^t:Y_0^t)$ does not vanish in a steady state and, then, ${\cal T}_{X \to Y}\ne -{\cal T}_{Y \to X}$, while $l_X=-l_Y$.  Here and throughout the paper, quantities without explicit time-dependence refer to a steady state.

Although this is rarely evoked in the stochastic thermodynamic literature, we recall that transfer entropy is essentially a non-linear extension of Granger causality (GC)~\cite{G1969}, which is a concept widely used in econometrics and neuroscience for analyzing the relationships between time series and inferring causal interactions (see {\it e.g.} \cite{AM2013} of a review). The general issue is whether the knowledge of one of the variables can help forecast another one.  In contrast with transfer entropy, GC is usually identified with a model-based viewpoint, for instance a vector autoregressive  modeling of the time series data~\cite{L2005}. 
It turns out that linear GC and transfer entropy rate are fully equivalent when the variables are Gaussian distributed, with a simple factor of $2$ relating the two quantities~\cite{note54}. This equivalence, first shown in the case of discrete-time random processes~\cite{BBS2009}, can be extended to the continuous-time version~\cite{BS2017}.

Since the TE rates  are conditioned on whole trajectories, they are very hard  to compute numerically~\cite{note55} and one often replaces  $X_0^t$ and $Y_0^t$ by the states $X_t$ and $Y_t$ at the latest time $t$.  One then defines~\cite{AJM2009,HBS2014,HBS2016,MS2018}
\begin{align}
 \overline {\cal T}_{X\to Y}(t)&\equiv\lim_{h\rightarrow0^+}\frac{1}{h}\:I(X_t:Y_{t+h}\vert Y_t)=\lim_{h\rightarrow0^+}\frac{1}{h}\:\Bigl\langle \ln \frac{P(Y_{t+h}\vert X_t,Y_t)}{P(Y_{t+h}\vert Y_t)}\Bigr\rangle\ ,
\end{align}
 which is called a ``single-time-step" TE rate in \cite{MS2018} to contrast with the ``multi-time-step" TE rate ${\cal T}_{X\to Y}(t)$. We will adopt this terminology hereafter.

\subsection{Inequalities and sufficient statistic}
\label{subsec:INEQ}

For a Markov bipartite process in a steady state, one has the two inequalities~\cite{HBS2014,HBS2016} 
\begin{numcases}{({\bf B}+{\bf S})\:\:\:\:}
  \label{Ineq1}
  {\cal T}_{X\to Y}\le  {\overline {\cal T}}_{X\to Y}\\
   l_Y\le  {\cal T}_{X\to Y}\,,
\label{Ineq2}
  \end{numcases}
and we do not report the demonstration here since more general inequalities will be derived in the next section. The second inequality expresses the intuitive idea that the instantaneous value of $Y$ is less informative about the instantaneous value of $X$ that the whole past trajectory of $Y$. Within the context of a sensory system, where $X_t$ and $Y_t$ denote the states of the  signal and the sensor, respectively, this  prompted the authors of \cite{HBS2016} to introduce a so-called ``sensory capacity" $C_Y= l_{Y}/{\cal T}_{X\to Y}$ as a tool to quantify the performance of the sensor (assuming that $ l_{Y}\ge 0$). In particular, $C_Y$ reaches its maximal value $1$ when inequality (\ref{Ineq2}) is saturated. As discussed in \cite{MS2018}, inequalities (\ref{Ineq1}) and (\ref{Ineq2}) are both saturated when the following condition is satisfied:
\begin{align} 
\label{Eqsufstat1}	
 P(X_t\vert Y_0^t)=P(X_t\vert Y_t)\ , 
\end{align}
which means that ``$Y_t$ is a sufficient statistic of $X_t$"~\cite{CT2006} and no more information about $X_t$ is contained in the trajectory $Y_0^t$ than in $Y_t$ alone. Interestingly, such an optimization of information transfer may occur in actual  biological signaling circuits~\cite{HSWIH2016,MS2018}. By construction, this condition is realized by the Kalman-Bucy filter~\cite{KB1961,KSH2000,A2006}, as will be illustrated later on.

As can be expected, things become more complicated when the bipartite assumption is dropped, and we show in the next section that this requires introducing additional information-theoretic measures.

\subsection{Entropy production and second law}
\label{subsec:SecLaw}

While the  conventional second law of thermodynamics deals with the irreversibility of the whole  process ${\bf Z}_t$, information measures can be used to formulate modified versions of the second law (which may then be called ``second laws of  information thermodynamics '') that assess the irreversibility of one subsystem alone, say $X_t$, in the presence of the coupling with the other subsystem. The key quantity is the (fixed-time) entropy production rate $\sigma_X(t)$ which is defined by considering $X$ as  an open system and $Y$ as just a fictitious external protocol (or idealized work source)~\cite{SU2012}. On general grounds (see, {\it e.g.}, \cite{VdBE2015}),  $\sigma_X(t)$ can be decomposed as  
\begin{align}
\label{Eqsigma12}
\sigma_X(t)\equiv d_tS_X(t)+ \sigma_X^B(t)\ ,
\end{align}
where $d_tS_X(t)$, the time derivative of the marginal Shannon entropy $S_X(t)=-k_B\int dx \: P_t(x)\ln P_t(x)$, is the  rate of change of the entropy of $X$, and $\sigma_X^B(t)$ is the rate of change of the entropy of the environment or the bath. (From now on the Boltzmann constant $k_B$ is set equal to $1$, so that we may use $S$ instead of $H$ as Shannon entropy.)    As is now standard in the framework of stochastic thermodynamics (see, {\it e.g.},  \cite{M2003}),  the cumulative entropy change  $\Sigma_X^B=\int_0^t ds\:  \sigma_X^B(s)$  can be expressed as the mean of the logratio of the probabilities to observe a trajectory in forward and backward ``experiments". As $Y$ is treated as an external protocol, one has
\begin{align}
\label{EqSigmaBX}
({\bf B})\:\:\:\: \Sigma_X^B=\Bigl\langle  \ln \frac{\widehat P_{Y_0^t}(X_0^t\vert X_0)}{\widehat P_{Y_t^0}(X_t^0\vert X_t)} \Bigr\rangle \  ,
\end{align}
where $\widehat P_{Y_0^t}(X_0^t\vert X_0)$  is the  probability of the trajectory of $X$  for a {\it fixed} trajectory of $Y$ and $\widehat P_{Y_t^0}(X_t^0\vert X_t)$ is the corresponding probability of the time-reversed trajectory $X_t^0$ for the fixed time-reversed trajectory $Y_t^0$~\cite{footnote_probdetached}. For a bipartite pure jump process, $\sigma_X^B(t)$ is  then given by
\begin{align}
\label{Eqsigma1Bjump}\sigma_{X}^{B}(t)\equiv\sum_{\mathbf{z},x'}P_{t}\left(\mathbf{z}\right)W_{t,y}(x,x')\ln \frac{W_{t,y}(x,x')}{W_{t,y}(x',x)}\,,
\end{align}
whereas for a bipartite diffusion process it is equal to 
\begin{align}
\label{EqsigmaBdiffus}
\sigma_X^B(t)\equiv \int d{\bf z} \,D_{XX,t}({\bf z})^{-1} J_{X,t}({\bf z})\widehat F_{X,t}({\bf z})\ ,
\end{align}
where the diffusion matrix $D_{XX,t}>0$ and the probability current $J_{X,t}$ have been defined above, and $\widehat F_{X,t}({\bf z})$ is the modified drift defined by $\widehat{F}_{X,t}\left(\mathbf{z}\right)\equiv F_{X,t}\left(\mathbf{z}\right)-\nabla_{x}.D_{XX,t}({\bf z})$~\cite{CG2008}.
In cases where the thermodynamics of subsystem $X$ can be defined and the environment is a single thermal bath at a given inverse temperature $\beta$, $\sigma_X^B$ identifies with the heat flow $\beta \dot {\cal Q}$ from $X$ to the bath. 

Since the two subsystems are coupled,  $\sigma_X(t)$ may become negative, but a lower bound is provided by including the information shared with $Y$.  For a bipartite Markov process, the various second-law-like inequalities proven in the literature~\cite{AJM2009,IS2013,HE2014,HBS2014,IS2015}  can be summarized by the following hierarchy of bounds,
\begin{equation}
\label{Eq:second_law}
\sigma_{X}(t) \geq l_{X}(t) \geq d_{t}I(X_{t}:Y_{0}^{t})-\mathcal{T}_{X\rightarrow Y}(t),
\end{equation}
or, for the time-integrated quantities, 
\begin{align}
\label{Eq:second_law1}
\Sigma_X\equiv \int_0^t ds\:  \sigma_X(s) \ge \int_0^t ds\: l_X(s)\ge \Delta I -\int_0^t ds\: {\cal T}_{X\to Y}(s)\ge  \Delta \overline I -\int_0^t ds\: {\cal T}_{X\to Y}(s)\ge \Delta \overline I -\int_0^t ds\: \overline {\cal T}_{X\to Y}(s)\ ,
\end{align}
where $ \Delta I=I(X_t:Y_0^t)-I(X_0,Y_0)$ and  $\Delta \overline I=I(X_t:Y_t)-I(X_0,Y_0)$, with $\Delta I\ge \Delta \overline I$ by marginalization.  
The key  fact  is that the tightest bound is provided by the learning rate.

\section{Information measures for non-bipartite processes}
\label{sec:Info}

\subsection{Learning rates}

We first search for a generalization of  Eq. (\ref{EqderivI}).  The decomposition of $d_tI(X_t:Y_t)$ introduced above in the bipartite case suggests to define the new rate
\begin{align}
\label{Iflowm}
l^-_X(t)&\equiv\lim_{h\rightarrow0^+}\frac{1}{h} [I(X_{t+h}:Y_{t+h})-I(X_t:Y_{t+h})]
\end{align}
in addition to $l_X(t)$. Accordingly,  $l_X(t)$ will be denoted $l^+_X(t)$ hereafter to make the notations more consistent. Indeed, $l^-_X(t)$ can be also written as $l^-_X(t)=-dI(X_{t-h}:Y_t)/dh\vert_{h=0^+}$ whereas $l^+_X(t)=dI(X_{t+h}:Y_t)/dh\vert_{h=0^+}$.  This also suggests to call $l_X^-(t)$ a {\it backward} learning rate, in contrast with the  {\it forward} rate $l^+_X(t)$. We stress that the definition of $l^-_X(t)$ as a derivative has nothing to do with stationarity but simply results from a Taylor expansion in $h$: Indeed, for any continuously differentiable function $f(s,t)$, one has $f(t+h,t+h)-f(t,t+h)=h(\partial_1 f)(t,t)+{\cal O}(h^2)=f(t,t)-f(t-h,t)+{\cal O}(h^2)$. Note also that  $l^-_X(t)\ne dI(X_t:Y_{t+h})/dh\vert_{h=0^+}$.

Thanks to the introduction of $l^-_X(t)$, and of the corresponding $l^-_Y(t)$, Eq. (\ref{EqderivI}) is now replaced by the two relations
\begin{align}
d_tI(X_t:Y_t)&=l^+_X(t)+l^-_Y(t)=l^+_Y(t)+l^-_X(t). 
\label{EqdIdt}
\end{align}

The distinction between the forward and backward learning rates in the general ({\it i.e.}, non-bipartite) case suggests to introduce the symmetric quantities 
\begin{align}
l_X^S(t)&\equiv \frac{1}{2}\:\big[l^+_X(t)+l^-_X(t)\big]\:, l_Y^S(t)\equiv  \frac{1}{2}\:\big[l^+_Y(t)+l^-_Y(t)\big] \ ,
\end{align}
such that Eqs. (\ref{EqdIdt}) now yield
\begin{align}
\label{EqdIdtsym}
d_tI(X_t,Y_t)&=l_X^S(t)+l_Y^S(t)\ .
\end{align}
As will be seen below, these symmetric learning rates play a natural role in the thermodynamics since they vanish when the joint system is at equilibrium, {\it i.e.}, when the condition of detailed balance is satisfied. On the other hand, the non-symmetric rates  vanish when the  processes $X_t$ and $Y_t$ are independent.  Note that we use the adjective ``symmetric" to qualify  $l_{X}^{S}(t)$ because it is the half-sum of the forward and backward learning rates, {\it i.e.}, of right and left derivatives.  There must be no confusion with the time-symmetric, kinetic (path-) observables such a ``frenesy" which also play a significant role in the study of non-equilibrium phenomena~\cite{M2019}.

In the general case, but in a {\it steady state}, there are  only two independent learning rates since $d_t I(X_t:Y_t)=0$ and Eqs. (\ref{EqdIdt}) yield the ``conservation" relations 
\begin{align}
\label{EqMIflows}
({\bf S})\:\:\:\: l^+_X&=-l^-_Y \:\:(\ne -l^+_Y)\nonumber\\
 l^+_Y&=-l^-_X\:\: (\ne -l^+_X)\ ,
\end{align}
and thus 
\begin{align}
({\bf S})\:\:\:\:  l^S_Y=-l^S_X\ .
\end{align}

 A basic feature of  the learning rates is that they can be expressed in terms of the two-point probability distribution $P({\bf z},t;{\bf z}',t')\equiv \langle \delta({\bf Z}_t-{\bf z})\delta({\bf Z}_{t'}-{\bf z}')\rangle$, with ${\bf z}\equiv(x,y)$, and the corresponding marginal distributions. (Hereafter, variables with a prime symbol such as ${\bf z}',x',y'$  will always refer to a time $t'\le t$ in the two-point probability distribution functions.) Starting from the definitions (\ref{Iflowp}) and (\ref{Iflowm}), and using the normalization condition $\int dx\:dy'\:P(x,t;y',t')=1$,  we get
\begin{subequations}
\label{Eqexpflow1:subeqns}
\begin{align}
l^+_X(t)&=\int dx\:dy'\:\frac{d}{dh}P(x,t+h;y',t)\vert_{h=0^+} \ln \frac{P_t(x,y')}{P_t(x)P_t(y')}
\label{Eqexpflow1:subeq1}\\
l^-_X(t)&=-\int dx'\:dy\:\frac{d}{dh}P(y,t; x',t-h)\vert_{h=0^+} \ln \frac{P_t(x',y)}{P_t(x')P_t(y)}\ ,
\label{Eqexpflow1:subeq2}
\end{align}
\end{subequations}
where we have used the notation $P_t(x,y)\equiv P(x,t;y,t)$  for the joint probability distribution at the same time $t$ (and $P_t(x)$, $P_t(y)$ for the marginal distributions); similar expressions are obtained for $l^+_Y(t)$ and $l^-_Y(t)$. (We recall that we use the same notations for continuous and discrete spaces. In the latter case, integrals must be replaced by sums.) From these equations, we readily see that the learning rates vanish when $P_t({\bf z})=P_t(x) P_t(y)$, which means that the processes $X$ and $Y$ are independent.
We stress  that these formulas are fully general and do not require the joint process ${\bf Z}$ to be Markovian. However,   further simplifications occur in the Markovian case,  as one can replace the derivative with respect to $h$ by using the Kolmogorov equation and introducing the Markovian generator $L_{t}({\bf z'},{\bf z})$,  which leads to
\begin{align}
({\bf M})\:\:\:\:  l_{X}^{+}(t)=\int d{\bf z}\:d{\bf z'}\:P_{t}({\bf z'})L_{t}\left({\bf z'},{\bf z}\right)\ln \frac{P_{t}(x,y')}{P_{t}(x)P_{t}(y')}.
\label{IF+}
\end{align}
Furthermore, by using the decomposition in Eq. (\ref{EqdIdt}) and the expression of the time derivative of the mutual information,
\begin{align}
d_tI_t(X_{t},Y_{t})=\int d{\bf z}\: \partial_tP_t({\bf z}) \ln \frac{P_t({\bf z})}{P_t(x)P_t(y)}\, , \
\label{EqDIM}
\end{align} 
$l_X^-(t)$ is obtained as 
\begin{align}
({\bf M})\:\:\:\:  l_{X}^{-}(t)=\int d{\bf z}\:d{\bf z'}\:P_{t}({\bf z'})L_{t}\left({\bf z'},{\bf z}\right) \ln \frac{P_{t}(x,y)P_{t}(x')}{P_{t}(x',y)P_{t}(x)}.
\label{IF-}
\end{align}
Finally, after using the conservation of probability $\int d{\bf z}L_{t}({\bf z'},{\bf z})=0$,  we obtain the symmetric learning rates as
\begin{align}
({\bf M})\:\:\:\:  l_{X}^{S}(t)=\frac{1}{4}\int d{\bf z}\:d{\bf z'}\:\left [(P_{t}({\bf z'})L_{t}\left({\bf z'},{\bf z}\right)-P_{t}({\bf z})L_{t}\left({\bf z},{\bf z'}\right)\right ]\ln\left(\frac{P_{t}(x,y')P_{t}({\bf z})}{P_{t}(x',y)P_{t}({\bf z}')}\left[\frac{P_{t}(x')}{P_{t}(x)}\right]^{2}\right)\,.
\label{EqlXS1}
\end{align}
From the above expression, one can immediately see that if $Z_{t}$ is an equilibrium process, such that the probability current $P_{t}({\bf z'})L_{t}\left({\bf z'},{\bf z}\right)-P_{t}({\bf z})L_{t}\left({\bf z},{\bf z'}\right)$ vanishes, the symmetric learning rates both vanish. In Sec. \ref{subsec:IflowsDiff}, we will provide more explicit expressions for these learning rates in the case of a Markovian diffusion process. The case of  a Markovian pure jump process in a discrete space is treated in Appendix \ref{sec:AppJump}.

If we now come back to the special situation of a bipartite process, due to the additive form of the Markovian generator [Eq. (\ref{BipG})] and the conservation of probability,  the formulas given in Eqs. (\ref{IF+}) and (\ref{IF-}) coincide and
\begin{align}
({\bf B})\:\:\:\:  l_{X}^{+}(t)=l_{X}^{-}(t)=\int d{\bf z'}dxP_{t}({\bf z'})L_{t,y'}(x',x)\ln \frac{P_{t}(x,y')}{P_{t}(x)P_{t}(y')}.
\label{IFB}
\end{align}
A similar relation holds for $l_{Y}^{+}(t)$ and $l_{Y}^{-}(t)$.  There are thus only two independent learning rates instead of four, and Eq. (\ref{EqdIdt}) then gives back Eq. (\ref{EqderivI}). Note that the present equalities between learning rates differ from those expressed in Eq. (\ref{EqMIflows}). More generally, one must carefully distinguish relations valid for a {\it bipartite} process from those valid for a non-bipartite process in a {\it steady state}~\cite{note1}. Of course, if the joint process is both {\it bipartite and stationary}, Eqs. (\ref{EqMIflows}) and (\ref{IFB}) imply that only one independent learning rate subsists, for instance $l^S_Y=-l^S_X=l^+_Y=l^-_Y=-l^+_X=-l^-_X$.

We conclude this part on the learning rates by briefly discussing their content in terms of information. The various quantities $l_{X}^{+}(t)$, $l_{X}^{-}(t)$, $l_{X}^{S}(t)$, and their counterparts for the $Y$ subsystem, all measure the change of mutual information between $X$ and $Y$ due to different aspects of an infinitesimal dynamical evolution of one subsystem or the other. When both $l_{X}^{+}(t)$ and $l_{X}^{-}(t)$ are strictly positive, and as a direct consequence $l_{X}^{S}(t)>0$, one can plausibly conclude that $X$ is ``learning about" $Y$ through its dynamics. However, the non-bipartite structure of the process allows cases with $l_{X}^{+}(t)>0$ and $l_{X}^{-}(t)<0$, which have no manifest interpretation in the context of learning.

\subsection{Transfer entropy rates}

Transfer entropy rates ${\cal T}_{X\to Y}(t)$ and ${\cal T}_{Y\to X}(t)$ can be defined by the same formulas as in the bipartite case: see Eq. (\ref{EqTErateXY}). They keep the same property of being non-negative and the same meaning as information-theoretic measures. However, whereas the generalization of  the decomposition of the variation of mutual information [Eq. (\ref{EqderivI})] to a non-bipartite process was straightforward, a similar operation for the pathwise mutual information [Eq. (\ref{EqdItraj})] turns out to be problematic. Indeed, using the second line of Eq. (\ref{EqdItraj}), one can write
\begin{align}
d_t I(X_0^t,Y_0^t)= {\cal T}_{X\to Y}(t)+{\cal T}_{Y\to X}(t)+ {\cal T}_{X.Y}(t) \ ,
\end{align}
where  ${\cal T}_{X.Y}(t)\equiv \lim_{h\to 0^+} h^{-1}I(X_{t+h}:Y_{t+h}\vert X_0^t,Y_0^t)$ is a symmetric quantity measuring  the ``instantaneous"  dependence of the two processes (see, {\it  e.g.}, \cite{C2011} for the discrete-time version). However,  there is a serious obstruction, at least for diffusion processes: ${\cal T}_{X.Y}(t)$ is either zero if ${\bf Z}$ is bipartite [cf. Eq. (\ref{EqdItraj}) above] or infinite otherwise! In other words, $d_tI(X_0^t:Y_0^t)$ and thus $I(X_0^t:Y_0^t)$ itself are  infinite for  non-bipartite diffusions.  Indeed, as noticed in \cite{N2016}, $X_t$ and $Y_t$ have a non-zero quadratic variation if the noises are correlated, which results in the singularity of their joint distribution with respect to the product of the corresponding marginals. Such a difficulty does not occur in discrete time, as briefly discussed in note~\cite{note4}, nor in discrete space.

We thus turn our attention to another class of rates which are well-defined  in the non-bipartite case and  will allow us to generalize the important inequality (\ref{Ineq2}). These rates are associated with  the mutual informations $I(X_t:Y_0^t)$ and $I(Y_t:X_0^t)$ that appear in filtering theory~\cite{K1962,LS2001}.  We introduce the TE rate, called ``filtered transfer entropy rate", 
\begin{align}
\label{EqTflitremhat}
 {\widehat {\cal T}}_{X\to Y}(t)&\equiv \lim_{h\rightarrow0^+}\frac{1}{h}\:[I(X_{t+h}:Y_0^{t+h})-I(X_{t+h}:Y_0^t)]=\lim_{h\rightarrow0^+}\frac{1}{h}I(X_{t+h}:Y_{t+h}\vert Y_0^t)\nonumber\\
&=\lim_{h\rightarrow0^+}\frac{1}{h} \Bigl\langle \ln \frac{P(X_{t+h},Y_{t+h}\vert Y_0^t)}{P(X_{t+h}\vert Y_0^t)P(Y_{t+h}\vert Y_0^t)}\Bigr\rangle\ ,
\end{align}
which quantifies how much the prediction of $X_{t+h}$, for $h$ infinitesimal, is improved by knowing $Y_{t+h}$ in addition to the trajectory $Y_0^t$. 
In general there is no simple relation between ${\widehat {\cal T}}_{X\to Y}(t)$ and ${\cal T}_{X\to Y}(t)$, except when the process is Markov bipartite, where
\begin{align}
\label{EqTfiltre5}
({\bf B})\:\:\:\: {\widehat {\cal T}}_{X\to Y}(t)={\cal T}_{X\to Y}(t)\ .
\end{align}
Indeed, from the definitions (\ref{EqTErateXY}) and (\ref{EqTflitremhat}), we have the general equation
\begin{align}
\label{EqTfiltre6}
 {\cal T}_{X\to Y}(t)- {\widehat {\cal T}}_{X\to Y}(t)&= \lim_{h\rightarrow0^+}\frac{1}{h}\:\Bigl\langle \ln \frac{P(Y_{t+h}\vert X_0^t,Y_0^t)}{P(Y_{t+h}\vert Y_0^t,X_{t+h})}\Bigl\rangle \ ,
\end{align}
and it can be proven that the right-hand side of this equation is equal to $0$ when the process is bipartite. The demonstration for jump processes is in Appendix C of \cite{HBS2014} and for diffusion processes it is in given in Appendix \ref{sec:App_proof} of the present paper.

There is of course a single-time-step TE rate corresponding to ${\widehat T}_{X\to Y}(t)$, which is defined as
\begin{align}
 {\widehat {\overline {\cal T}}}_{X\to Y}(t)&\equiv \lim_{h\rightarrow0^+}\frac{1}{h}\:I(X_{t+h}:Y_{t+h}\vert Y_t)\ .
\label{EqTflitrehatbar}
\end{align} 
Then, 
 \begin{align} 
\label{EqDiff3}
 {\overline {\cal T}}_{X \to Y}(t)- {\widehat {\overline {\cal T}}}_{X\to Y}(t)= \lim_{h \to 0^+} \frac{1}{h}\langle \ln \frac{P(Y_{t+h}\vert X_t,Y_t)}{P(Y_{t+h}\vert X_{t+h},Y_t)}\rangle \ ,
   \end{align} 
and
\begin{align}
\label{EqTildeT1}
({\bf B})\:\:\:\:   {\widehat {\overline {\cal T}}}_{X\to Y}(t)= {\overline {\cal T}}_{X\to Y}(t)
\end{align}
in the bipartite case, as will be illustrated below for diffusion processes [see Eq. (\ref{EqThatmoins})].
 
As for the  learning rates,  the single-time-step TE rates can be expressed in terms of the two-point probability distribution $P({\bf z},t;{\bf z}',t')\equiv \langle \delta({\bf Z}_t-{\bf z})\delta({\bf Z}_{t'}-{\bf z}')\rangle$, with ${\bf z}\equiv(x,y)$, and the corresponding  marginal distributions:   
\begin{subequations}
\label{EqTXYhatbar:subeqns}
\begin{align}
{\overline {\cal T}}_{X\to Y}(t)&=\lim_{h\rightarrow0^+}\frac{1}{h}\: \int dy\: d{\bf z}'\: P(y,t+h;{\bf z}',t)\ln \frac{P(y,t+h\vert {\bf z}',t)}{P(y,t+h\vert y',t)}
\label{EqTXYhatbar:subeq1}\\
{\widehat {\overline {\cal T}}}_{X\to Y}(t)&=\lim_{h\rightarrow0^+}\frac{1}{h}\: \int d{\bf z}\: dy'\: P({\bf z},t+h;y',t)\ln \frac{P({\bf z},t+h\vert y',t)}{P(x,t+h\vert y',t)P(y,t+h\vert y',t)}\ .
\label{EqTXYhatbar:subeq2}
\end{align}
\end{subequations}
In contrast with the learning rates [Eqs. (\ref{Eqexpflow1:subeqns})],  one cannot generally introduce derivatives with respect to $h$ in these expressions. On the other hand, we  shall derive explicit expressions for Markov diffusion processes: see Sec. \ref{subsec:SSPD} below.

\subsection{Backward transfer entropy rates }

\label{Subsec:BTE}

Finally, we add to our list of  information-theoretic measures another TE rate which can be used to assess the directionality of information transfer  (see Sec. \ref{sec:Kiviet}) and which will  play an important role in the generalization of the second law (see Sec. \ref{sec:StochMech}). 
To this aim, we slightly change our notations by assuming that the trajectories  of $X$ and $Y$ are now observed in the time interval $[0,T]$. We then define  
\begin{align} 
\label{EqTEback1}
  {\cal T}^\dag_{X\to Y}(t)& \equiv \lim_{h \to 0^+} \frac 1{h}[I(X_{t+h}^T:Y_t^{T})-I(X_{t+h}^T:Y_{t+h}^T)] = \lim_{h \to 0^+} \frac{1}{h} I(X_{t+h}^T:Y_t\vert Y_{t+h}^T) \nonumber\\&
  =\lim_{h \to 0^+} \frac{1}{h} \Bigl\langle \ln \frac{P(Y_t\vert X_{t+h}^T,Y_{t+h}^T)}{P(Y_t\vert Y_{t+h}^T)}\Bigr\rangle \ ,
   \end{align} 
where $X_{t+h}^T$ and $Y_{t+h}^T$  denote the trajectories of $X$  and $Y$ in the time interval $[t+h,T]$.  ${\cal T}^\dag_{X\to Y}(t)$ has clearly the meaning and the properties of a TE rate, but it involves the {\it future} trajectories of $X$ and  $Y$ instead of their past. It may thus be called a {\it backward} TE  (BTE) rate and regarded as the continuous-time version of the BTE introduced in \cite{I2016} in the discrete-time framework (see also \cite{HNMN2013,V2015,W2016} for the introduction of time-reversed Granger causality).  It is actually much simpler to consider continuous time from the outset as this  makes the generalization  of the relations derived  in \cite{I2016} to the non-bipartite case a straightforward operation. We draw attention to the fact  that the BTE defined  in this way has no relation with the  time-reversed  transfer entropy considered in \cite{CS2016,SLP2016,SLP2018}.

When ${\bf Z}$ is a Markov process,  the definition (\ref{EqTEback1}) can be also rewritten as
 \begin{align} 
\label{EqTEback2}
({\bf M})\:\:\:\:  {\cal T}^\dag_{X\to Y}(t)& \equiv \lim_{h \to 0^+} \frac{1}{h} \Bigl\langle \ln \frac{P(Y_t\vert X_{t+h},Y_{t+h})}{P(Y_t\vert Y_{t+h}^T)}\Bigr\rangle \ .
 \end{align} 
As before, we  also define a  single-time-step BTE  rate as 
\begin{align} 
\label{EqsTEback}
 {\overline {\cal T}}^\dag_{X\to Y}(t)&\equiv \lim_{h \to 0} \frac{1}{h} I(X_{t+h}:Y_t\vert Y_{t+h}) =\lim_{h \to 0} \frac{1}{h} \Bigl\langle \ln \frac{P(Y_t\vert X_{t+h},Y_{t+h})}{P(Y_t\vert Y_{t+h})}\Bigr\rangle \ ,
  \end{align}
which will play a useful role in Sec. \ref{sec:StochMech}~\cite{note5}. Note however that this does not add a new independent measure of information to our list since it can be easily seen from the definitions (\ref{Iflowm} ) of $l_Y^-(t)$ and (\ref{EqTflitrehatbar}) of $\widehat{\overline {\cal T}}_{X\to Y}(t)$ that
 \begin{align} 
\label{EqsTEback1}
 {\overline {\cal T}}^\dag_{X\to Y}(t)=\widehat{\overline {\cal T}}_{X\to Y}(t)-l_Y^-(t)\ . 
  \end{align}
Combining Eq. (\ref{EqsTEback1}) with Eq. (\ref{EqDiff3})  also yields
  \begin{align} 
\label{EqDiff7}
 {\overline {\cal T}}_{X \to Y}(t)-{\overline {\cal T}}^\dag_{X\to Y}(t)=l_Y^-(t) + \lim_{h \to 0^+} \frac{1}{h} \Bigl\langle \ln \frac{P(Y_{t+h}\vert X_t,Y_t)}{P(Y_{t+h}\vert X_{t+h},Y_t)}\Bigr\rangle \ ,
   \end{align} 
which, in the bipartite case, implies that 
\begin{align} 
\label{EqDiff8}
({\bf B})\:\:\:\:  {\overline {\cal T}}_{X \to Y}(t)-{\overline {\cal T}}^\dag_{X\to Y}(t)= l_Y(t). 
   \end{align} 
 This is in  agreement with Eq. (18)  in \cite{I2016,note30}.  

Furthermore, when the joint process $\bf Z_t$ is Markovian, simple manipulations yield
\begin{align}
\label{EqDiff2}
({\bf M})\:\:\:\:  [{\cal T}_{X\to Y}(t)- {\cal T}^\dag_{X\to Y}(t)]-[{\overline {\cal T}}_{X\to Y}(t)- {\overline {\cal T}}^\dag_{X\to Y}(t)]&=\lim_{h \to 0^+} \frac{1}{h} \Bigl\langle \ln \frac{P(Y_t\vert Y_{t+h}^T)}{P(Y_{t+h}\vert Y_0^t)} \frac{P(Y_{t+h})}{P(Y_t)}\Bigr\rangle \nonumber\\
 &=\frac{d}{dt}[S(Y_0^t)+S(Y_t^T)-S(Y_t)]\ ,
   \end{align} 
where we have used $Y_0^{t+h}\sim (Y_0^t,Y_{t+h})$ and $Y_t^T\sim (Y_t,Y_{t+h}^T) $ for $h$ infinitesimal to obtain the second equality.  By integrating from $0$ to $T$, the left-hand side of this equation vanishes since $[S(Y_0^t)+S(Y_t^T)-S(Y_t)]_0^T=S(Y_0^T)-S(Y_0)+S(Y_T)-S(Y_0^T)-S(Y_T)+S(Y_0)=0$.We finally obtain a simple relation (not evident from the outset) involving  the  forward and backward time-integrated TEs,
\begin{align}
\label{EqDiff5}
({\bf M})\:\:\:\:  \int_0^T dt \:[ {\cal T}_{X\to Y}(t)- {\cal T}^\dag_{X\to Y}(t)]=\int_0^T dt\: [{\overline {\cal T}}_{X\to Y}(t)- {\overline {\cal T}}^\dag_{X\to Y}(t)]\ .
   \end{align} 
Although this may be regarded as the continuous-time analog of Eq.  (17) in \cite{I2016}, we stress that the derivation of this relation does {\it not} require the joint process to be bipartite.  
Eq. (\ref{EqDiff5}) also leads to the interesting steady-state relation~\cite{note15}  
\begin{align}
\label{EqDiff4}
({\bf M}+ {\bf S})\:\:\:\:  {\cal T}_{X\to Y}- {\cal T}^\dag_{X\to Y}={\overline {\cal T}}_{X\to Y}- {\overline {\cal T}}^\dag_{X\to Y}\ .
 \end{align}

\subsection{Inequalities}
\label{Subsec:ineq}

We are now in position to  generalize the standard inequalities (\ref{Ineq1}) and (\ref{Ineq2}) obtained in the bipartite case.

1) First, one easily obtains from the definitions that the single-time-step TE rate is an upper bound on the forward learning rate,
 \begin{align} 
 \label{EqIneq1}	
l^+_Y(t)\le  {\overline {\cal T}}_{X\to Y}(t)\ ,
\end{align}
and that the single-time-step filtered TE rate is an upper bound on the backward learning rate,
\begin{align} 
\label{EqIneq5}	
 l^-_Y(t)\le {\widehat {\overline T}}_{X\to Y}(t)\ . 
\end{align}
Indeed,  one has ${\overline T}_{X\to Y}(t)-l^+_Y(t)=\lim_{h\rightarrow0^+}h^{-1}[S(X_t\vert Y_{t+h})-S(X_t\vert Y_t,Y_{t+h})]$ and $ {\widehat {\overline {\cal T}}}_{X\to Y}(t)-l^-_Y(t)=\lim_{h\rightarrow0^+}h^{-1}[S(X_{t+h}\vert Y_{t+h})-S(X_{t+h}\vert Y_t, Y_{t+h})]$, and  Shannon entropy never increases by conditioning. These inequalities hold for a general (non-stationary and possibly non-Markovian) process.

2) If the joint process is Markovian, the single-time-step TE rate is an upper bound on the multi-time-step TE rate,
\begin{align} 	
\label{EqIneq2}
({\bf M})\:\:\:\:\ {\cal T}_{X\to Y}(t)\le  {\overline {\cal T}}_{X\to Y}(t)\ ,
\end{align}
since ${\overline {\cal T}}_{X\to Y}(t)-{\cal T}_{X\to Y}(t)=\lim_{h\rightarrow0^+}h^{-1}[S(Y_{t+h}\vert Y_t)-S(Y_{t+h}\vert Y_0^t)]\ge 0$. 
On the other hand, note that ${\widehat {\overline {\cal T}}}_{X\to Y}(t)$ is {\it not} an upper bound on ${\widehat {\cal T}}_{X\to Y}(t)$.

3) In a steady state, the second inequality (\ref{Ineq2})  is replaced by 
 \begin{align} 
\label{EqIneq3}	
({\bf M}+{\bf S})\:\:\:\:\ l^-_Y\le \widehat {\cal T}_{X\to Y}\ ,
\end{align}
which is a special case of the more general inequality
\begin{align} 
\label{EqIneq4}	
({\bf M})\:\:\:\:\   l^+_X(t)\ge \frac{d}{dt}I(X_t:Y_0^t)-\widehat {\cal T}_{X\to Y}(t) ,
\end{align}
since  $l^+_X=-l^-_Y$ and $I(X_t:Y_0^t)$ is not time extensive, {\it i.e.} $\lim_{t \to \infty} t^{-1} I(X_t:Y_0^t)=0$ [in contrast with $I(X_0^t:Y_0^t)$].  Inequality  (\ref{Ineq2}) is then recovered  for  a bipartite process since $\widehat {\cal T}_{X\to Y}= {\cal T}_{X\to Y}$ in this case, as we have seen before [Eq. (\ref{EqTfiltre5})]. We  prove  the inequality in Eq. (\ref{EqIneq4}) in Appendix \ref{sec:App_proof}.
 \begin{figure}[hbt]
\begin{center}
\includegraphics[width=14cm]{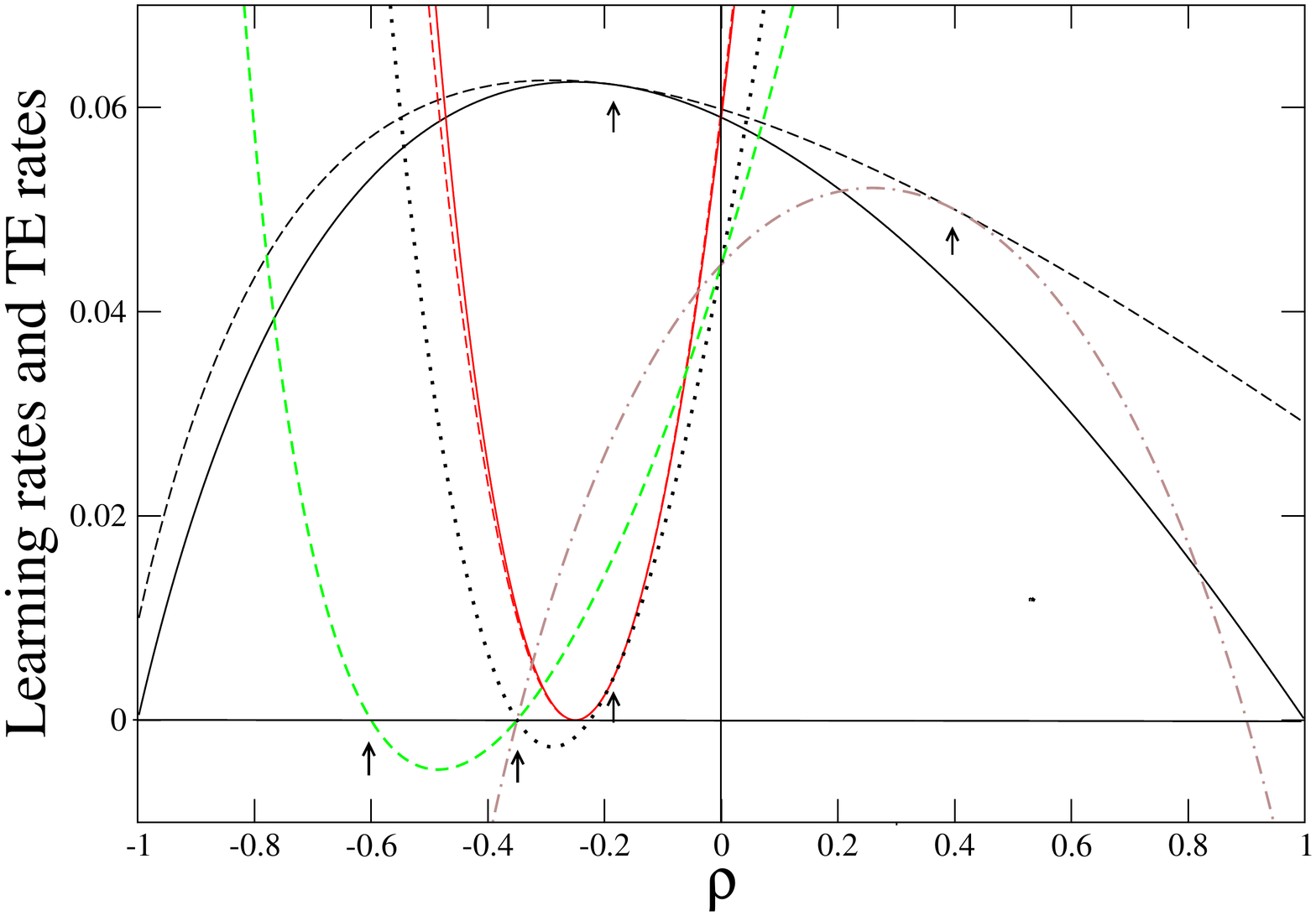}
\caption{\label{Fig1} (Color on line)  Steady-state transfer entropy and learning rates for a stationary bi-dimensional Ornstein-Uhlenbeck process as a function of the parameter $-1\le \rho\le 1$ that  quantifies the  correlations between the noises: $ {\cal T}_{X \to Y}$ (solid black line), $ {\overline  {\cal T}}_{X\to Y}$ (long-dashed black line), $ {\widehat  {\cal T}}_{X\to Y}$ (solid red line), ${\widehat {\overline  {\cal T}}}_{X\to Y}$ (long-dashed red line), $l^+_Y$ (dashed-dotted brown line), $l^-_Y=-l^+_X$ (black dotted line), and $l^S_Y=(l^+_Y+l^-_Y)/2$ (short dashed  green line).  One has $l_Y^+=l_Y^-=l_Y^S$, $\widehat{\cal T}_{X\to Y}= {\cal T}_{X\to Y}$, ${\widehat {\overline  {\cal T}}}_{X\to Y}= {\overline  {\cal T}}_{X\to Y}$ in the bipartite case ($\rho=0$) and $l^+_Y\le {\overline  {\cal T}}_{X\to Y}$, $ l^-_Y\le {\widehat {\overline  {\cal T}}}_{X\to Y}$, $ {\cal T}_{X\to Y}\le {\overline  {\cal T}}_{X\to Y}$, $ l^-_Y\le \widehat  {\cal T}_{X\to Y}$ more generally, as predicted by inequalities (\ref{EqIneq1}-\ref{EqIneq3}) (but ${\widehat {\overline {\cal T}}}_{X\to Y}$ is not an upper bound on ${\widehat {\cal T}}_{X\to Y}$).  Special values of $\rho$ are indicated by little arrows: a)  $\rho=-0.6$: the joint system is at equilibrium, so that $l^S_Y=0$; b) $\rho=-0.35$: the subprocesses  are  independent, so that $l_Y^+=l_X^+=l^S_Y=0$;  c) $\rho\approx -0.185$: $Y_t$ is a sufficient statistic of $X_t$ [Eq. (\ref{Eqsufstat1})], so that inequalities (\ref{EqIneq2})) and (\ref{EqIneq3}) are saturated and ${\widehat {\cal T}}_{X\to Y}={\widehat {\overline {\cal T}}}_{X\to Y}$;  d) $\rho=0.4$:  inequality (\ref{EqIneq1}) is saturated. The model parameters are $a_{11}=1,a_{22}=0.25, a_{12}=-0.05, a_{21}=-0.5$ and $D_{11}=D_{22}=2$.}
\end{center}
\end{figure}

To summarize all these inequalities and be more concrete, let us give a numerical illustration. Anticipating the calculations performed in Sec. \ref{Sec:OU} for a bi-dimensional Ornstein-Uhlenbeck process, we show in Fig. \ref{Fig1} the behavior of the various information measures as a function of the parameter $-1\le \rho\le 1$  that quantifies the  correlations between the noises affecting the two Langevin subprocesses. The numerical values of the other parameters of the model are chosen in such a way that $X$ may be considered as a source signal measured by $Y$ (see the discussion in Sec. \ref{Sec:OU}). 

The small arrows in the figure indicate values of $\rho$ for which the system has a non-generic but remarkable behavior. 

i) The first  one on the left  side ($\rho= -0.6$) indicates that the symmetric learning rate $l^S_Y$ (and thus also $l_X^S=-l_Y^S$) vanishes. As seen from Eq. (\ref{EqlXS1}), this occurs when the joint system is at equilibrium (in the sense of satisfied detailed balance and zero probability currents). 

ii) The next arrow ($\rho= -0.35$) indicates that  $l_Y^+=l_X^+=l^S_Y=0$, which occurs when the two subprocesses $X$ and $Y$ become independent: see Eqs. (\ref{IF+}) and (\ref{IF-}). 

iii) The third arrow ($\rho\approx -0.185$) indicates that   $ {\cal T}_{X\to Y}= {\overline  {\cal T}}_{X\to Y}$ and $ l^-_Y= \widehat  {\cal T}_{X\to Y}={\widehat {\overline {\cal T}}}_{X\to Y}$.  Extending the analysis performed in \cite{MS2018} and taking the continuous-time limit from the outset, we  show in Appendix \ref{sec:AppSuffStat} that this occurs when $Y_t$ is a sufficient statistic of $X_t$, as expressed by Eq. (\ref{Eqsufstat1}). Thanks to Eqs. (\ref{EqsTEback1}) and (\ref{EqDiff4}), this also implies that 
 $ {\cal T}^\dag_{X\to Y}= {\overline {\cal T}}^\dag_{X\to Y}=0$.

iv) Finally, the last arrow on the right side ($\rho= 0.4$) indicates that  inequality (\ref{EqIneq1}) is  saturated and $l^+_Y(t)= {\overline {\cal T}}_{X\to Y}(t)$. This generally does not coincide with the saturation of inequalities (\ref{EqIneq2}) and (\ref{EqIneq3}). Indeed, since ${\overline T}_{X\to Y}(t)-l^+_Y(t)=\partial_h I(X_t:Y_t\vert Y_{t+h})\vert_{h=0^+}$,  equality is obtained when $\lim_{h\rightarrow0^+}h^{-1}\:[ P(X_t\vert Y_t,Y_{t+h})-P(X_t\vert Y_{t+h})] =0$, which differs from condition (\ref{Eqsufstat1}). 
The case of  a bipartite process considered in \cite{MS2018} is an exception, as inequalities (\ref{EqIneq1}) and  (\ref{EqIneq3}) then coincide (${\overline {\cal T}}_{X\to Y}={\widehat {\overline {\cal T}}}_{X\to Y}$ and $l^-_Y=l_Y^+$).

Although inequality  (\ref{EqIneq3}), which appears as the generalization of inequality (\ref{Ineq2}), has no intuitive interpretation [in contrast with (\ref{Ineq2})], the fact that it becomes an equality if $Y_t$ is a sufficient statistic of $X_t$ may suggest to generalize the concept of a sensory capacity as
\begin{align} 
\label{EqSensCap}	
({\bf M}+{\bf S})\:\:\:\: C_Y=\frac{ l^-_Y}{{\widehat {\cal T}}_{X\to Y}}\ .
\end{align}
Likewise,  we may define a ``single-time-step"  capacity,
\begin{align} 
\label{EqssSensCap}	
({\bf S})\:\:\:\: {\overline C}_Y=\frac{l^-_Y}{ {\widehat {\overline {\cal T}}}_{X\to Y}}=\frac{l^-_Y}{l^-_Y+  {\overline {\cal T}}^\dag_{X\to Y}}\ ,
\end{align}
which is also bounded by $1$  thanks to inequality (\ref{EqIneq5}), and  equal to $1$ if $Y_t$ is a sufficient statistic of $X_t$ (see Appendix \ref{sec:AppSuffStat}). Note that the joint dynamics needs not be Markovian. Moreover, since ${\overline C}_Y$ involves the single-time-step TE's ${\overline {\cal T}}_{X\to Y}$  or ${\overline {\cal T}}^\dag_{X\to Y}$ instead of ${\widehat {\cal T}}_{X\to Y}$,  it is much simpler to obtain this quantity from experimental time series than $C_Y$. Of course, it remains to be seen on specific examples of sensory systems  if these quantities  are helpful  to estimate the performance of the sensor in the presence of correlations between the observation and signal noises, a situation classically treated in the framework of filtering theory~\cite{K1962,LS2001}.

\section{Markov diffusion processes and second law of information thermodynamics}

\label{sec:DIF}

To make all the above definitions and relations more explicit and to derive a second law, we now focus on  Markov diffusion processes as defined in Eq. (\ref{eq:BDP}). 
To reduce the amount of notation, we  consider the case where $X_{t}$ and $Y_{t}$  are unidimensional processes.  The vector fields $F_{X,t}$,  $F_{Y,t}$ and  the matrix fields $D_{XX,t}$, $D_{XY,t}$, and $D_{YY,t}$ are now all scalar fields.  The general case can be easily extrapolated from this one.

\subsection{Learning rates}

\label{subsec:IflowsDiff}

As we have already pointed out, the expressions  (\ref{Eqexpflow1:subeqns}) of the learning rates can be simplified when the joint process ${\bf Z}_t$ is Markovian. From the forward Kolmogorov equation (\ref{KFE}) and the Markovian generator, we readily obtain
\begin{align}
\label{EqKFP1}
\frac{d}{dh}P({\bf z},t+h\vert {\bf z}',t)\vert_{h=0^+}&= L_t^{FP}({\bf z}) \delta({\bf z}-{\bf z'})\ 
\end{align}
for $t'\le t$.   Integrating over $y$ and using Eq. (\ref{eq:GFP}) then yields
\begin{align}
\label{EqKFP2}
\frac{d}{dh}P(x,t+h\vert {\bf z}',t)\vert_{h=0^+}&= -\frac{\partial}{\partial x}[F_{X,t}(x,y')\delta(x-x')]+\frac{\partial^2}{\partial x^2}[D_{XX,t}({\bf z})\delta(x-x')]\ ,
\end{align}
as all terms involving derivatives with respect to $y$ vanish at the boundaries (assuming natural boundary conditions). As a result,
\begin{align}
\label{EqKFP3}
&\frac{d}{dh}P(x,t+h;y',t)\vert_{h=0^+}= \int dx'\: \frac{d}{dh}P(x,t+h\vert {\bf z}',t)\vert_{h=0}P_t({\bf z}')\nonumber\\
&= -\frac{\partial}{\partial x}[F_{X,t}(x,y')P_t(x,y')]+\frac{\partial^2}{\partial x^2}[D_{XX,t}({\bf z})P_t(x,y')]\ ,
\end{align}
and, after integration by parts, we transform Eq. (\ref{Eqexpflow1:subeq1}) into
\begin{align}
l^+_X(t)&=\int d{\bf z}\:\left (F_{X,t}({\bf z})P_t({\bf z}) - \partial_{x}[D_{XX,t}({\bf z})P_t({\bf z})]\right ) \partial_{x}\ln P_t(y\vert x)
\label{EqflowD1}
\end{align}
where $P_t(y\vert x)\equiv P_t({\bf z})/P_t(x)$ is the conditional probability distribution function. Since we only consider Markov processes in this section, we no longer add the bold letter ${\bf M}$ on the left of the equations.

The rate $l_X^-(t)$ is obtained by using the relation  $d_tI_t(X_{t},Y_{t})= l_Y^+(t)+l_X^-(t)$ [Eq. (\ref{EqdIdt})],  the expression of the time derivative of the mutual information [Eq. (\ref{EqDIM})], and the FP equation. It reads
\begin{align}
l^-_X(t)&=\int d{\bf z}\:\left (F_{X,t}({\bf z})P_t({\bf z}) - \partial_{x}[D_{XX,t}({\bf z})P_t({\bf z})]- 2\partial_{y}[D_{XY,t}({\bf z})P_t({\bf z})]\right ) \partial_{x}\ln P_t(y\vert x)\ .
\label{EqflowD3}
\end{align}
We immediately see that $l^+_X(t)=l^-_X(t)$ and $l^+_Y(t)=l^-_Y(t)$ when the process is bipartite ({\it i.e.}, $D_{XY,t}({\bf z})=0$) in agreement with Eq. (\ref{IFB}). 

One may also rewrite these expressions in terms of the probability currents defined by Eq. (\ref{eq:CourrantDiffT-1}). This yields
\begin{align}
l^{\pm}_X(t)&=\int d{\bf z}\:\left (J_{X,t}({\bf z}) \pm  \partial_{y}[D_{XY,t}({\bf z})P_t({\bf z})]\right ) \partial_{x}\ln P_t(y\vert x)\,,
\label{EqflowD4}
\end{align}
and the symmetric learning rate $l^S_X(t)=(1/2)[l^+_X(t)+l^-_X(t)]$  is then simply given by
\begin{align}
\label{eq:symefPD}
l^S_X(t)=\int d{\bf z}\:J_{X,t}({\bf z})\partial_{x}\ln P_t(y\vert x)\,.
\end{align}
As announced before, we  observe that the  two symmetric  rates $l^S_X(t)$ and $l^S_Y(t)$ vanish in a steady state when the two probability currents are zero, which corresponds to equilibrium. On the other hand, the non-symmetric rates remain finite.  All the learning rates vanish when the two subprocesses $X$ and $Y$ are independent, {\it i.e.}, when $P_t({\bf z})=P_t(x)P_t(y)$.

The above equations generalize the expressions in the current literature obtained for bipartite processes and additive noises~\cite{AJM2009,HS2014,HBS2016} (note that the sign convention may differ). These expressions are immediately recovered  by setting $D_{XY,t}=0$ and taking  $D_{XX,t}$ and $D_{YY,t}$ independent of $x$ and $y$. Finally, we note that the learning rates obtained above are finite, at least if the integrals in the right-hand sides of (\ref{EqflowD4}) and (\ref{eq:symefPD}) are finite, for all  diffusion processes described by Eq. (\ref{eq:BDP}). This will not necessarily be the case of the  single-time-step transfer entropy rates that we consider in the following.

\subsection{Single-time-step transfer entropy rates}
\label{subsec:SSPD}
 
We now derive the expressions of the various single-time-step TE rates. 
 It turns out that it suffices to compute the rate ${\overline  {\cal T}}_{X\to Y}(t)$ given by  Eq. (\ref{EqTXYhatbar:subeq1}) (a similar calculation was presented in \cite{LNTRL2017}, but it is worth repeating it for completeness). The backward rate ${\overline {\cal T}}^\dag_{X\to Y}(t)$  is then deducible from ${\overline  {\cal T}}_{X\to Y}(t)$, and  ${\widehat {\overline {\cal T}}}_{X\to Y}(t)$ is  finally  obtained from Eq. (\ref{EqsTEback1}).
 
 We start from the expression of the infinitesimal transition probability (or propagator), 
\begin{align}
\label{Eqzzp}
P({\bf z},t+h\vert {\bf z}',t)=\delta({\bf z}-{\bf z}')+h L^{FP}_{t}({\bf z})\left[\delta({\bf z}-{\bf z'})\right]   + {\rm O}(h^2)\ ,
\end{align}
with $L^{FP}_{t}$ the Focker-Planck operator appearing in Eq. (\ref{EqKFP1}) and defined by Eq. (\ref{eq:GFP}).

Integrating the propagator  over $x$ and then integrating $P(y,t+h; {\bf z}',t)$ over $x'$, we  obtain
\begin{align}
P(y,t+h\vert {\bf z}',t)&=\delta(y-y')- h\big[F_{Y,t}({\bf z'}) \frac{\partial}{\partial y}-D_{YY,t}({\bf z'})\frac{\partial^2}{\partial y^2}\big]\delta(y-y')+ {\rm O}(h^2)\nonumber\\
\label{Eqpyyp}
P(y,t+h\vert y',t)&= \delta(y-y')-h\big [{\overline F}_{Y,t}(y')\frac{\partial}{\partial y}-{\overline D}_{YY,t}(y')\frac{\partial^2}{\partial y^2}\big]\delta(y-y')+ {\rm O}(h^2)\ ,
\end{align}
where
\begin{align}
{\overline F}_{Y,t}(y)&\equiv \int dx\: P_t(x\vert y) F_{Y,t}({\bf z})
\end{align}
and 
\begin{align}
{\overline D}_{YY,t}(y)\equiv \int dx\: P_t(x\vert y) D_{YY,t}({\bf z})\ .
\end{align}

To compute the logarithms of the transition probabilities, we need to replace Eqs (\ref{Eqpyyp}) by their Gaussian  small-time expressions,~\cite{R1989}
\begin{align}
P(y,t+h\vert {\bf z}',t)&=\frac{1}{\sqrt{4\pi hD_{YY,t}({\bf z}')}}e^{-\frac{1}{4hD_{YY,t}({\bf z}')}[y-y'-h F_{Y,t}({\bf z}')]^2}\nonumber\\
\label{EqPropa1}
P(y,t+h\vert y',t)&=\frac{1}{\sqrt{4\pi h{\overline D}_{YY,t}(y')}}e^{-\frac{1}{4h{\overline D}_{YY,t}(y')}[y-y'-h {\overline F}_{Y,t}(y')]^2}\,,
\end{align}
when $h\to 0^+$, using the convention that the argument of the functions $D_{YY,t}$  and $F_{Y,t}$ is ${\bf z}'$. Inserting Eq. (\ref{EqPropa1}) into Eq. (\ref{EqTXYhatbar:subeq1}), we readily see that ${\overline{\mathcal T}}_{X\to Y}(t)$ diverges if ${\overline D_{YY,t}}(y) \neq D_{YY,t}({\bf z})$, that is if $D_{YY,t}$ is also a function of $x$. We thus recover the conditions specified in \cite{D1971} for the mutual information $I(X_0^t,Y_0^t)$ to be finite in the case of multiplicative independent noises. To proceed, we thus assume that $D_{YY,t}$ only depends on $y$. Then, 
\begin{align}
\ln \frac{P(y,t+h\vert {\bf z}',t)}{P(y,t+h\vert y',t)}=\frac{1}{2D_{YY,t}(y')}\Big\{(y-y')[F_{Y,t}({\bf z}')-{\overline F}_{Y,t}(y)]-\frac{h}{2}[F_{Y,t}^2({\bf z}')-{\overline F}^2_{Y,t}(y')]\Big\}+ {\rm O}(h^2)\ ,
\end{align}
and using Eq. (\ref{EqTXYhatbar:subeq1}), we finally obtain after some manipulations 
\begin{equation}
{\overline  {\cal T}}_{X\to Y}(t)=
\begin{cases}
\frac{1}{4}\int d{\bf z}\:\frac{P_t({\bf z})}{D_{YY,t}(y)} [F_{Y,t}({\bf z})-{\overline F}_{Y,t}(y)]^{2}\ & \textrm{if   } {D_{YY,t}({\bf z})= D_{YY,t}(y)}  
\\
\infty & \textrm{otherwise}.
\end{cases}
\label{EqTbarXfin}
\end{equation}

This generalizes the expression  given in  \cite{HBS2016} for a bipartite system with additive noises (an explicit calculation is also performed in \cite{IS2015} for a bi-dimensional Ornstein-Uhlenbeck model). Note  that, contrary to the learning rates, the single-time-step TE rate is infinite when $D_{YY,t}=0$ (or in the multi-dimensional case when the matrix $D_{YY,t}$ is not invertible), which implies for instance that it is not well-suited for underdamped processes. The same is true for the other TE rates considered below.

To obtain the expression of the BTE rate ${\overline {\cal T}}^\dag_{X\to Y}(t)$ defined by Eq. (\ref{EqsTEback}),  a possible method is to use Bayes theorem to modify the argument of the logarithm and recast Eq. (\ref{EqsTEback}) as 
\begin{align} 
\label{EqsTEback2}
 {\overline {\cal T}}^\dag_{X\to Y}(t)&=\lim_{h \to 0} \frac{1}{h} \Bigl\langle \ln \frac{P(X_{t+h},Y_{t+h}\vert Y_t)}{P(X_{t+h}\vert Y_{t+h})P(Y_{t+h}\vert Y_t)}\Bigr\rangle \ .
  \end{align}
The calculation would then follow the same lines as above. However, it is  more instructive to use the fact that $ {\overline {\cal T}}^\dag_{X\to Y}(t)$ is the TE rate at time $T-t$ corresponding to the process ${\bf Z}_{T-t}$, which is the time reversal of the process ${\bf Z}_t$ (as the state at time $t$ along a forward trajectory is now conditioned on the state at time $t+h$).  As is well known,  ${\bf Z}_{T-t}$ is also a diffusion process, under some mild conditions (see, {\it e.g.}, \cite{HP1986,C2005}). The covariance (diffusion) matrix and  drift coefficients of the time-reversed process  are given respectively by  $D_{t}^*({\bf z})=D_{T-t}({\bf z})$ and  
\begin{align}
\label{Eqbackdrift}
F^*_{X,t}({\bf z})&=-F_{X,{T}-t}({\bf z})+\frac{2}{P_{T-t}({\bf z})}\Bigl(\partial_{x}[D_{XX,T-t}({\bf z})P_{T-t}({\bf z})]+\partial_{y}[D_{XY,T-t}({\bf z})P_{T-t}({\bf z})]\Bigr)=F_{X,T-t}({\bf z})-2\frac{J_{X,T-t}({\bf z})}{P_{T-t}({\bf z})}\nonumber\\
F^*_{Y,t}({\bf z})&=-F_{Y,{T}-t}({\bf z})+\frac{2}{P_{T-t}({\bf z})}\Bigl(\partial_{x}[D_{YX,T-t}({\bf z})P_{T-t}({\bf z})]+\partial_{y}[D_{YY,T-t}({\bf z})P_{T-t}({\bf z})]\Bigr)=F_{Y,T-t}({\bf z})-2\frac{J_{Y,T-t}({\bf z})}{P_{T-t}({\bf z})}\ .
\end{align}
Denoting the single-time-step TE associated to $D_{t}^*$ and $F^*_{X,t}, F^*_{Y,t}$ by ${\overline {\cal T}}^*_{X \to Y}$, we have by definition of the single-time-step backward TE rate that
\begin{align}
\label{EqHP2}
 {\overline {\cal T}}^\dag_{X\to Y}(t)&={\overline {\cal T}}^*_{X \to Y}(T-t)\ .
\end{align} 
 Of course, the same is true for the  multi-time-step TE rate ${\cal T}^\dag_{Y\to X}(t)$ which identifies with  ${\cal T}^*_{Y \to X}(T-t)$.
 
After some algebra described in Appendix \ref{sec:App_proof}, we obtain
\begin{align}
{\overline {\cal T}}^\dag_{X\to Y}(t)=
 {\overline {\cal T}}_{X\to Y}(t)-\int d{\bf z} \: J_{Y,t}({\bf z})\left (\partial_{y}\ln P_t(x\vert y)+\frac{1}{D_{YY,t}(y)P_t({\bf z})}\partial_{x}[D_{XY,t}({\bf z})P_t({\bf z})]\right )
\:\:\:\: \textrm{if   }  D_{YY,t}({\bf z})= D_{YY,t}(y)
\label{T12backsup1}
\end{align}
and ${\overline {\cal T}}^\dag_{X\to Y}(t)=\infty$ otherwise. Using Eq. (\ref{EqflowD4}), we see that relation in Eq. (\ref{EqDiff8}) is recovered in the  bipartite case ($D_{XY,t}=0$), as expected. Moreover, if the joint system ${\bf Z}$ is at equilibrium ({\it i.e.}, if the probability currents vanish), we also have ${\overline {\cal T}}^\dag_{X\to Y}(t)={\overline {\cal T}}_{X\to Y}(t)$, which is not obvious from Eq. (\ref{EqDiff7}). From Eq. (\ref{EqDiff4}), this also implies that ${\cal T}^\dag_{X\to Y}= {\cal T}_{X\to Y}$. In other words, the TE rate is time-symmetric at equilibrium, as it should be.

Finally, after using Eq. (\ref{EqsTEback1}) and the expression  of $l^-_Y$, {\it i.e.}, Eq. (\ref{EqflowD3}) with $X$ and $Y$ interchanged, we find  
\begin{align}
{\widehat {\overline  {\cal T}}}_{X\to Y}(t)=
 {\overline {\cal T}}_{X\to Y}(t)-\int d{\bf z}\: \frac{1}{D_{YY,t}({y})P_t({\bf z})}\Big[J_{Y,t}({\bf z})+\partial_{y}[D_{YY,t}(y)P_t({\bf z})]\Big]\partial_{x}[D_{XY,t}({\bf z})P_t({\bf z})]
\:\:\:\: \textrm{if   }  D_{YY,t}({\bf z})= D_{YY,t}(y)
\label{EqThatmoins}
\end{align}
and ${\widehat {\overline  {\cal T}}}_{X\to Y}(t)=\infty$ otherwise. This also immediately shows  that ${\widehat {\overline {\cal T}}}_{X\to Y}(t)={\overline {\cal T}}_{X\to Y}(t)$ in the bipartite case [Eq. (\ref{EqTildeT1})]. 

We reiterate that for diffusion processes with multiplicative noises, one must have that $D_{YY,t}({\bf z})= D_{YY,t}(y)$, and similarly $D_{XX,t}({\bf z})= D_{XX,t}(x)$, otherwise the various TE rates  are infinite, even in the bipartite case: see also  Appendix \ref{sec:App_proof}.

\subsection{Multi-time-step transfer entropy rates}
\label{subsec:multi}

It turns out that an expression somewhat similar to Eq. (\ref{EqTbarXfin}) can also be obtained for the multi-time-step TE rate $\mathcal T_{X\to Y}(t)$. To do this one has to generalize the preceding derivation for ${\overline  {\cal T}}_{X\to Y}(t)$. A new ingredient is the presence of an infinitesimal propagator $P(y,t+h\vert y'\hspace{0.1mm}_0^t)$ that is conditioned by a whole path $y'\hspace{0.1mm}_0^t$ from $0$ to $t$ instead of a single value at time $t$. By using Bayes theorem and the Markovian property, we rewrite it as
\begin{equation}
\label{eq:multi_propag}
P(y,t+h\vert y'\hspace{0.1mm}_0^t)=\int dx'\,P(y,t+h\vert {\bf z}',t)P(x',t\vert y'\hspace{0.1mm}_0^t),
\end{equation}
where by convention $y'\hspace{0.1mm}_0^t$ includes the endpoint $y'$ at time $t$. From Eq. (\ref{Eqpyyp}) we then obtain
\begin{equation}
P(y,t+h\vert y'\hspace{0.1mm}_0^t)=\delta(y-y')-h\left [ \widetilde{F}_{Y,t}(y'\hspace{0.1mm}_0^t)\frac{\partial}{\partial y}-\widetilde{D}_{YY,t}(y'\hspace{0.1mm}_0^t) \frac{\partial^2}{\partial y^2}\right]\delta(y-y') +{\rm O}(h^{2}),\label{eq:Py}
\end{equation}
with 
\begin{align}
\label{eq:deft}
&\widetilde{F}_{Y,t}(y_0^t)\equiv \int dx\, P(x,t \vert y_0^t)F_{Y,t}({\bf z}) \nonumber \\&
\widetilde{D}_{YY,t}(y_0^t)\equiv \int dx\,P(x,t\vert y_0^t)D_{YY,t}({\bf z})\,.
\end{align}
The above infinitesimal propagator can also be cast in the form of a Gaussian small-time expression as in Eq. (\ref{EqPropa1}).

The derivation of the expression of $\mathcal T_{X\to Y}(t)$ from its definition (\ref{EqTErateXY}) then directly follows that of the single-time-step TE rate $\overline{\mathcal T}_{X\to Y}(t)$ with the final result
\begin{align}
\mathcal T_{X\rightarrow Y}(t)=
\frac{1}{4}\int dx\:{\cal D}[y_0^t]\:\frac{P(x,t;y_0^t)}{D_{YY,t}(y)}[F_{Y,t}({\bf z})-\widetilde{F}_{Y,t}(y_0^t) ]^2 \:\:\:\:\textrm{if } D_{YY,t}({\bf z})=D_{YY,t}(y)
\label{eq:multi_final}
\end{align}
and $\mathcal T_{X\rightarrow Y}(t)=\infty$ otherwise. As found for the single-time-step TE rate, $\mathcal T_{X\rightarrow Y}(t)$ is only finite if the diffusion coefficient $D_{YY,t}({\bf z})$ only depends on the variable $y$. As it should be,  Eq. (\ref{EqTbarXfin}) is recovered and ${\cal T}_{X\to Y}(t)={\overline  {\cal T}}_{X\to Y}(t)$ when $Y_t$ is a sufficient statistic of $X_t$ [Eq. (\ref{Eqsufstat1})], as $\widetilde{F}_{Y,t}(y_0^t)={\overline F}_{Y,t}(y)$ and $\widetilde{D}_{YY,t}(y_0^t)=\overline{D}_{YY,t}(y)$ in this case.
Note also that the above formula has been derived for simplicity for unidimensional processes $X_t$ and $Y_t$ but {\it mutatis mutandis} it is easily extended to multidimensional processes. For instance, the result in Eq. (\ref{eq:multi_final}) can be generally rewritten as 
\begin{align}
\mathcal T_{X\rightarrow Y}(t)= \frac{1}{4}\Bigl\langle [F_{Y,t}(X_{t},Y_{t})-\widetilde{F}_{Y,t}(Y_0^{t})].D_{YY,t}(Y_{t})^{-1}. [F_{Y,t}(X_{t},Y_{t})-\widetilde{F}_{Y,t}(Y_0^{t}) ]\Bigr\rangle \ ,
\label{eq:multi_final1}
\end{align}
which expresses the transfer entropy rate in terms of the minimum mean-square error (MMSE) of the causal estimation. This generalizes  the relation obtained in \cite{WKP2013} for a bipartite diffusion process with additive noise and extends the  classical and beautiful result of Duncan~\cite{D1970,D1971} linking information theory and estimation theory: see \cite{MWZ1985, AVW2013} for more on this theme.

\subsection{Second law for non-bipartite processes}

\label{sec:StochMech}

So far we have discussed  learning rates and transfer entropy rates from the strict viewpoint of information exchange between two interacting systems.  We now wish to use these concepts to discuss non-equilibrium thermodynamics. In particular, we want to investigate how the second-law inequality involving the learning rate, which provides the tightest lower bound for bipartite processes (see Sec. \ref{subsec:SecLaw})  is modified when the bipartite assumption is dropped. We  stress that we are interested in the average entropy production (EP) during a finite time interval $[0,t]$ and not only in the stationary state or in the limit $t\to \infty$.  

Let us again focus on subsystem $X$. At the ensemble level, an entropy balance equation can be obtained as usual by decomposing the time-derivative of the marginal Shannon entropy $S_X(t)$,
\begin{align}
\label{Eqbalance}
d_tS_X(t)&=-\int d{\bf z} \:J_{X,t}({\bf z})\partial_{x} \ln P_t(x)\ ,
\end{align}
which can be rewritten as  
\begin{align}
\label{Eqbalance}
d_tS_X(t)=l^S_X(t)-\int d{\bf z} J_{X,t}({\bf z})\partial_{x}\ln P_t({\bf z})\ ,
\end{align}
after using Eq. (\ref{eq:symefPD}) to introduce  the symmetric learning rate $l^S_X(t)$. Inserting the Fokker-Planck equation and performing a few manipulations, we then obtain
\begin{align}
\label{Eqbalance}
d_tS_X(t)=l^S_X(t)+\sigma_X^{irr}(t)-\int d{\bf z} \frac{J_{X,t}({\bf z})}{D_{XX,t}({ \bf z})}[\widehat F_{X,t}({\bf z})-D_{XY,t}({\bf z})\partial_{y}\ln P_t({\bf z})]\ ,
\end{align}
where  we have defined the non-negative quantity
\begin{align}
\sigma_X^{irr}(t)\equiv \int d{\bf z} \: \frac{(J_{X,t}({\bf z}))^{2}}{D_{XX,t}({ \bf z})P_t({\bf z})}\ ,
\label{Eqsigmairr}
\end{align}
(traditionally referred to as the ``irreversible" EP), and the modified drift $\widehat F_{X,t}({\bf z})$ is defined by~\cite{CG2008}
\begin{align}
\widehat F_{X,t}({\bf z})\equiv F_{X,t}({\bf z})-\partial_{x} D_{XX,t}( {\bf z})-\partial_{y} D_{XY,t}({\bf z})\ .
\end{align}
Eq. (\ref{Eqbalance}) can be further transformed as 
\begin{align}
\label{EqdotSigma1}
\sigma_X^{irr}(t)&=\sigma_X(t)-l_X^S(t)-\int d{\bf z} \: \frac{D_{XY,t}({\bf z})}{D_{XX,t}({\bf z})}J_{X,t}({\bf z}) \partial_{y}\ln P_t({\bf z})\ ,\end{align}
where  $\sigma_X(t)$ is  defined as in the bipartite case by Eq. (\ref{Eqsigma12}), {\it i.e.}, $\sigma_X(t)\equiv d_tS_X(t)+ \sigma_X^B(t)$, 
with $\sigma_X^B(t)$ formally defined by~\cite{note8}
\begin{align}
\label{Eqsigma12B}
\sigma_X^B(t)\equiv \int d{\bf z} \frac{J_{X,t}({\bf z})}{D_{XX,t}({\bf z})}\widehat F_{X,t}({\bf z})\ .
\end{align}

Finally, exchanging  $X$ and $Y$ in Eq. (\ref{T12backsup1}) and using again Eq. (\ref{eq:symefPD}), we obtain [provided that  $D_{XX,t}({\bf z})=D_{XX,t}({x})$]
\begin{align}
\label{EqdotSigma2}
\sigma_X^{irr}(t)=\sigma_X(t)-\: [{\overline {\cal T}}_{Y\to X}(t)-{\overline {\cal T}}^\dag_{Y\to X}(t)]+\int d{\bf z} \: \frac{J_{X,t}({\bf z})}{D_{XX,t}(x)}\partial_{y}D_{XY,t}({\bf z}) \ .
\end{align}
So far, we have been only performing mathematical substitutions.  However, as already mentioned in Sec.  \ref{subsec:SecLaw}, connection to physics is possible if the thermodynamics of subsystem $X$ can  be defined and $\sigma^B_X(t)$  identified as the heat flow from $X$ to the environment, {\it e.g.}, a thermal  bath at a given temperature. In this regard, the existence of correlations between the noises is not an obstacle (see also the related discussion in \cite{SLP2018}).  For an external observer monitoring only $X$, the quantity $\sigma_X(t)$ defined by Eq. (\ref{Eqsigma12}) would  then be interpreted  as the EP of the thermodynamic system $X$, ignoring that $X$ also influences the dynamics of  $Y$. Note  that Eq. (\ref{EqdotSigma1}) implies that $\sigma_X=0$ at equilibrium, since $\sigma_X^{irr}=0$ from Eq. (\ref{Eqsigmairr})  (as $J_X=0$) and $l_X^S=0$, as pointed out in the preceding section. 

Since the two subsystems  are  coupled,  $\sigma_X(t)$ may be negative, but  thanks to Eqs. (\ref{EqdotSigma1}) and (\ref{EqdotSigma2}) there is  a  lower bound,
\begin{align}
\sigma_X(t)\ge l_X^S(t)+\int d{\bf z} \: \frac{D_{XY,t}({\bf z})}{D_{XX,t}({\bf z})}J_{X,t}({\bf z}) \partial_{y}\ln P_t({\bf z})\ ,
\end{align}
which is conveniently rewritten as 
\begin{align}
\label{Eqfinala}
\sigma_X(t)\ge{\overline {\cal T}}_{Y\to X}(t)-{\overline {\cal T}}^\dag_{Y\to X}(t)-\int d{\bf z} \: \frac{J_{X,t}({\bf z})}{D_{XX,t}(x)}\partial_{y}D_{XY,t}({\bf z})\ .
\end{align}
This  inequality takes a remarkably simple form in the case  where  $D_{XY,t}({\bf z})=D_{XY,t}(x)$, which includes the important case of additive noises, as it reduces to
 \begin{align}
\label{Eqfinalb}
  \sigma_X(t)\ge {\overline {\cal T}}_{Y\to X}(t)-{\overline {\cal T}}^\dag_{Y\to X}(t),
  \end{align}
or, after integration from $0$ to $T$, 
\begin{align}
\label{Eqfinalc}
 \Sigma_X\equiv \int_0^T dt\: \sigma_X(t) \ge \int_0^T dt\:[{\overline {\cal T}}_{Y\to X}(t)-{\overline {\cal T}}^\dag_{Y\to X}(t)]= \int_0^T dt\: [ {\cal T}_{Y\to X}(t)- {\cal T}^\dag_{Y\to X}(t)]\ ,
  \end{align}
where we have used Eq. (\ref{EqDiff5}) to write the second equality. In practice, the first equality may  be  more useful since the  single-time-step TE rates can be extracted more easily from time series than the  multi-time-step TE's.

Since ${\overline {\cal T}}_{Y\to X}(t)-{\overline {\cal T}}^\dag_{Y\to X}(t)=l_X(t)$ in the bipartite case  [Eq. (\ref{EqDiff8})], inequality (\ref{Eqfinalb}) [or its time-integrated version (\ref{Eqfinalc})] may be considered as the natural generalization of the  second law of information thermodynamics involving the learning rate. As far as we know, it is a new result, which represents a significant outcome of the present work~\cite{note9}. 
 It can be easily extended to non-bipartite Markov diffusion processes in which the two subprocesses are multidimensional. 
 
\begin{figure}[hbt]
\begin{center}
\includegraphics[width=10cm]{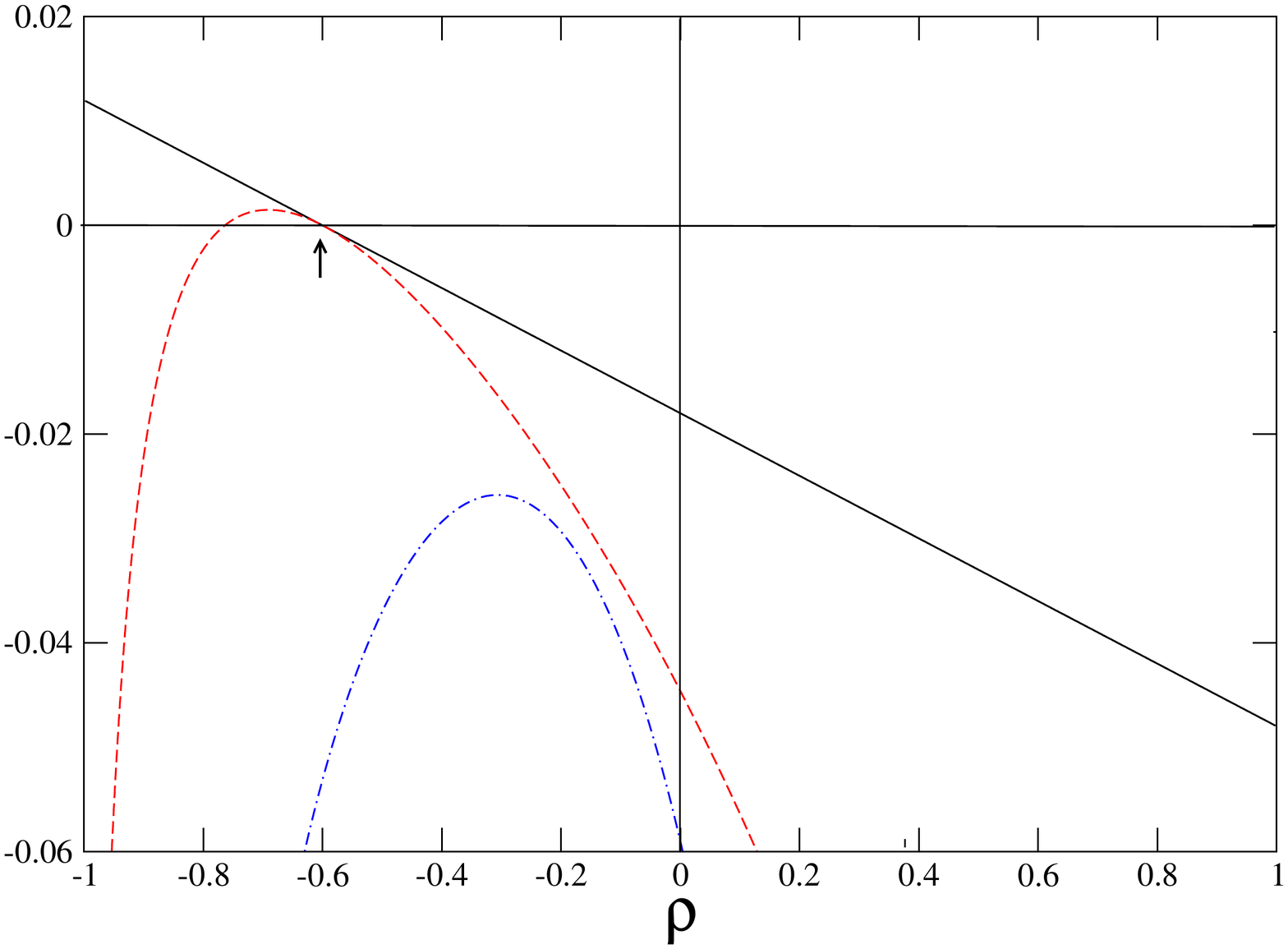}
\caption{\label{Fig2ndlaw} (Color on line) Numerical illustration of the second-law inequality (\ref{Eqfinalb}) for a stationary bi-dimensional Ornstein-Uhlenbeck  process:  $ \sigma_X$ (black solid line), ${\cal T}_{Y\to X}- {\cal T}^\dag_{Y\to X}={\overline {\cal T}}_{Y\to X}-{\overline {\cal T}}^\dag_{Y\to X}$ (red dashed line). The blue dashed-dotted line represents $-[{\cal T}_{X\to Y}+\Delta \dot I^{int}_{X\to Y}]$ (see Eq. (87) in \cite{SLP2018}). Note that  $\sigma_X$ varies linearly with $\rho$ (the explicit expression is given by Eq. (\ref{EqsigmaxOU}). One has $\sigma_X=0$ and ${\cal T}_{Y\to X}= {\cal T}^\dag_{Y\to X}$  for $\rho=-0.6$, which corresponds to equilibrium, as indicated by the small arrow. The model parameters are the same as in Fig. \ref{Fig1}.}
\end{center}
\end{figure}

Anticipating again the calculations performed in Sec. \ref{Sec:OU} for a bi-dimensional Ornstein-Uhlenbeck process, we illustrate in Fig. \ref{Fig2ndlaw} the generalized second law  and show the  variations of  $\sigma_X$ and  of the difference ${\cal T}_{T\to X}- {\cal T}^\dag_{Y\to X}={\overline {\cal T}}_{Y\to X}-{\overline {\cal T}}^\dag_{Y\to X}$ with the parameter $\rho$  quantifying  the  correlations between the noises affecting $X$ and $Y$. The figure  also displays the lower bound $-[{\cal T}_{X\to Y}+\Delta \dot I^{int}_{X\to Y}]$ recently obtained in Ref. \cite{SLP2018} by interpreting  the various contributions to the entropy production  in terms of  computational irreversibilities. The additional term  $\Delta \dot I^{int}_{X\to Y}$ vanishes for a bipartite dynamics and  the  bound is then known to be weaker than the one with the learning rate [see Eq. (\ref{Eq:second_law})]. For the specific case shown in Fig. \ref{Fig2ndlaw}, we see that this  remains true for $\rho\ne 0$ (but this is not a general feature). Note also that  ${\cal T}_{X\to Y}+\Delta \dot I^{int}_{X\to Y}\ne 0$ when the joint system is at equilibrium, in contrast with $\sigma_X$. Incidentally, we see  that one may have ${\cal T}_{Y\to X}= {\cal T}^\dag_{Y\to X}$ even in a non-equilibrium steady state, which occurs when the second term of Eq. (\ref{T12backsup1}) is zero.

\section{Stationary bi-dimensional Ornstein-Uhlenbeck process}
\label{Sec:OU}

To make one further step to derive explicit analytical expressions and make the discussion of the consequences of dropping the bipartite property even more concrete, we restrict ourselves to a Gaussian Markov process, more specifically a bi-dimensional stationary Ornstein-Uhlenbeck (OU) process with additive noises. This allows us to obtain simple analytical expressions for {\it all} the information measures, including the multi-time-step TE rates. In the following, we only explain how these quantities can be computed,  the details being given  in Appendix \ref{sec:AppOU}. In this Appendix, we also list  the expressions of the  learning rates and the single-time-step TE rates, which are easily obtained from the general formulas derived previously. To alleviate the notations we now denote the two (one-dimensional) subprocesses by $X_1$ and $X_2$ in place of $X$ and $Y$ and the combined process by $\bf X$ instead of $\bf Z$: $l_X$ will for instance be replaced by $l_1$ and $\mathcal T_{X\to Y}$ by $\mathcal T_{1\to 2}$. 

The stochastic dynamics is governed by the coupled Langevin equations
\begin{align} 
\label{EqOU}
\dot{\bf X}_t=-{\bf A}{\bf X}_t+{\boldsymbol \xi}_t\ ,
\end{align}
where ${\bf A}=[a_{ij}]$ is a $2$ by $2$ matrix and  $\boldsymbol \xi=\{\xi_i\}$  is a vector formed by two Gaussian noises with zero mean and covariances $\langle \xi_i(t)\xi_j(t')\rangle=2D_{ij}\delta(t-t')$. We assume that  all eigenvalues of ${\bf A}$ have a positive real part so that a stable steady-state solution exists~\cite{R1989,G2004}. Since we only focus on this regime hereafter, we may assume that the process has started at $t=-\infty$ and  forget about the initial condition [accordingly, the past history of $X_i(t)$ up to time $t$ is now denoted by $X_i^-(t)\equiv \{X_i(s):s\le t\}$ instead of  $(X_i)_0^t$]. The solution of Eq. (\ref{EqOU}) then reads
\begin{align} 
\label{EqH}
{\bf X}(t)=\int_{-\infty}^t ds\:{\bf H}(t-s){\boldsymbol \xi}(s)\, ,
\end{align}
where ${\bf H}(t)=[e^{-{\bf A}t}]_{ij}$ is the response (or Green's or transfer) functions matrix, and the power-spectrum matrix whose elements are the Fourier transform of the stationary  correlation functions $\phi_{ij}(t)=\langle X_i(t')X_j(t'+t)\rangle$ is  given by 
\begin{align} 
\label{EqSpectrum}
{\bf S}(\omega)={\bf H}(\omega)(2{\bf D}) {\bf H}^T(-\omega)\, ,
\end{align}
where ${\bf H}(\omega)=\int_{-\infty}^{\infty} dt\: e^{i\omega t} {\bf H}(t)=[{\bf A}(\omega)-i\omega {\bf I}]^{-1}$ and  $2{\bf D}$ is the diffusion matrix with elements $2D_{ij}$.

\subsection{ Expression of the multi-time-step TE rates}

An expression of ${\cal T}_{i \to j}$ for a non-bipartite stationary OU process has already been given in \cite{SLP2018}, but it turns out that it contains an error, as explained below. Moreover, the derivation is convoluted. We thus believe that it is worth presenting an alternative and much simpler route which has also been recently used for computing the TE rate in the presence of time delay~\cite{RTM2018}. This is actually a mere  application  of the formalism presented in \cite{BS2017}  for computing Granger causality for discrete and continuous-time autoregressive processes. As has already been mentioned, Granger causality and transfer entropy are identical (up to a factor $1/2$) when the random variables are Gaussian distributed~\cite{BBS2009}, which is the case here. 

By definition,  ${\cal T}_{i \to j}$ is  the slope  at $h=0$ of the finite-horizon TE defined by~\cite{note0}
\begin{align} 
\label{EqT12ha}
T_{i \to j}(h)&\equiv \Bigl\langle \ln  \frac{P(X_j(t+h)\vert {\bf X}^-(t))}{P(X_j(t+h)\vert X_j^-(t)) } \Bigr\rangle \ .
\end{align}
  From the expression of the entropy of Gaussian distributions in terms of their covariance matrix, we then readily obtain
\begin{align} 
\label{EqT12hb}
T_{i\to j}(h)=\frac{1}{2}\ln\frac{\sigma_{jj,j}(h)}{\sigma_{jj}(h)}\ ,
\end{align}
where 
\begin{align} 
\label{Eqsigmajj}
\sigma_{jj}(h)\equiv \Bigl\langle  [X_j(t+h)-\langle X_j(t+h)\vert {\bf X}^-(t)\rangle]^2 \Bigr\rangle
\end{align}
and 
\begin{align} 
\label{Eqsigmaprimejjj}
\sigma_{jj,j}(h)\equiv  \Bigl\langle [X_j(t+h)-\langle X_j(t+h)\vert X_j^-(t)\rangle]^2 \Bigr\rangle
\end{align}
are the mean of the variances of the conditional probabilities $P(X_j(t+h)\vert {\bf X}^-(t))$ and $P(X_j(t+h)\vert X_j^-(t))$, respectively. In the language of estimation theory, the conditional expectations $\langle X_j(t+h)\vert {\bf X}^-(t)\rangle$ and $\langle X_j(t+h)\vert X_j^-(t)\rangle$ represent the minimum mean-square error (MMSE) estimates of $X_j(t+h)$ when the trajectory up to time $t$ of either the full process ${\bf X}$ or of only $X_j$ is known: They are the orthogonal projections  onto ${\bf X}^-(t)$ and $X_j^-(t)$, respectively. As detailed in  Appendix \ref{sec:AppOU}, the calculation of ${\cal T}_{i \to j}$ amounts to computing first the conditional expectations, then the mean-square errors  $\sigma_{jj}(h)$ and $\sigma_{jj,j}(h)$, and finally expanding around $h=0$. In particular, this requires to determine the causal factor of the function $S_{jj}(\omega)$, which is a simple task since it is a rational function~\cite{A2006}. The final expression of  ${\cal T}_{1 \to 2}$ is 
\begin{align} 
\label{EqTE12new}
{\cal T}_{1 \to 2}&=\frac{1}{2}[r_2-a_{11}+\frac{D_{12}}{D_{22}}a_{21}]\ ,
\end{align}
where $r_2=[a_{11}^2+(D_{11}/D_{22})a_{21}^2-2(D_{12}/D_{22})a_{11}a_{21}]^{1/2}$; ${\cal T}_{2\to 1}$ is of course obtained by interchanging the roles of $1$ and $2$.  As it should be, one can verify that the same result is obtained from Eq. (\ref{eq:multi_final}), which is actually no more straightforward because  the calculation of the effective drift $\widetilde F_2$ requires  to compute the MMSE estimate $\langle X_1(t)\vert X_2^-(t)\rangle$.

As noticed above, Eq. (\ref{EqTE12new}) differs from the expression  of ${\cal T}_{1 \to 2}$ obtained in \cite{SLP2018} after a lengthy calculation (see also \cite{SL2018}).  In this expression, $a_{11}$ is replaced by $\vert a_{11}\vert$ (cf. Eq. (88) in \cite{SLP2018}),  which is erroneous and  may  even lead to negative values of ${\cal T}_{1 \to 2}$ for $a_{11}<0$.  (Having $a_{11}<0$  does not preclude the existence of a stable steady state so long as  $a_{11}+a_{22}>0$ and $a_{11}a_{22}-a_{12}a_{21}>0$.)  More generally, we stress that one must be careful in using a spectral representation of the TE rate, as done for instance in \cite{ HS2014} in the bipartite case. Indeed, as discussed in \cite{RTM2018}, the spectral expression has a limited range of validity (specifically, one must have $a_{11}-(D_{12}/D_{22}) a_{21}>0$). Otherwise, it underestimates the actual value of the TE rate, as was already pointed out in \cite{G1982} in the case of  discrete-time Granger causality.

From Eq. (\ref{EqTE12new}), one obtains the expression  of  the BTE rate ${\cal T}_{1\to 2}^\dag$ by modifying the drifts coefficients according to Eq. (\ref{Eqbackdrift}).  For linear Gaussian processes, this simply amounts to changing the matrix ${\bf A}$ into   ${\bf A}^*=-{\bf A}+2{\bf D}{\boldsymbol \Sigma}^{-1}={\boldsymbol \Sigma} {\bf A}^T {\boldsymbol \Sigma}^{-1}$~\cite{LK1976,A1982}, where ${\boldsymbol \Sigma}$ is the stationary covariance matrix, solution of the Lyapunov equation ${\bf A}.{\boldsymbol \Sigma}+{\boldsymbol \Sigma}.{\bf A}^T=2{\bf D}$~\cite{R1989,G2004}.  The quantity $r_2$ is invariant under the transformation ${\bf A}\to{\bf A}^*$,  and we obtain 
\begin{align} 
\label{EqTE12backnew}
{\cal T}_{1\to 2}^\dag&=\frac{1}{2}[r_2-a^*_{11}+\frac{D_{12}}{D_{22}}a^*_{21}]\ ,
\end{align}
with  $a^*_{11}$ and $a^*_{21}$ given by Eqs. (\ref{Eqastar}). It can be checked that this is in agreement with the expression  obtained from Eq. (\ref{EqDiff4}), with ${\overline {\cal T}}_{1 \to 2}$ and ${\overline {\cal T}}^\dag_{1\to 2}$ given by Eqs. (\ref{EqOUT12supp}) and (\ref{T12backsupa}), respectively.

On the other hand, the calculation of the filtered TE rate ${\widehat {\cal T}}_{1 \to 2}$ is more involved. For Gaussian processes,  Eq. (\ref{EqTflitremhat}) yields
\begin{align} 
{\widehat {\cal T}}_{1 \to 2}=\lim_{h \to 0} \frac{1}{2h} \ln \frac{\sigma_{11,2}(h)\sigma_{22,2}(h)}{\sigma_{11,2}(h)\sigma_{22,2}(h)-[\sigma_{12,2}(h)]^2}\ ,
\label{EqOUT12hatm0}
\end{align}
where  $\sigma_{11,2}(h)\equiv \langle \big[X_1(t+h)-\langle X_1(t+h)\vert X_2^-(t)\rangle\big]^2\rangle$, $\sigma_{22,2}(h)$ is defined above by Eq. (\ref{Eqsigmaprimejjj}), and $\sigma_{12,2}(h)\equiv \langle \big[X_1(t+h)-\langle X_1(t+h)\vert X_2^-(t)\rangle\big]\big[X_2(t+h)-\langle X_2(t+h)\vert X_2^-(t)\rangle\big]\rangle$. The calculation, which uses the fact that  the conditional expectation $\langle X_1(t+h)\vert X_2^-(t)\rangle$ is the orthogonal projection of $X_1(t+h)$ onto the trajectory $X_2^-(t)$, is detailed in Appendix \ref{sec:AppOU}. After some algebra this leads to the  expression
 \begin{align} 
\label{EqOUT12hatm}
{\widehat {\cal T}}_{1 \to 2}=D_{22} \frac{a_{11}(r_2-a_{11})^2}{a_{21}^2D_{11}-D_{22}(r_2-a_{11})^2}\ .
\end{align}

Finally, as also shown in Appendix \ref{sec:AppOU}, one may recast the non-Markovian Langevin equation for the marginal process $X_2$ as
\begin{align} 
\label{EqCG}
\dot X_2(t)=-(\omega_++\omega_--r_2)X_2(t)-(r_2-\omega_+)(r_2-\omega_-)\int_{-\infty}^t e^{-r_2(t-s)}X_2(s) ds +\xi'_2(t) \ ,
\end{align}
where $\omega_{\pm}$ are the eigenvalues of the matrix ${\bf A}$ whose expressions are given after  Eq. (\ref{EqS22}).  This formulation  is instructive because it shows that a remarkable simplification occurs  when $r_2= \omega_{\pm}$, that is when
\begin{align} 
\label{EqMcond}
(r_2-a_{11})(r_2-a_{22})-a_{12}a_{21}=0\ .
\end{align}
The second term in the right-hand side of Eq. (\ref{EqCG}) then vanishes and the  equation describes a Markovian dynamics.  Accordingly, one has $P(X_2(t+h)\vert X_2^-(t))=P(X_2(t+h)\vert X_2(t))$ and  ${\cal T}_{1\to 2}$  becomes equal to its upper bound ${\overline {\cal T}}_{1\to 2}$, as discussed in Appendix \ref{sec:AppSuffStat}.  We show in Appendix \ref{sec:AppOU} that from the perspective of the Kalman-Bucy filter~\cite{KB1961,KSH2000,A2006}, where $X_1(t)$ is a state variable that is not fully observed and the equation for $X_2(t)$ describes the dynamics of the filter state,  this means that the observer gain is optimal.  Eq. (\ref{EqMcond}) generalizes the conditions for sufficient statistic discussed in Ref.~\cite{MS2018} to the non-bipartite case. This may be viewed either as an optimal condition for the set of parameters $a_{ij}$'s for given noise intensities $D_{ij}$'s or as an optimal condition for the noises for a given set of the $a_{ij}$'s. Note that there are  in general two distinct solutions when the eigenvalues $\omega_{\pm} $ of the matrix ${\bf A}$ are real. Otherwise, there is no real value of $r_2$ that satisfies Eq. (\ref{EqMcond}) and the statistic is never sufficient. 
 
\subsection{Numerical illustration} 

\label{SubSec:OUnum}

We now give a numerical illustration which will allow us to discuss the behavior of the various information measures in the presence of correlated noises. To this aim, we consider a situation with a quasi-unidirectional coupling between the two random variables $X_1$ and $X_2$. Indeed, the  presence of both bidirectional interactions {\it and} correlated noises would make the interpretation of the information exchanges almost impossible. Actually, as far as we know, this case is seldom considered in the literature on Granger causality. Specifically, we choose the parameters of the model (see the caption of Fig. 1) such that the coupling in the direction $1\to 2$ is significantly larger than that in the opposite direction, so that one may consider $X_1$ as the source signal (the driver) and $X_2$ as the receiver. 

\begin{figure}[hbt]
\begin{center}
\includegraphics[width=10cm]{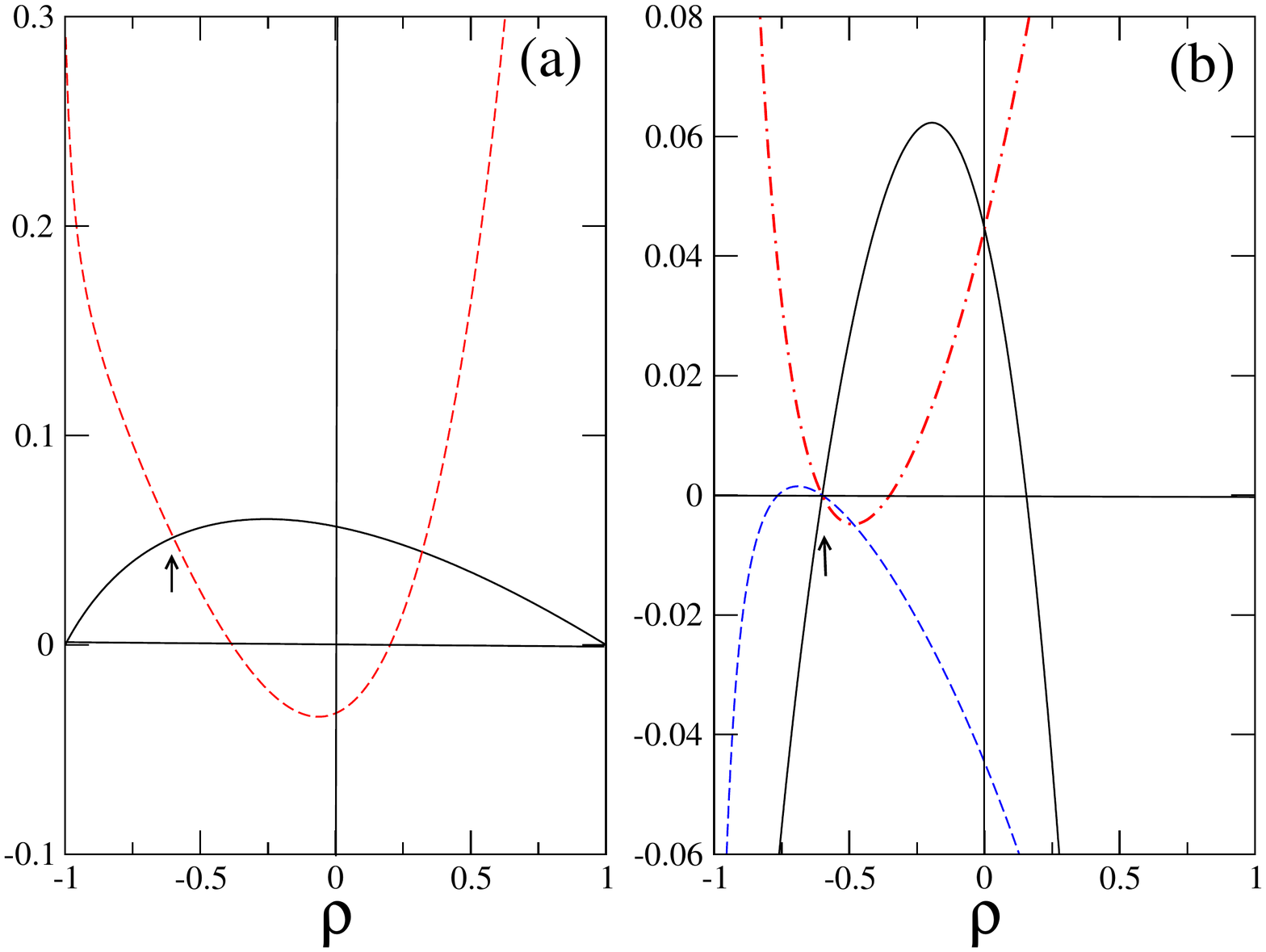}
\caption{\label{Fig2} (Color on line) Forward and backward TE rates  for a stationary bi-dimensional Ornstein-Uhlenbeck  process: (a): ${\cal T}^{(net)}_{1\to 2}\equiv {\cal T}_{1\to 2}-{\cal T}_{2\to 1}$ (black solid line), ${\cal T}^{\dag (net)}_{1\to 2}\equiv {\cal T}_{1\to 2}^\dag -{\cal T}^\dag _{2\to 1}$(red dashed line). (b): ${\cal T}_{1\to 2}-{\cal T}_{1\to 2}^\dag$ (black solid line), ${\cal T}_{2\to 1}-{\cal T}^\dag_{2\to 1}$ (blue dashed line),  $l_2^S=-l_1^S$ (red dashed-dotted line). The joint system is at equilibrium for $\rho=-0.6$, as indicated by the small arrow. The model parameters are given in Fig. 1.}
\end{center}
\end{figure}

On general ground, one is more interested in the {\it net} information flow in the system than in the magnitude of the flows in each direction.  Inspired by the recent literature on Granger causality~\cite{W2016}, and  replacing Granger causality by transfer entropy, we  then consider various  differences of the TE rates, such as ${\cal T}^{(net)}_{1\to 2}=-{\cal T}^{(net)}_{2\to 1}\equiv {\cal T}_{1\to 2}-{\cal T}_{2\to 1}$,  ${\cal T}^{\dag (net)}_{1\to 2}=-{\cal T}^{\dag (net)}_{2\to 1}\equiv {\cal T}_{1\to 2}^\dag -{\cal T}^\dag _{2\to 1}$, and the individual differences ${\cal T}_{1\to 2}- {\cal T}_{2\to 1}^\dag$, ${\cal T}_{2\to 1}- {\cal T}_{2\to 1}^\dag$. We also compare with the behavior of the symmetric learning rate $l_2^S=-l_1^S$. The results are shown in  Fig. \ref{Fig2}.

In the bipartite case ($\rho=0$), all quantities behave as expected and indicate that the information mainly flows from $1$ to $2$, due to  the fact that the interaction  $1\to 2$ predominates ($a_{21}/a_{12}=10$).  More precisely, we observe in Fig. 3a that ${\cal T}^{(net)}_{1\to 2}>0$ and ${\cal T}^{\dag (net)}_{1\to 2}<0$. The rationale given in the literature (see, {\it e.g.},  \cite{HNMN2013,V2015,W2016}) for the second inequality is that the directed information should be reduced (if not reversed) when the temporal order is reversed. We also verify in Fig. 3b that  ${\cal T}_{1\to 2}- {\cal T}_{1\to 2}^\dag=-({\cal T}_{2\to 1}-{\cal T}_{2\to 1}^\dag)=l_2^S>0$. As expected, $X_2$ is ``learning about" $X_1$ through its dynamics.

One observes that introducing correlations between the noises  has a strong effect on most quantities, even if the properties found for $\rho=0$  remain valid in some interval of $\rho$ around $0$ (for instance, ${\cal T}^{(net)}_{1\to 2}$ remains larger than ${\cal T}^{\dag (net)}_{1\to 2}$ in the whole interval $-0.6\lesssim\rho \lesssim0.3$). What appears robust whatever the correlation of the noises is that the net information flow measured by ${\cal T}^{(net)}_{1\to 2}$ always detects the correct dominant interaction. All of the other differences, as well as the learning rates, wildly vary and change sign as $\rho$ varies. There is clearly a complex competition between  the feedback (delayed) effects and the instantaneous  influence  generated by the correlation of the noises. The downside is that this competition  sensitively depends on the quantity under study and on other details of the dynamics (for instance on the relative magnitudes of the intrinsic time scales of each subprocesses, {\it i.e.}, $a_{11}^{-1}$ and $a_{22}^{-1}$ in the present model). The upside is that one can infer the presence of a strong correlation between the noises if one of these quantities has not the expected sign when ${\cal T}^{(net)}_{1\to 2}$ is positive, which may be a useful piece of information. 

\begin{figure}[hbt]
\begin{center}
\includegraphics[width=10cm]{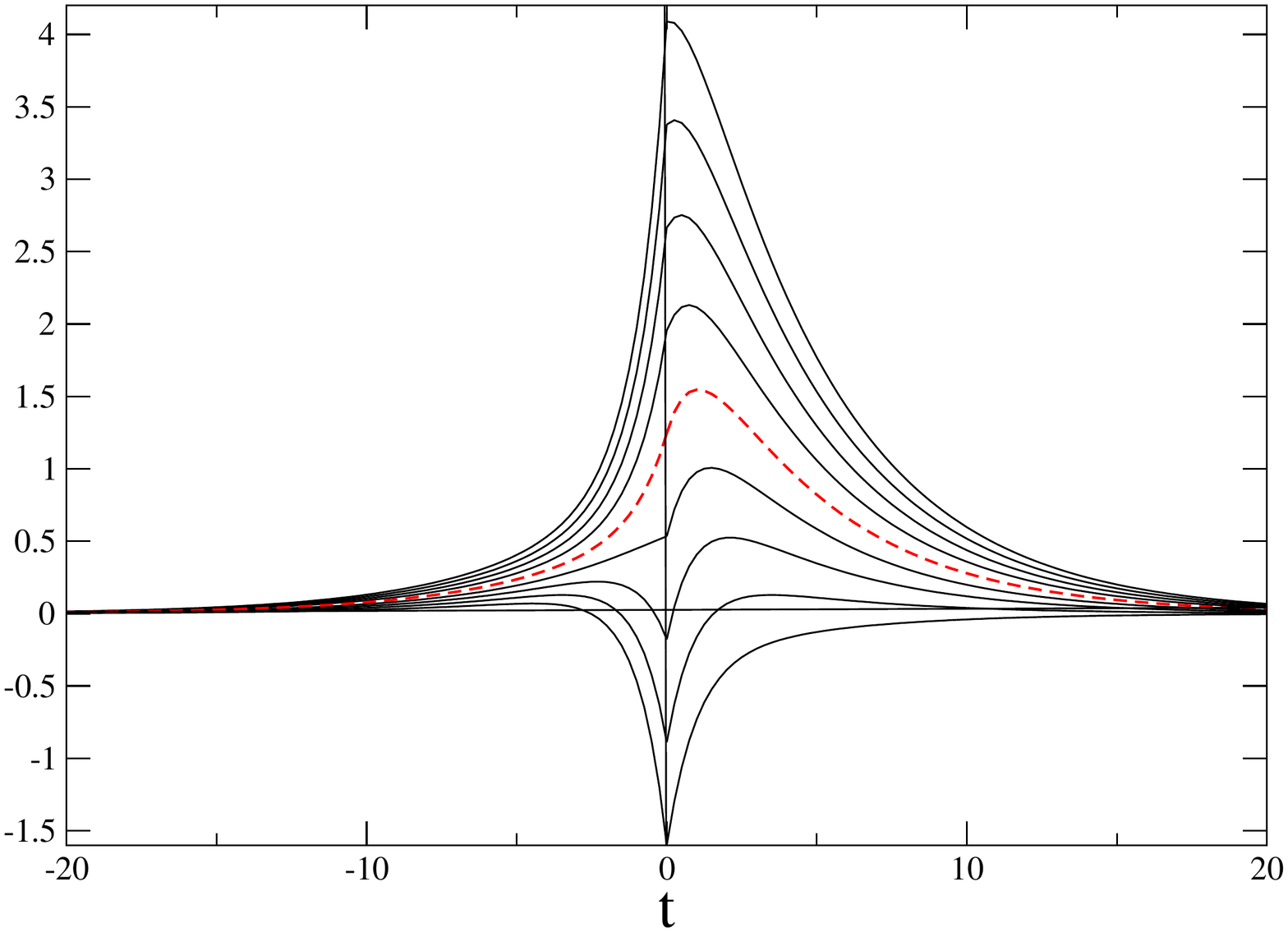}
\caption{\label{Fig3} (Color on line) Cross-correlation function $\phi_{12}(t)=\langle X_1(0)X_2(t)\rangle$ for a stationary bi-dimensional Ornstein-Uhlenbeck  process for different values of $\rho$ (from top to bottom: $\rho=0.8,0.6,0.4,0.2,0,-0.2,-0.4,-0.6,-0.8$). The red dashed curve corresponds to $\rho=0$. The model parameters are the same as in Fig. \ref{Fig1}.}
\end{center}
\end{figure}

It is interesting to draw a comparison with the information about the  dynamics of the system that could  be extracted from the cross-correlation function $\phi_{12}(t)=\langle X_1(0)X_2(t)\rangle$. This  is  indeed a widely used method to infer the directional influence between biological processes~\cite{DCLME2008}, as will be evoked in the next section. This function is shown in Fig. \ref{Fig3} for several values of the parameter $\rho$.

The two random variables are generally positively correlated, except at short times when the noises are strongly anti-correlated (and for $\rho \gtrsim -0.8$, $\phi_{12}(t)<0$ for all $t>0$). For $\rho \gtrsim -0.3$, the peak at $t>0$ indicates that $X_2(t)$ correlates more strongly with the value of $X_1$ at an earlier time,  suggesting that $X_1$ drives $X_2$.  On the other hand, the presence of a maximum for both $t>0$ and $t<0$ in the interval $-0.3\gtrsim \rho \gtrsim -0.7$ is hard to interpret and could perhaps erroneously suggest the presence of bidirectional coupling. This again illustrates the subtle competition between the correlation in the noises and the actual feedback.

 \section{Generalization to a non-Markovian process and application to the study of directional influence between cellular processes }
 
 \label{sec:Non-Markov}
 
We now consider the application of the general formalism for non-bipartite processes presented in Sec. \ref{sec:Info} to a class of non-Markovian processes.  We have in mind a situation that is commonly encountered in biological networks in which one is interested in the information exchanges between two  random variables, say $X_1$ and $X_2$ (typically one-dimensional), but other  random variables - intrinsic or extrinsic to the system under study - come into play (see, {\it e.g.}, Fig. 1 in \cite{BS2012}). Within the linear-noise approximation, the network dynamics can be described by a set of chemical Langevin equations where $X_i(t)$ is the deviation of the concentration of species $i$  from its mean value  (see, {\it e.g.}, \cite{G2000,TWW2006,HT2014,H2016}). The network dynamics corresponds to a multi-dimensional Markov process, but the (coarse-grained) dynamics of the combined process $(X_1,X_2)$ (and not only the dynamics of the individual processes)  is non-Markovian. From this coarse-grained perspective, the system therefore corresponds to a  process that is non-bipartite and furthermore non-Markovian in general. This is the case on which we focus below.

\subsection{Extension to a bivariate non-Markovian process}
\label{subsec:nonMarkov}

For a bivariate non-Markovian process $(X_1,X_2)$ obtained as discussed above by coarse-graining a multi-dimensional Markov process over all the extraneous random variables, all the general definitions and relations of information measures introduced in Sec. \ref{sec:Info} directly apply. Furthermore, the Markov character of the underlying network brings a drastic simplification for the derivation of explicit expressions for the information measures, and the latter  can be obtained by some form of averaging over the extraneous dynamical variables. Skipping details, and considering only the case of additive noises for simplicity, we find
\begin{align}
l_1^{\pm}(t)&=\int dx_1 dx_2 [\bar J_{1,t}(x_1,x_2)\pm D_{12}\partial_{x_2} P_t(x_1,x_2)]\partial_{x_1}\ln P_t(x_2\vert x_1)\ ,
\end{align}
\begin{align}
\label{EqTE12multi}
{\overline {\cal T}}_{1 \to 2}(t)=\frac{1}{4}\int dx_1dx_2\: \frac{P_t(x_1,x_2)}{D_{22}} [{\overline F}_{2,t}^2(x_1,x_2)-{\overline{\overline F}}_{2,t}^2(x_2)]\ ,
\end{align}
\begin{align}
\label{EqTE12backmulti}
{\overline {\cal T}}^\dag_{1 \to 2}(t)&={\overline {\cal T}}_{1 \to 2}(t)-\int dx_1dx_2\:\bar{J}_{2,t}(x_1,x_2)[\partial_{x_2} \ln P_t(x_1\vert x_2)+\frac{D_{12}}{D_{22}}\partial_{x_1}\ln P_t(x_1,x_2)]\ ,
\end{align}
with  ${\bar J}_{i,t}(x_1,x_2)=\int (\prod_{j\ne 1,2}dx_j)\: J_{i,t}({\bf x})$ ($i=1,2$), ${\overline F}_{2,t}(x_1,x_2)=\int (\prod_{i\ne 1,2}dx_i)\: F_{2,t}({\bf x})P_t({\bf x}\vert x_1,x_2)$,  and ${\overline {\overline F}}_{2,t}(x_2)=\int dx_1\: {\overline F}_{2,t}(x_1,x_2)P_t(x_1\vert x_2)$, where ${\bf x}$ denotes all the variables of the network, and $F_{i,t}({\bf x})$ and $J_{i,t}({\bf x})$ denote the drift and the probability current for species $i$ in the full multi-dimensional Markov  process.

The  calculation of the multi-time-step TE rates is more involved and is detailed in Appendix \ref{sec:AppKiviet} in the special case of  the three-dimensional  model considered in Sec. \ref{sec:Kiviet}.

We also do not discuss here the extension of the second law inequalities (\ref{Eqfinala}) or  (\ref{Eqfinalb}) as this requires a more extensive and delicate analysis which we defer to future investigations.  Indeed, one must first decide which components or processes must be taken into account in the theoretical description, how information is transmitted throughout the network, and identify the sources of stochasticity~\cite{BS2012}.  This is a nontrivial task which is better done on a case by case basis. In particular,  the non-bipartite character of the dynamics, {\it i.e.}, the existence of  transitions affecting simultaneously the states of the subsystems, may strongly depend on the level of coarse-graining of the description (see, {\it e.g.}, \cite{LEM2017} for a recent and detailed experimental and theoretical study of a chemical nanomachine).

\subsection{Application to the study of directional influence between cellular processes}

\label{sec:Kiviet}

In this final section, we apply the previous framework to revisit and complement the study performed in Ref. \cite{LNTRL2017} about the information transmission  between single cell growth rate and gene expression in the metabolism of  {\it E. coli}. Understanding how fluctuations in gene expression can affect the growth stability of a cell and, in turn, how  the growth noise affects gene expression is an important issue that was initially investigated  in  Ref. \cite{K2014}. The purpose of  Ref. \cite{LNTRL2017} -- which actually prompted our concern for the problem of correlated noises -- was to show that  transfer entropy is a versatile and model-free tool that can be used to infer directional interactions in biochemical networks, in addition to (or possibly as a substitute for) the standard method based on time-delayed cross-correlation functions~\cite{DCLME2008}. 

A characteristic feature of the biochemical processes studied in \cite{K2014, LNTRL2017}  is the presence of a common extrinsic noise source affecting both the {\it lac} enzyme concentration $E(t)$ and the  growth rate $\mu(t)$, which makes the stochastic system under study non-bipartite.  Specifically, the stochastic model proposed in \cite{K2014} to account for the experimental data leads to the following  equations within the linear response approximation:
\begin{align}
\label{EqLKiviet0}
\dot X_1&=\mu_0[(T_{EE}-1)X_1+ T_{E\mu}X_2 +T_{EG}N_G+N_E]\nonumber\\
X_2&=T_{\mu E}X_1+T_{\mu G}N_G+N_{\mu}\ ,
\end{align}
where $X_1(t)\equiv (E(t)-E_0)/E_0$ and $X_2(t)\equiv (\mu(t)-\mu_0)/\mu_0$ quantify the deviations of $E(t)$ and $\mu(t)$ from their mean $E_0$ and $\mu_0$, $N_l(t)$ ($l=E,\mu,G$) are three independent Ornstein-Uhlenbeck (OU) noises that are generated by the auxiliary equations $\dot N_l=-\beta_l N_l+\xi_l$ where the $\xi_l$'s are zero-mean Gaussian white noises with amplitudes $\theta_l=\eta_l\sqrt{2\beta_l}$) and $T_{ll'}$ are logarithmic gains representing how a variable $l$ responds to the fluctuations of a source $l'$ (with $T_{E \mu }=-1$ and $T_{\mu G}=1$. In fact, as shown in  \cite{LNTRL2017}, the intrinsic noise $N_E(t)$ affecting the enzyme concentration may be replaced by a  delta-correlated noise with the same intensity $D_E=\eta_E^2/\beta_E$ without deteriorating the quality of the fit to the experimental time-dependent correlation functions for $E(t)$ and $\mu(t)$. Then, after some simple manipulations and eliminating the variable $N_G$, Eqs. (\ref{EqLKiviet0})  become equivalent to three coupled linear Langevin equations~\cite{LNTRL2017}, 
\begin{align}
\label{EqLKiviet}
\dot X_1&=-[\mu_E+\mu_0 T_{\mu E}(T_{EG}-1)]X_1+ \mu_0 (T_{EG}-1)X_2 -\mu_0T_{EG}X_3+\xi_1\nonumber\\
\dot X_2&=T_{\mu E}[\beta_G -\mu_E-\mu_0T_{\mu E}(T_{EG}-1)]X_1+[\mu_0 T_{\mu_E}(T_{EG}-1)-\beta_G]X_2+[\beta_G-\beta_{\mu}-\mu_0T_{\mu E}T_{EG}]X_3+\xi_2\nonumber\\
\dot X_3&=-\beta_{\mu} X_3+\xi_3\end{align}
where the third equation describes the dynamics of the OU noise $X_3(t)\equiv N_{\mu}(t)$ affecting the growth rate [Eqs (\ref{EqLKiviet}) correspond to  Eqs. (59) in \cite{LNTRL2017} where $X_1,X_2$ and $X_3$ are denoted $x,y$ and $v$, respectively]. The rate $\mu_E=\mu_0(1+T_{\mu E}-T_{EE})$ sets the time scale of $E$  fluctuations and the three Gaussian white noises $\xi_1\equiv\mu_0N_E,\xi_2\equiv\xi_{\mu}+\xi_G+\mu_0 T_{\mu E}N_E,\xi_3\equiv\xi_{\mu}$ have covariances $D_{11}=D_E\mu_0^2$, $D_{22}=\beta_{\mu}\eta_{\mu}^2+\beta_{G}\eta_{G}^2+D_E\mu_0^2T_{\mu E}^2$, $D_{33}=\beta_{\mu}\eta_{\mu}^2$, $D_{12}=D_E\mu_0^2T_{\mu E}$, $D_{13}=0$, $D_{23}=\beta_{\mu}\eta_{\mu}^2$. The numerical values of all these  parameters are given in Table S1 of \cite{K2014}.  

Eqs. (\ref{EqLKiviet}) define a Markov process for a set of $3$ interacting random variables, but we are interested in analyzing  the information transfer between $X_1$ and $X_2$ which together form a joint non-Markovian process, as discussed just above. We can thus study the information-theoretic quantities previously introduced. This task was partially accomplished in \cite{LNTRL2017} where the single-time-step TE  rates $\overline {\cal T}_{1 \to 2},\overline {\cal T}_{2 \to 1}$  and the  learning rates $l_1^+,l_2^+$ -- dubbed information flows -- were computed (see note \cite{note1} that explains a regrettable confusion in the definition of the learning rates). To complement this  analysis,  we consider the multi-time-step TE rates  ${\cal T}_{1 \to 2},{\cal T}_{2 \to 1}$ and  the backward rates ${\cal T}_{1\to 2}^\dag, {\cal T}_{2\to 1}^\dag$ and $\overline {\cal T}_{1\to 2}^\dag, {\overline{\cal T}}^\dag_{2 \to 1}$. To  the best of our knowledge, this is the first time that the backward TE rates are used to infer the direction of information exchanges in a real biochemical network. Whereas $\overline {\cal T}_{1\to 2}^\dag$ and ${\overline{\cal T}}^\dag_{2 \to 1}$ are obtained from Eq. (\ref{EqTE12backmulti}), the calculation of the multi-time-step TE rates is more involved and is detailed in Appendix \ref{sec:AppKiviet}.
\begin{table*}[!h]
  \centering
  \renewcommand{\arraystretch}{1.5}
  \begin{tabular}{|c|c|c|c|}
    \hline
    {\bf Conc. of IPTG} & Low & Intermediate & High \\
    \hline
    $\rho_{12}$ &$ 0.408$ & $0.362$& 0\\
   $\rho_{23}$ &$ 0.537$ & $0.498$& 0.135\\
    \hline
    $ {\cal T}_{1\to 2}$ ($\overline{{\cal T}}_{1\to 2})$ & $0.053 \:(0.053)$ & $0.020\: (0.020)$ & $0 \: (0)$\\
    ${\cal T}_{2\to 1}$ ($\overline{{\cal T}}_{2\to 1})$& $0.008 \: (0.008)$ & $0.001\: (0.001)$ & $0.036 \: (0.036)$\\
    \hline
    ${\cal T}_{1\to 2}^\dag$ & $0.020$ & $ 0.011$ & $0.026$ \\   
    ${\cal T}_{2\to 1}^\dag$ & $0.001$ & $< 10^{-4}$ & $0.010$ \\ 
      \hline
     $\overline{\cal T}_{1\to 2}^\dag$ & $0.021$ & $0.011$ & $0.026$\\
    $ \overline{\cal T}_{2\to 1}^\dag$ & $0.002$ & $< 10^{-4} $ & $0.010$ \\   
     \hline
     $l_1^+=-l_2^-$ & $-0.327$ & $-0.219$ & $0.026$\\
    $ l_2^+=-l_1^-$ & $-0.198$ & $-0.177$ & $-0.026$ \\ 
    \hline
     \end{tabular}
  \caption{ Theoretical values of the various TE rates for three IPTG concentrations (given in Table S1 of \cite{K2014}).  The time unit is the inverse of  the average growth rate $\mu_0$. The values of $\overline{{\cal T}}_{i\to j}$  (numbers in brackets) and $l_i^+$ computed in \cite{LNTRL2017}  are also reported. The parameters $\rho_{12}=D_{12}/\sqrt{D_{11}D_{22}}$  and $\rho_{23}=D_{23}/\sqrt{D_{22}D_{33}}$ quantify the correlations between the white noises $\xi_1$, $\xi_2$ , and $\xi_2$, $\xi_3$ in Eqs. (\ref{EqLKiviet}).}
\label{Table1}
\end{table*}

The numerical results are presented in Tables \ref{Table1} and \ref{Table2} where the inverse of  the average growth rate $\mu_0$ is taken as the time unit. For comparison, the values of $\overline {\cal T}_{i\to j}$ and $l_i^+$ computed in  \cite{LNTRL2017} are also reported in Table \ref{Table1}.  As in \cite{K2014, LNTRL2017}, we consider three different concentrations (low, intermediate, and high) of the inducer IPTG, which allows one to explore different regimes of noise transmission.  

\begin{figure}[hbt]
\begin{center}
\includegraphics[width=10cm]{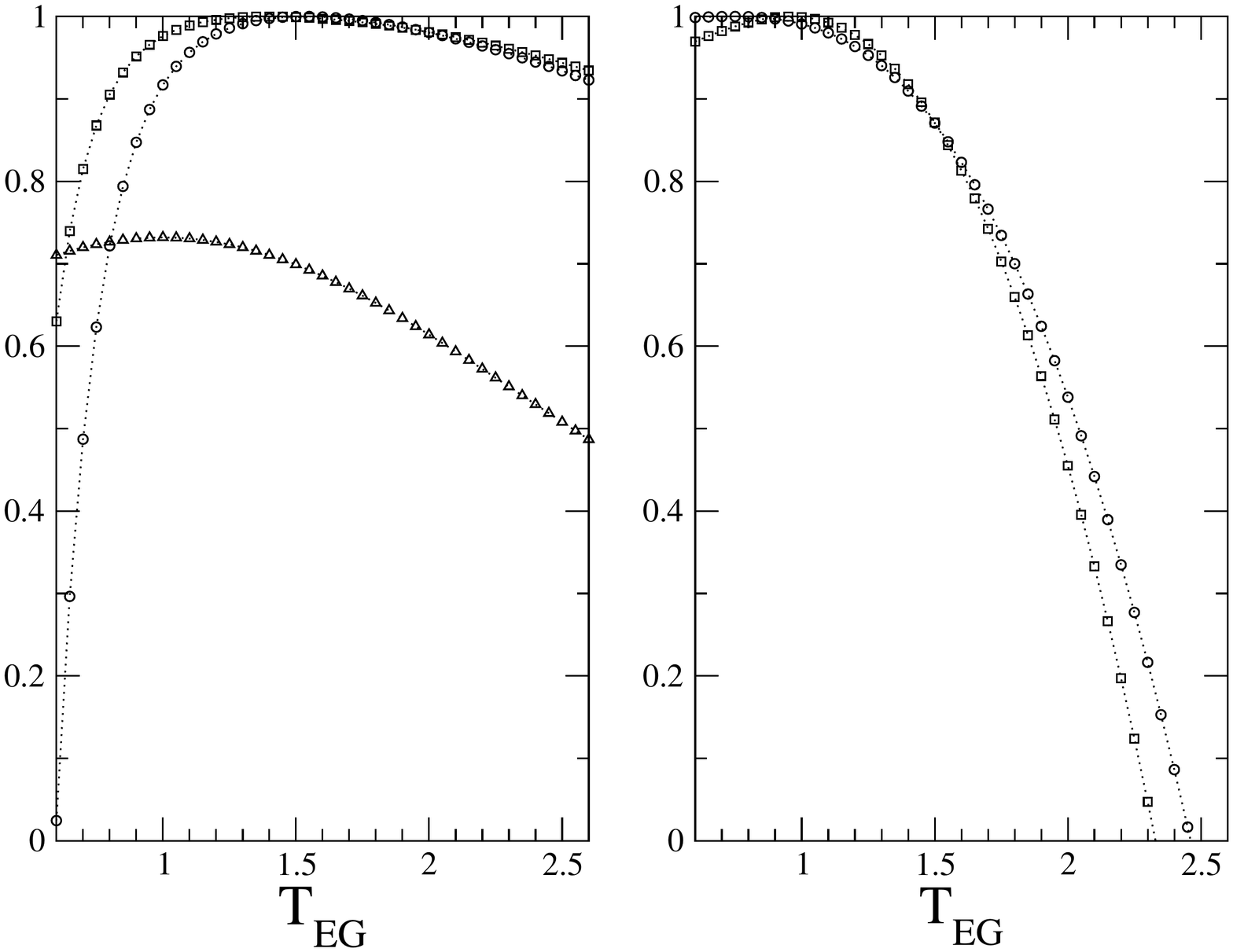}
\caption{\label{Fig5} (Color on line) Single-time-step sensory capacities  $\overline{C}_1$ (left) and $\overline{C}_2$ (right) as a function of the noise transmission coefficient $T_{EG}$. Circles, squares, and triangles refer to the low, intermediate, and high IPTG concentrations, respectively. $\overline{C}_2\to -\infty$ in the latter case because  $l_2^-<0$ and $\widehat {\overline{{\cal T}}}_{1\to 2}=l_2^-+ {\overline{{\cal T}}}_{1\to 2}=0$ as $T_{\mu E}=0$ in Eqs. (\ref{EqLKiviet0}) and $X_1$ no longer influences  $X_2$.}
\end{center}
\end{figure}

Comparing the results for  ${\cal T}_{1\to 2}$ and $ {\cal T}_{2\to 1}$ in Table \ref{Table1} to the corresponding results for $\overline {\cal T}_{1\to 2}$ and $\overline {\cal T}_{1\to 2}$ obtained in \cite{LNTRL2017} (shown in brackets in the Table), we observe that the numerical values are almost identical. The same is true for the backward rates. In fact, we  find numerically that $T_{ij}(h)\approx \overline T_{ij}(h)$ for all values of the prediction horizon $h$ (recall that the TE rates are given by the slopes of the finite-horizon curves at the origin). More precisely, $\langle X_i(t+h)\vert X_1^-(t),X_2^-(t)\rangle\approx \langle X_i(t+h)\vert X_1(t),X_2(t)\rangle$ and $\langle X_i(t+h)\vert X_i^-(t)\rangle\approx \langle X_i(t+h)\vert X_i(t)\rangle$ separately, which means that the joint process $(X_1,X_2)$ and also the marginal processes $X_1$ and $X_2$  can be reformulated (in the stationary regime) as Markov processes to a very good approximation.  More details are given in Appendix \ref{sec:AppKiviet}.
This is an intriguing result that suggests some kind of optimization in the transmission of information. 
Another indication is the behavior of  the single-time-step sensory capacity  $\overline{C}_1$  defined by Eq. (\ref{EqssSensCap}) as one varies the transmission coefficient $T_{EG}$  describing the response of {\it lac} expression to the common noise $N_G(t)$ (see Eqs. (\ref{EqLKiviet0})).  As shown in Fig. \ref{Fig5},  $\overline{C}_1$ at low and intermediate IPTG concentrations  reaches the maximal value $1$ for $T_{EG}\approx 1.5$, which is close to the value $T_{EG}=1.3$ used in \cite{K2014}  to fit  the experimental cross-correlation functions  (the fit is actually satisfactory for $T_{EG}$ in the range $1.3-1.5$). On the other hand,  the maximum of $\overline{C}_2$  occurs for a smaller value of $T_{EG}$, significantly below  the acceptable range~\cite{note14}.

Is this a real feature of the metabolic network?  Giving a definite answer to this question is difficult because there are many ingredients in the stochastic description and it is not easy to identify those which are responsible for such a behavior. We leave the discussion of this interesting issue to a future investigation.

Finally, we complement  the analysis performed in \cite{LNTRL2017} by exploiting our determination of the backward TE rates.  As suggested in Refs. \cite{HNMN2013,V2015,W2016} and tested on time series generated by multivariate autoregressive processes, time-reversed Granger causality, or backward TE in the present framework, may lead to a better estimate of the directionally of information flows. As in Section \ref{SubSec:OUnum}, we thus  consider the quantities $ {\cal T}^{(net)}_{1\to 2}= {\cal T}_{1\to 2}- {\cal T}_{2\to 1}$, ${\cal T}^{\dag (net)}_{1\to 2}= {\cal T}_{1\to 2}^\dag- {\cal T}_{2\to 1}^\dag$, and  the individual differences $ {\cal T}_{1\to 2}- {\cal T}_{1\to 2}^\dag$ and  $ {\cal T}_{2\to 1}- {\cal T}_{2\to 1}^\dag$. (We could also consider the corresponding quantities built from the single-time-step TE rates but they are almost identical.) The results are presented in Table \ref{Table2} where we also indicate the values of the symmetric learning rates $l_1^S=-l_2^S$. 
\begin{table*}[!h]
  \centering
  \renewcommand{\arraystretch}{1.5}
  \begin{tabular}{|c|c|c|c|}
    \hline
    {\bf Conc. of IPTG} & Low & Intermediate & High \\
     \hline
    $ {\cal T}^{(net)}_{1\to 2}= {\cal T}_{1\to 2}- {\cal T}_{2\to 1}$ & $0.045$& $0.019$ & $-0.036$\\
    ${\cal T}^{\dag (net)}_{1\to 2}= {\cal T}_{1\to 2}^\dag- {\cal T}_{2\to 1}^\dag$ & $0.019$& $0.011$ & $0.016$\\
    $ {\cal T}_{1\to 2}-{\cal T}_{1\to 2}^\dag$ & $0.033$ & $0.009$ & $-0.026$\\
    $ {\cal T}_{2\to 1}-{\cal T}_{2\to 1}^\dag$ & $0.007$ & $ 0.001$ & $0.026$ \\   
    $l_1^S=-l_2^S$ &$ -0.064$ & $-0.021$& $0.026$\\
    \hline
    \end{tabular}
  \caption{Differences in the TE rates and symmetric learning rates (the time unit is $\mu_0^{-1}$).}
\label{Table2}
\end{table*}
  
We first observe  that  $ {\cal T}^{(net)}_{1\to 2}>{\cal T}^{\dag (net)}_{1\to 2}$, ${\cal T}_{1\to 2}-{\cal T}_{1\to 2}^\dag>0$  and $l_2^S>0$ at  low and intermediate ITPG concentrations, which strengthens the  conclusion  of \cite{LNTRL2017} (based solely on the positivity of  $\overline {\cal T}^{(net)}_{1\to 2}$) that  information  flows from $ X_1$ to $X_2$ in these two cases. That {\it lac} fluctuations propagate  through the metabolic network and perturb growth was also the conclusion of \cite{K2014}  based on the corresponding time-correlation functions. Note however that ${\cal T}_{2\to 1}- {\cal T}_{2\to 1}^\dag$ is also slightly positive. As discussed in Section \ref{SubSec:OUnum}, this may result from the competition between the correlation in the noises and the direct interactions between the variables $X_1$ and $X_2$.  We must also take into account  that the process described by Eqs. (\ref{EqLKiviet}) is  significanltly more complicated to than the one discussed in Section \ref{SubSec:OUnum} because of the presence of bidirectional interactions.  

At high  IPTG concentration, all scores consistently indicate that there is a backward transmission from $X_2$ to $X_1$, {\it i.e.}, from growth to expression, which is again  in line with the conclusions of \cite{K2014} and \cite{LNTRL2017}: see Table 1 in  \cite{LNTRL2017} where $ {\overline {\cal T}}^{(net)}_{1\to 2}$ is computed directly from the experimental time series collected in \cite{K2014}.  Of course, one has $ {\cal T}^{(net)}_{1\to 2} <0$ and ${\cal T}_{1\to 2}-{\cal T}_{1\to 2}^\dag<0$ in this case simply because $X_1$ no longer influences  $X_2$ and thus  ${\cal T}_{1\to 2}=0$. Indeed, the transmission coefficient $T_{\mu E}$ describing  the response of $\mu$ to $E$ fluctuations is taken equal to $0$  in the stochastic model (see Table S1 in \cite{K2014}). On the other hand, a less trivial observation is that  the difference $ {\cal T}_{2\to 1}-{\cal T}_{2\to 1}^\dag$  is significantly more positive than at  low and intermediate ITPG concentrations. It would certainly  be useful  in future investigations to also estimate  the backward TE rates directly from the experimental time series.

In passing, note  that the numerical results for ${\cal T}_{1\to 2}-{\cal T}_{1\to 2}^\dag$, $-({\cal T}_{2\to 1}-{\cal T}_{2\to 1}^\dag)$ and $l_2^S=-l_1^S$ are equal in the last column of Table \ref{Table2}. One could think that this directly results from the fact that (i) the white noises $\xi_1$ and $\xi_2$ are mutually independent in this case ($D_{12}=0$) and (ii) the noise $N_{\mu}\equiv X_3$ no longer affects $X_2$ (as $\beta_\mu=\beta_G$ in the model of Ref. \cite{K2014}). The reason turns out to be less straightforward. Indeed, the colored noises $\xi_1^{eff}$ and $\xi_2^{eff}$ affecting $X_1$ and $X_2$ are still correlated because $X_3$ affects $X_1$ and $D_{23}\ne 0$ [see Eqs. (\ref{EqEffnoises}) and (\ref{EqEffcor})].  But in the end, when  the $3$ coupled Langevin equations (\ref{EqLKiviet}) are replaced by  the equivalent Langevin representation for $X_1$ and $X_2$ given by Eqs. (\ref{EqLnew3}), one discovers that in this case the joint process $(X_1,X_2)$ is both bipartite {\it and} quasi-Markov. Accordingly, one has  ${\cal T}_{1\to 2}-{\cal T}_{1\to 2}^\dag\approx l_2^S$ and ${\cal T}_{2\to 1}-{\cal T}_{2\to 1}^\dag\approx l_1^S=-l_2^S$.

\section{Summary}

In this paper, we have considered information and thermodynamic exchanges between two coupled  stochastic systems that do not satisfy the bipartite property. This should correspond to the generic situation  in biochemical networks, in which noises can be correlated (for diffusion processes) or transitions of the two systems can simultaneously take place (for jump processes). The generalization of information quantities, such as learning rates and transfer entropy rates,  to the non-bipartite situation is non-trivial and involves introducing several additional rates. We have described how this can be achieved in the framework of continuous-time Markov processes. We have also derived several inequalities  that are valid beyond the bipartite assumption and   generalized  a classical relationship between mutual information and causal estimation error. We have illustrated our general formalism on the case of Markov diffusion processes and obtained a new formulation of the second law of information thermodynamics  in which the  learning rate is replaced by a difference of transfer entropy rates in the forward and backward time directions. Explicit analytical expressions of all information measures have been derived for  the special case of  a bivariate Ornstein-Uhlenbeck process, allowing a discussion of  the influence  of the correlation between the noises on the sign and/or the relative magnitude of the various information measures.  Finally, we have applied our formalism to the analysis of the directional influence between cellular processes in a concrete example, which required considering the case of a non-bipartite non-Markovian process. An  intriguing ``optimal" transmission of information  has been observed, which calls for future investigations. More generally, it remains as an interesting future work to extend the present study to describe the fluctuations of pathwise thermodynamic and information-theoretic quantities. 

\section*{Acknowledgements} T.S. is supported by JSPS KAKENHI Grant Number JP16H02211. 

\appendix 

\section{Expression of the learning rates for a pure jump Markov process in discrete space}

 \label{sec:AppJump}

\renewcommand{\theequation}{A\arabic{equation}} 

In this Appendix, we give the expression of the learning rate for non-bipartite Markov pure jump processes. In this case, the Markovian generator takes the form (\ref{eq:GPS}) in terms of the transition rates $W_{t}\left(z,z'\right)$. 

After some algebra, the general Markovian expressions (\ref{IF+}) and (\ref{IF+}) of the learning rates can be cast in the form
\begin{equation}
\begin{cases}
 l_{X}^{+}(t)=\sum_{z,z'}P_{t}(z')W_{t}\left(z',z\right)\ln \frac{P_{t}(x')P_{t}(x,y')}{P_{t}(x)P_{t}(x',y')}\\
 l_{X}^{-}(t)=\sum_{z,z'}P_{t}(z')W_{t}\left(z',z\right) \ln \frac{P_{t}(x,y)P_{t}(x')}{P_{t}(x',y)P_{t}(x)}
\end{cases}\label{eq:LRJP}.
\end{equation}
For a bipartite pure jump process, the bipartite transition-rate relation [below Eq. (\ref{BipG})] allows one to show that the two formulas in Eq. \ref{eq:LRJP} coincide,  (see also ~\cite{HBS2014,SS2015}) 
\begin{equation}
l_{X}^{+}(t)=l_{X}^{-}(t)=\sum_{x,z'}P_{t}(z')W_{t,y'}\left(x',x\right)\ln \frac{P_{t}(x')P_{t}(x,y')}{P_{t}(x)P_{t}(x',y')}=\sum_{x,z'}P_{t}(z')W_{t,y'}\left(x',x\right)\ln\frac{P_{t}(y'\vert x )}{P_{t}(y'\vert x')}.
\label{eq:lrpm-1-1}
\end{equation}

\section{Proof of various relations of the main text}

\label{sec:App_proof}

\renewcommand{\theequation}{B\arabic{equation}} 

\textit{1. Proof  of Eq. (\ref{EqTfiltre5}) for a bipartite Markov diffusion process}
\\

To prove Eq. (\ref{EqTfiltre5}) we start with the first line of Eq. (\ref{EqTfiltre6}) and consider $<\ln [P(Y_{t+h}\vert X_0^t,Y_0^t)/P(Y_{t+h}\vert Y_0^t,X_{t+h})]>$ which we rewrite as 
\begin{align}
\label{Eq:app_bip1}
&\Bigl\langle \ln \frac{P(Y_{t+h}\vert X_0^t,Y_0^t)}{P(Y_{t+h}\vert Y_0^t,X_{t+h})}\Bigr\rangle = \int dx\, dy\, {\cal D}[x'\hspace{0.1mm}_0^t] \,{\cal D}[y'\hspace{0.1mm}_0^t]\: P({\bf z},t+h;x'\hspace{0.1mm}_0^t,y'\hspace{0.1mm}_0^t)\ln \frac{P(y,t+h\vert {\bf z}',t)P(x,t+h\vert y'\hspace{0.1mm}_0^t)}{P({\bf z},t+h\vert y'\hspace{0.1mm}_0^t)} \nonumber \\&
= \int dx\: dy\: {\cal D}[x'\hspace{0.1mm}_0^t] \,{\cal D}[y'\hspace{0.1mm}_0^t]\:P({\bf z},t+h\vert {\bf z}',t)P(x'\hspace{0.1mm}_0^t,y'\hspace{0.1mm}_0^t)\ln \frac{P(y,t+h\vert {\bf z}',t)\int dx'' P(x,t+h\vert x'',y',t)P(x'',t\vert y'\hspace{0.1mm}_0^t)}{\int dx'' P({\bf z},t+h\vert x'',y',t)P(x'',t\vert y'\hspace{0.1mm}_0^t)}
\end{align}
where $x'\hspace{0.1mm}_0^t$ (resp. $y'\hspace{0.1mm}_0^t$) is a short-hand notation to indicate paths of $X$ (resp. $Y$) between $0$ and $t$ and we have used the fact that the joint process is Markovian. For convenience of notation, integration with the measure ${\cal D}[x'\hspace{0.1mm}_0^t]$ (resp. ${\cal D}[y'\hspace{0.1mm}_0^t]$) also includes the integration over the initial and final states, this latter being noted $x'$ (resp. $y'$). 
We can now use the properties of a bipartite process, {\it i.e.}, the factorization of the infinitesimal Markovian propagator (or transition probability) in Eq. (\ref{EqBip1}) and the decomposition of the Markovian generator in Eq. (\ref{BipG}), to write
\begin{align}
\label{Eq:app_bip2}
P({\bf z},t+h\vert {\bf z}',t)&=P(x,t+h\vert {\bf z}',t)P(y,t+h\vert {\bf z}',t) \nonumber \\&
=\delta(x-x')\delta(y-y') +h[\delta(y-y')L_{t,y'}(x',x)+\delta(x-x')L_{t,x'}(y',x)] + {\cal O}(h^2).
\end{align}

The specific problem to be handled in the case of diffusion processes in the appearance of singular delta functions in the numerator and denominator of the argument of the logarithm. This can be conveniently bypassed by considering the Gaussian expressions of the infinitesimal propagators, as done in Eq. (\ref{EqPropa1}). So, for instance, up to a ${\cal O}(h^2)$,
\begin{align}
\label{Eq:app_bip3}
&\int dx'' P({\bf z},t+h\vert x'',y',t)P(x'',t\vert y'\hspace{0.1mm}_0^t)\nonumber \\&
=\int dx'' P(x'',t\vert y'\hspace{0.1mm}_0^t)[\delta(x-x'')+h L_{t,x''}(y',x)]\frac{1}{\sqrt{4\pi hD_{YY,t}(x'',y')}}e^{-\frac{[y-y'-h F_{Y,t}(x'',y')]^2}{4hD_{YY,t}(x'',y')}}\nonumber \\&
= \frac{P(x,t\vert y'\hspace{0.1mm}_0^t)}{\sqrt{4\pi hD_{YY,t}(x,y')}}e^{-\frac{[y-y'-h F_{Y,t}(x,y')]^2}{4hD_{YY,t}(x,y')}}+h \int dx''   \frac{P(x'',t\vert y'\hspace{0.1mm}_0^t)L_{t,x''}(y',x)}{\sqrt{4\pi hD_{YY,t}(x'',y')}}e^{-\frac{[y-y'-h F_{Y,t}(x'',y')]^2}{4hD_{YY,t}(x'',y')}}
\end{align}
where there was no need to consider the Gaussian expression for $P(x,t+h\vert x'',y',t)$ because the associated delta function is integrated over.

In the following we consider the case where $D_{YY,t}({\bf z})$ is independent of  $x$. Otherwise, the TE rates from $X\to Y$ are  infinite, as shown in Sec. \ref{subsec:multi}. One therefore has
\begin{align}
\label{Eq:app_bip4}
&\ln \frac{P(y,t+h\vert {\bf z}',t)}{\int dx'' \: P({\bf z},t+h\vert x'',y',t)P(x'',t\vert y'\hspace{0.1mm}_0^t)}
= - \ln P(x,t\vert y'\hspace{0.1mm}_0^t) -\frac{[y-y'-h F_{Y,t}(x',y')]^2-[y-y'-h F_{Y,t}(x,y')]^2}{4hD_{YY,t}(y')}\nonumber \\&
- h \frac 1{P(x,t\vert y'\hspace{0.1mm}_0^t)} \int dx'' \:  P(x'',t\vert y'\hspace{0.1mm}_0^t)L_{t,x''}(y',x) e^{-\frac{[y-y'-h F_{Y,t}(x'',y')]^2-[y-y'-h F_{Y,t}(x,y')]^2}{4hD_{YY,t}(y')}} +{\cal O}(h^2)\ ,
\end{align}
and
\begin{align}
\label{Eq:app_bip5}
\ln \int dx'' \: P(x,t+h\vert x'',y',t)P(x'',t\vert y'\hspace{0.1mm}_0^t)&=\ln \int dx'' \: [\delta(x-x'')+h L_{t,x''}(y',x)] P(x'',t\vert y'\hspace{0.1mm}_0^t) \nonumber \\&
= \ln P(x,t\vert y'\hspace{0.1mm}_0^t) + h \frac 1{P(x,t\vert y'\hspace{0.1mm}_0^t)} \int dx'' \:  P(x'',t\vert y'\hspace{0.1mm}_0^t)L_{t,x''}(y',x)  +{\cal O}(h^2)\ , 
\end{align}
so that
\begin{align}
\label{Eq:app_bip6}
&\ln \frac{P(y,t+h\vert {\bf z}',t)\int dx'' \: P(x,t+h\vert x'',y',t)P(x'',t\vert y'\hspace{0.1mm}_0^t)}{\int dx'' \: P({\bf z},t+h\vert x'',y',t)P(x'',t\vert y'\hspace{0.1mm}_0^t)}
=  (y-y') \frac{[F_{Y,t}(x',y')- F_{Y,t}(x,y')]}{2D_{YY,t}(y')}  \nonumber \\& 
-h\,\frac{[F_{Y,t}({\bf z}')^2- F_{Y,t}(x,y')^2]}{4D_{YY,t}(y')}  
+ h  \int dx'' \:  \frac{P(x'',t\vert y'\hspace{0.1mm}_0^t)L_{t,x''}(y',x)}{P(x,t\vert y'\hspace{0.1mm}_0^t)}\left (1 -e^{-(y-y') \frac{[F_{Y,t}(x',y')- F_{Y,t}(x'',y')]}{2D_{YY,t}(y')}}\right ) +{\cal O}(h^2)\,.
\end{align}
One can check that the above expression reduces to a ${\cal O}(h^2)$  when $x=x'$ and $y=y'$. As a consequence when combining Eqs. (\ref{Eq:app_bip1}), (\ref{Eq:app_bip2}) and (\ref{Eq:app_bip6}), it only remains, up to a ${\cal O}(h^2$),
\begin{align}
\label{Eq:app_bip7}
&\Bigl\langle \ln \frac{P(Y_{t+h}\vert X_0^t,Y_0^t)}{P(Y_{t+h}\vert Y_0^t,X_{t+h})}\Bigr\rangle \nonumber \\&
=h \int {\cal D}[x'\hspace{0.1mm}_0^t] \: {\cal D}[y'\hspace{0.1mm}_0^t]\: P(x'\hspace{0.1mm}_0^t,y'\hspace{0.1mm}_0^t) \int dx\, dy\, [\delta(y-y')L_{t,y'}(x',x)+\delta(x-x')L_{t,x'}(y',x)](y-y')\frac{[F_{Y,t}({\bf z}')- F_{Y,t}(x,y')]}{2D_{YY,t}(y')} \,,
\end{align}
and it is straightforward to see that the above term linear in $h$ exactly vanishes. One can therefore conclude from Eq. (\ref{EqTfiltre6}) that the equality in Eq. (\ref{EqTfiltre5}) is satisfied. Note that we have considered unidimensional processes $X_t$ et $Y_t$ for simplicity but the demonstration is easily generalized to multidimensional diffusion processes.
\\

\textit{2. Proof  of  inequality (\ref{EqIneq4})}\\

To prove  inequality (\ref{EqIneq4}) we consider the difference 
\begin{align} 
\label{Eqdemineq}
l^+_X(t)-[ \frac{d}{dt}I(X_t:Y_0^t)-\widehat {\cal T}_{X\to Y}(t)]&=\lim_{h\rightarrow0}\frac{1}{h}\:\Big[I(X_{t+h}:Y_t)-I(X_t:Y_t)-I(X_{t+h}:Y_0^t)+I(X_t:Y_0^t)\Big]\nonumber\\
&=\lim_{h\rightarrow0}\frac{1}{h}\:[\Bigl\langle \ln \frac{P(X_t\vert Y_0^t)}{P(X_{t+h}\vert Y_0^t)}\Bigr\rangle - \Bigl\langle \ln \frac{P(X_t\vert Y_t)}{P(X_{t+h}\vert Y_t)}\Bigr\rangle]\ ,
\end{align}
where we have used  the definitions of $l_X^+(t)$, $\widehat {\cal T}_{X\to Y}(t)$,  and the derivative of $I(X_t:Y_0^t)$. Since $P(X_{t+h}\vert X_t,Y_t)=P(X_{t+h}\vert X_t,Y_0^t)$ if the joint process is Markovian, we  can write
\begin{align} 
({\bf M})\:\:\:\:\ \frac{P(X_t\vert Y_0^t)}{P(X_{t+h}\vert Y_0^t)}&=\frac{P(X_t,Y_0^t)}{P(X_{t+h},Y_0^t)}\frac{P(X_{t+h}\vert X_t,Y_0^t)}{P(X_{t+h}\vert X_t,Y_t)}=\frac{P(X_{t+h},X_t,Y_0^t)}{P(X_{t+h},Y_0^t)P(X_{t+h}\vert X_t,Y_t)}\ .
\end{align}
 Furthermore, the integral analogue of the log-sum inequality applied to the trajectory $Y_0^{t-0^{+}}$ yields
 \begin{align} 
\Bigl\langle \ln \frac{P(X_{t+h},X_t,Y_0^t)}{P(X_{t+h},Y_0^t)}\Bigr\rangle \ge \Bigl\langle \ln  \frac{P(X_{t+h},X_t,Y_t)}{P(X_{t+h},Y_t)} \Bigr\rangle\ .
\end{align}
As a result,
 \begin{align} 
({\bf M})\:\:\:\:\ \Bigl\langle \ln \frac{P(X_t\vert Y_0^t)}{P(X_{t+h}\vert Y_0^t)}\Bigr\rangle \ge \Bigl\langle \ln \frac{P(X_{t+h},X_t,Y_t)}{P(X_{t+h},Y_t)P(X_{t+h}\vert X_t,Y_t)}\Bigr\rangle=\Bigl\langle \ln \frac{P(X_t\vert Y_t)}{P(X_{t+h}\vert Y_t)}\Bigr\rangle\ ,
\end{align}
which leads to inequality (\ref{EqIneq4}).
\\

\textit{3. Proof of the expression (\ref{T12backsup1}) for the single-time-step BTE rates for a Markov diffusion process}
\\

For the same reason as ${\overline {\cal T}}_{X\to Y}(t)$, the BTE rate ${\overline {\cal T}}^\dag_{X\to Y}(t)$ is infinite when  $D_{YY,t}({\bf z}) \neq {\overline D_{YY,t}(y)}$. We thus consider in the following the case where $D_{YY,t}({\bf z})={\overline D_{YY,t}}(y) = D_{YY,t}(y)$.

Inserting the expression of $F^*_{Y,t}({\bf z})$ given in Eq. (\ref{Eqbackdrift}) into Eq. (\ref{EqTbarXfin}), we obtain
\begin{align}
\label{Eq:app_BTE1}
F_{Y,T-t}^{*}({\bf z})^{2}-[\overline F_{Y,T-t}^{*}(y)]^{2}&=F_{Y,t}({\bf z})^{2}-[{\overline F}_{Y,t}(y)]^{2} \nonumber\\
&-4\frac{J_{Y,t}({\bf z})}{P_t({\bf z})^2}\big (\partial_{x}[D_{XY,t}({\bf z})P_t({\bf z})]+\partial_{y}[D_{YY,t}(y)P_t({\bf z})]\big )+4\frac{{\bar J}_{Y,t}(y)}{P_t(y)^{2}}\partial_{y}[D_{YY,t}(y)P_t(y)]\ ,
\end{align}  
where ${\bar J}_{Y,t}(y)={\overline F}_{Y,t}(y)P_{t}(y)-\partial_{y}[D_{YY,t}(y)P_t(y)]$. After combining Eqs. (\ref{EqTbarXfin}), (\ref{EqHP2}) and (\ref{Eq:app_BTE1}), we arrive at
\begin{align}
\label{Eq:app_BTE2}
&{\overline {\cal T}}^\dag_{X\to Y}(t)-{\overline {\cal T}}_{X\to Y}(t)  \nonumber \\&
=-\int d{\bf z}\,\frac{J_{Y,t}({\bf z})}{D_{YY,t}(y) P_t({\bf z})} \partial_{x}[D_{XY,t}({\bf z})P_t({\bf z})]-\int d{\bf z}\,\frac{J_{Y,t}({\bf z})}{D_{YY,t}(y) P_t({\bf z})}[D_{YY,t}(y)\partial_{y}P_t({\bf z})+P_t({\bf z})\partial_{y}D_{YY,t}(y)]\nonumber \\&
+\int dy\,\frac{\bar J_{Y,t}(y)}{D_{YY,t}(y) P_t(y)}[D_{YY,t}(y)\partial_{y}P_t(y)+P_t(y)\partial_{y}D_{YY,t}(y)]\ ,
\end{align}  
where we have used the fact that $D^*_{YY,T-t}(y)=D_{YY,t}(y)$ and $P^*_{T-t}({\bf z})=P_t({\bf z})$. With a few manipulations and the help of the relation $\int dx\, J_{Y,t}({\bf z})=\bar J_{Y,t}(y)$, the above equation can be rewritten as
\begin{align}
\label{Eq:app_BTE3}
&{\overline {\cal T}}^\dag_{X\to Y}(t)-{\overline {\cal T}}_{X\to Y}(t) 
=-\int d{\bf z}\,\frac{J_{Y,t}({\bf z})}{D_{YY,t}(y) P_t({\bf z})} \partial_{x}[D_{XY,t}({\bf z})P_t({\bf z})]-\int d{\bf z}\,J_{Y,t}({\bf z})\partial_{y} \ln P_t({\bf z})
+\int dy\,\bar J_{Y,t}(y)\partial_{y} \ln P_t(y) \nonumber \\&
=-\int d{\bf z}\,\frac{J_{Y,t}({\bf z})}{D_{YY,t}(y) P_t({\bf z})} \partial_{x}[D_{XY,t}({\bf z})P_t({\bf z})]-\int d{\bf z}\,J_{Y,t}({\bf z})\partial_{y} \ln P_t(x\vert y)\,,
\end{align}  
which corresponds to Eq. (\ref{T12backsup1}) of the main text.
\\

\section{Sufficient statistic for a non-bipartite dynamics}

\label{sec:AppSuffStat}

\renewcommand{\theequation}{C\arabic{equation}}

In this appendix, we show the consequences of the sufficient statistic condition $P(X_t\vert Y_0^t)=P(X_t\vert Y_t)$ [(Eq. (\ref{Eqsufstat1}) in the main text] for the various inequalities obtained in Sec. \ref{Subsec:ineq}. First, this implies that 
\begin{align} 	
 P(Y_{t+h}\vert Y_0^t)&=\int dX_t\:P(X_t,Y_{t+h}\vert Y_0^t)=\int dX_t\:P(Y_{t+h}\vert X_t,Y_0^t)P(X_t\vert Y_0^t)\nonumber\\
({\bf M})\:\:\:\:\ &=\int dX_t\:P(Y_{t+h}\vert X_t,Y_t)P(X_t\vert Y_0^t)=\int dX_t\:P(Y_{t+h}\vert X_t,Y_t)P(X_t\vert Y_t)\nonumber\\
({\bf M})\:\:\:\:\ &=\int dX_t\:P(X_t,Y_{t+h}\vert Y_t)=P(Y_{t+h}\vert Y_t)\ .
\end{align}
Therefore, ${\overline T}_{X\to Y}(t)-T_{X\to Y}(t)=\lim_{h\to 0^+}h^{-1}\langle \ln [P(Y_{t+h}\vert Y_0^t)/P(Y_{t+h}\vert Y_t)]\rangle=0$ and inequality (\ref{EqIneq2}) is saturated.

Since condition (\ref{Eqsufstat1}) also implies that $ P(X_{t+h}\vert Y_{t+h},Y_0^{t})=P(X_{t+h}\vert Y_{t+h})=P(X_{t+h}\vert Y_{t+h},Y_t)$, we  have
\begin{align} 	
 P(X_{t+h}\vert Y_0^t)&=\int dY_{t+h}\:P(X_{t+h},Y_{t+h}\vert Y_0^t)=\int dY_{t+h}\:P(X_{t+h}\vert Y_{t+h},Y_0^t)P(Y_{t+h}\vert Y_0^t)\nonumber\\
({\bf M})\:\:\:\:\ &=\int dY_{t+h}\:P(X_{t+h}\vert Y_{t+h},Y_t)P(Y_{t+h}\vert Y_t)\nonumber\\
({\bf M})\:\:\:\:\ &=P(X_{t+h}\vert Y_t)\ , 
\end{align}
which  yields from Eq. (\ref{Eqdemineq})
 \begin{align}
({\bf M})\:\:\:\:\ l^+_X(t)-[ \frac{d}{dt}I(X_t:Y_0^t)-\widehat {\cal T}_{X\to Y}(t)]=\lim_{h\rightarrow0}\frac{1}{h}\: \Bigl\langle \ln \frac{P(X_t\vert Y_0^t)P(X_{t+h}\vert Y_t)}{P(X_{t+h}\vert Y_0^t)p(X_t\vert Y_t)} \Bigr\rangle = 0\ .
\end{align}
Therefore, the steady-state inequality (\ref{EqIneq3}) is also saturated.

Finally, since $P(X_{t+h},Y_{t+h}\vert Y_0^t)=P(X_{t+h},Y_{t+h}\vert Y_t)$, one also has 
${\widehat {\cal T}}_{X\to Y}={\widehat {\overline {\cal T}}}_{X\to Y}$ and inequality (\ref{EqIneq5}) is saturated (note that the joint process does not need to be Markovian in this case).

\section{Analytical expressions for a stationary bi-dimensional Ornstein-Uhlenbeck process}

\label{sec:AppOU}
\renewcommand{\theequation}{D\arabic{equation}} 

 In this appendix, we give the analytical expressions of the learning rates and various TE rates for a stationary bi-dimensional OU process. 
\\
 
 \textit{1. Learning rates and single-time-step TE rates}
 \\
 
 These rates are obtained from the general formulas for Markov diffusion processes derived in Sec. \ref{sec:DIF}. To this aim, we use the expression of the stationary probability distribution function 
 \begin{align} 
P_{st}({\bf x})=\frac{1} {2\pi\vert \Sigma\vert^{1/2}} e^{-\frac{1}{2}{\bf x}^T{\boldsymbol \Sigma}^{-1}{\bf x}}\ ,
\end{align}
where ${\boldsymbol \Sigma}$ is the covariance matrix whose elements $\sigma_{ij}\equiv \langle X_i(0)X_j(0)\rangle$ are obtained by solving the Lyapunov equation. This gives
\begin{align}
\sigma_{11}&=\frac{(a_{22}^2+a_{11}a_{22}-a_{12}a_{21})D_{11}
 +a_{12}(a_{12}D_{22}-2a_{22}D_{12})}{(a_{11}a_{22}-a_{12}a_{21})(a_{11}+a_{22})}\nonumber\\
 \sigma_{22}&=\frac{(a_{11}^2+a_{11}a_{22}-a_{12}a_{21})D_{22}
 +a_{21}(a_{21}D_{11}-2a_{11}D_{12})}{(a_{11}a_{22}-a_{12}a_{21})(a_{11}+a_{22})}\nonumber\\
\sigma_{12}&=\sigma_{21}=\frac{2a_{11}a_{22}D_{12}-a_{22}a_{21}D_{11}-a_{11}a_{12}D_{22}}{(a_{11}a_{22}-a_{12}a_{21})(a_{11}+a_{22})}\ .
\end{align}
 One can check that $\sigma_{11}$ and $\sigma_{22}$ are  positive when the two conditions $a_{11}a_{22}-a_{12}a_{21}>0$ and $D_{11}D_{22}-D_{12}^2\ge 0$ are satisfied. Some useful relations  among these quantities are obtained by multiplying the stationary Fokker-Planck equation by $x_1^2, x_2^2$ and $x_1x_2$, respectively, and integrating over ${\bf x}$. This gives
\begin{subequations}
\label{EqAij:subeqns}
\begin{align} 
a_{11}\sigma_{11}+a_{12}\sigma_{12}&=D_{11}
\label{EqAij:subeq1}\\
a_{22}\sigma_{22}+a_{21}\sigma_{21}&=D_{22}
\label{EqAij:subeq2}\\
(a_{11}+a_{22})\sigma_{12}+a_{21}\sigma_{11}+a_{12}\sigma_{22}&=2D_{12}\ .
\label{EqAij:subeq3}
\end{align}
\end{subequations}
From Eq. (\ref{EqflowD4}), we then find
\begin{align} 
l_1^+=-l_2^-&=-\frac{\sigma_{12}}{\sigma_{11}}(a_{12}+\frac{\sigma_{12}}{\vert  {\boldsymbol \Sigma}\vert}D_{11})\nonumber\\
&=a_{11}-\frac{\sigma_{22}}{\vert  {\boldsymbol \Sigma}\vert}D_{11}\ ,
\end{align}
where we have used relation (\ref{EqAij:subeq1}) to go from the first to the second line. Likewise,  the symmetric learning rate $l_1^S=-l_2^S$  given by Eq. (\ref{eq:symefPD}) reads
\begin{align} 
l_1^S&=a_{11}-\frac{\sigma_{22}D_{11}-\sigma_{12}D_{12}}{\vert  {\boldsymbol \Sigma}\vert}\  .
\end{align}

The TE rate ${\overline {\cal T} }_{1 \to 2}$ is obtained from Eq. (\ref{EqTbarXfin}), using $\overline F_2(x_2)=-a_{21}\int dx_1\:x_1\: p(x_1\vert x_2)-a_{22}x_2=-(a_{21} \sigma_{12}/\sigma_{22}+a_{22})x_2$, and thus $F_2({\bf x})-\overline F_2(x_2)=-a_{21}x_1 +a_{21} (\sigma_{12}/\sigma_{22})x_2$. This yields
\begin{align} 
\label{EqOUT12supp}
{\overline {\cal T} }_{1 \to 2}=\frac{1}{4D_{22}}\frac{a_{21}^2\vert  {\boldsymbol \Sigma}\vert}{\sigma_{22}}\ ,
\end{align}
and for $D_{12}=0$ one recovers the expressions already given in the literature~\cite{HBS2016,MS2018}.

The BTE rate $ {\overline {\cal T}}^\dag_{1\to 2}$ is obtained by modifying the drifts according to Eq. (\ref{Eqbackdrift}), which amounts to replacing ${\bf A}$ by ${\bf A}^*={\boldsymbol \Sigma} {\bf A} {\boldsymbol \Sigma}^{-1}$~\cite{A1982}. The modified drifts are
\begin{align}
\label{Eqastar}
a_{11}^*&=\frac{(a_{11}\sigma_{11}+a_{12}\sigma_{12})\sigma_{22}-(a_{21}\sigma_{11}+a_{22}\sigma_{12})\sigma_{12}}{\vert {\boldsymbol \Sigma}\vert}\nonumber\\
a_{21}^*&=\frac{a_{12}\sigma_{22}^2-a_{21}\sigma_{12}^2+(a_{11}-a_{22})\sigma_{22}\sigma_{12}}{\vert {\boldsymbol \Sigma}\vert}
\end{align}
with $a_{22}^*$ and $a_{12}^*$ obtained by interchanging $1$ and $2$. Inserting these expressions into Eq. (\ref{EqOUT12supp}) and noting that the determinant $\vert \Sigma\vert$ of the covariance matrix is invariant under the transformation ${\bf A} \to{\bf A}^*$, we obtain after some  algebra 
\begin{align}
\label{T12backsupa}
  {\overline {\cal T}}^\dag_{1\to 2}&=\frac{(a_{21}\vert {\boldsymbol \Sigma}\vert +2D_{22}\sigma_{12}-2D_{12}\sigma_{22})^2}{4D_{22}\vert {\boldsymbol \Sigma}\vert \sigma_{22}}\ . 
 \end{align}

Finally, Eq. (\ref{EqsTEback1})  yields
\begin{align} 
\label{EqOUT12hatsupm}
{\widehat {\overline {\cal T}}}_{1 \to 2}=\frac{\big(2D_{12}\sigma_{22}-a_{21}\vert  {\boldsymbol \Sigma}\vert\big)^2}{4D_{22}\sigma_{22}\vert {\boldsymbol \Sigma} \vert}\ .
\end{align}
\\
 
\textit{2. Multi-time-step TE rates}
\\

We now detail the calculation of the  multi-time-step TE rates ${\cal T}_{1\to 2}$ and $\widehat {{\cal T}}_{1 \to 2}$.  As explained in the main text, this requires to compute the conditional expectations $\langle X_2(t+h)\vert {\bf X}^-(t)\rangle$, $\langle X_2(t+h)\vert X_2^-(t)\rangle$, $\langle X_1(t+h)\vert X_2^-(t)\rangle$, and the corresponding mean-square  errors $\sigma_{22}(h)$, $\sigma_{22,2}(h)$, and  $\sigma_{11,2}(h)$, $\sigma_{12,2}(h)$.
\vspace{0.15cm}

a) We first compute $\langle X_2(t+h)\vert {\bf X}^-(t)\rangle$ and $\sigma_{22}(h)$, which is is straightforward. Starting from  
\begin{align} 
X_2(t+h)=\int_{-\infty}^{t+h}ds\: [H_{21}(t+h-s)\xi_1(s)+H_{22}(t+h-s)\xi_2(s)]\, ,
\end{align}
we readily get 
\begin{align} 
\label{EqMMSE}
\Bigl\langle X_2(t+h)\vert {\bf X}^-(t)\Bigr\rangle=\int_{-\infty}^{t}ds\: [H_{21}(t+h-s)\xi_1(s)+H_{22}(t+h-s)\xi_2(s)]\, ,
\end{align}
since the noises are fixed for $s\le t$ by Eq. (\ref{EqOU}) and average to zero in the time interval $[t,t+h]$.
Eq. (\ref{Eqsigmajj}) then yields 
\begin{align} 
\label{Eqsigma22}
\sigma_{22}(h)&=2\int_0^h dt\:[D_{11}H_{21}^2(t)+D_{22}H_{22}^2(t)+2D_{12}H_{21}(t)H_{22}(t)]\, .
\end{align}
Of course, since the joint process is Markovian, ${\bf X}^-(t)$ can be replaced by ${\bf X}(t)$ in Eqs. (\ref{EqT12ha}) and (\ref{Eqsigmajj}), and  the same result could be obtained from the well-known expression of the transition probability of the OU process~\cite{R1989,G2004}.

\vspace{0.15cm}

b) The calculation of $\langle X_2(t+h)\vert X_2^-(t)\rangle$ and thus $\sigma_{22,2}(h)$ is less straightforward because  fixing only the past of the process $X_2$  up to time $t$ is not sufficient to fix the past of the noises $\xi_1$ and $\xi_2$. However, this difficulty is bypassed by determining  a {\it causal}  function $H'_{22}(t)$  such that 
\begin{align} 
\label{EqMVA2}
X_2(t)=\int_{-\infty}^{t}ds\: H'_{22}(t-s)\xi'_2(s)\ ,
\end{align}
where $\xi'_2(t)$ is another Gaussian white noise with variance $2D_{22}$.  Then, following the same reasoning as above, we get
\begin{align} 
\label{EqX2h}
\langle X_2(t+h)\vert X_2^-(t)\rangle=\int_{-\infty}^t ds\: H'_{22}(t+h-s)\xi'_2(s)\ ,
\end{align}
and in turn 
\begin{align} 
\label{Eqsigma22p}
\sigma_{22,2}(h)=2D_{22}\int_0^h dt \: H_{22}^{'2}(t)\, .
\end{align}
Since Eq. (\ref{EqMVA2}) implies that $S_{22}(\omega)\equiv\langle X_2(\omega)X_2(-\omega)\rangle=2D_{22}H'_{22}(\omega)H'_{22}(-\omega)$, the Fourier transform of $H'_{22}(t)$ is simply obtained by identifying  the  component of $S_{22}(\omega)$ that is analytic in the upper-half  plane $\mbox {Im}(\omega)>0$. Specifically,
\begin{align} 
\label{EqS22}
S_{22}(\omega)&=\langle \vert H_{21}(\omega)\xi_1(\omega)+H_{22}(\omega)\xi_2(\omega)\vert^2\rangle \ ,
\end{align}
where  $H_{21}(\omega)=-a_{21}/[(\omega_+-i\omega)(\omega_--i\omega)]$ and $H_{22}(\omega)=(a_{11}-i\omega)/[(\omega_+-i\omega)(\omega_--i\omega)]$ with  $\omega_{\pm}=\frac{1}{2}(a_{11}+a_{22}\pm \sqrt{\Delta})$ and $\Delta=(a_{11}-a_{22})^2+4a_{12}a_{21}$ ($\omega_{\pm}$ are the eigenvalues of ${\bf A}$). This yields
\begin{align} 
\label{PSD22}
S_{22}(\omega)=2D_{22}\frac{r_2^2+\omega^2}{(\omega_+^2+\omega^2)(\omega_-^2+\omega^2)}\ ,
\end{align}
where
 \begin{align} 
\label{Eqr2}
r_2=\sqrt{a_{11}^2+ \frac{D_{11}}{D_{22}}a_{21}^2-2 \frac{D_{12}}{D_{22}}a_{11}a_{21}}\, .
\end{align}
It then comes that 
\begin{align} 
\label{EqH22p}
H'_{22}(\omega)= \frac{r_2 -i\omega}{(\omega_+-i\omega)(\omega_--i\omega)}\, .
\end{align}
By construction, $H'_{22}(\omega)$ has no poles in the upper-half plane $\mbox {Im}(\omega)>0$ and therefore $H'_{22}(t)$  vanishes for $t<0$ [specifically, $H'_{22}(t)=(u_+e^{-\omega_+t}-u_-e^{-\omega_-t})\Theta(t)$ where  $\Theta(t)$ is the Heaviside step function
and $u_{\pm}=(\omega_{\pm}-r_2)/\sqrt{\Delta}$]. Moreover, we have chosen $r_2=\sqrt{r_2^2}>0$ so that $H'_{22}(\omega)$ is also zero-free in this region. As stressed in \cite{RTM2018}, this  condition (which  corresponds to the so-called ``minimum-phase" condition in the language of control theory~\cite{AM2008,B2005}) ensures that there is a one-to-one correspondence between the process and the corresponding forcing white noise: Fixing the history of the process $X_2$ up to time $t$ is equivalent to fixing the history of the noise $\xi'_2(t)$ and vice versa.

The TE rate ${\cal T}_{1 \to 2}$ is finally obtained by expanding $\sigma_{22}(h)$ and $\sigma_{22,2}(h)$  in powers of $h$. Using $H_{21}(t=0^+)=0$ and $H_{22}(t=0^+)=H'_{22}(t=0^+)=1$, we get
\begin{align} 
T_{1 \to 2}(h)=\frac{1}{2}\ln \frac{2D_{22}[h+\dot H'_{22}(t=0^+)h^2+{\cal O}(h^3)]}{2D_{22}h+2[D_{22}\dot H_{22}(t=0^+)+D_{12}\dot H_{21}(t=0^+)]h^2+{\cal O}(h^3)}\, ,
\end{align}
and then
\begin{align} 
\label{EqT1hderiv}
{\cal T}_{1 \to 2}&=\frac{1}{2}[\dot H'_{22}(t=0^+)-\dot H_{22}(t=0^+)-\frac{D_{12}}{D_{22}}\dot H_{21}(t=0^+)]\ .
\end{align}
Since $\dot H_{22}(t=0^+)=-a_{22}$, $\dot H_{21}(t=0^+)=-a_{21}$, and $\dot H'_{22}(t=0^+)=r_2-(\omega_++\omega_-)=r_2-a_{11}-a_{22}$,  we finally arrive at Eq. (\ref{EqTE12new}) in the main text. 
Moreover, inserting  the expression of $H'_{22}(t)$ into Eq. (\ref{EqMVA2}) and performing a few manipulations, we can transform this equation into the (non-Markovian) Langevin equation (\ref{EqCG}).

\vspace{0.15cm}

c) Finally, we consider the filtered TE rate $\widehat {{\cal T}}_{1 \to 2}$. This  requires to calculate the conditional mean $\langle X_1(t+h)\vert X_2^-(t)\rangle$.  To this aim, we use the fact that $\langle X_1(t+h)\vert X_2^-(t)\rangle $ is the orthogonal projection of $ X_1(t+h)$ onto   the trajectory $X_2^-(t)$.  It is a linear functional of $X_2^-(t)$, 
\begin{align} 
\label{Eqproj2}
\langle X_1(t+h)\vert X_2^-(t)\rangle=f(h) X_2(t)+\int_{-\infty}^t ds\: g(t-s,h)X_2(s) \, ,
\end{align}
where $f(h)$ and $g(t,h)$ are unknown functions to be determined by satisfying the orthogonality condition,
\begin{align} 
\label{Eqproj1}
\bigl\langle [X_1(t+h)-\langle X_1(t+h)\vert X_2^-(t)\rangle]X_2(s)\bigr\rangle=0 \: \: \: \forall s\le t\, .
\end{align}
Note that we have singled out the dependence on the value of $X_2$ at time $t$ from the dependence on the trajectory  up to to time $t$. Plugging Eq. (\ref{Eqproj2}) into Eq. (\ref{Eqproj1}), we  then obtain the integral equation
\begin{align} 
\label{EqWH}
\phi_{21}(t+h)=f(h)\phi_{22}(t)+ \int_{-\infty}^{t} ds\: g(t-s,h) \phi_{22}(s)\: \: \: \: \forall \: t\ge 0\ ,
\end{align}
where $\phi_{ij}(t)\equiv \langle X_i(t')X_j(t'+t)\rangle$. This equation is easily solved by using the fact that the stationary correlation functions $\phi_{ij}(t)$ of an OU process are sums of exponentials~\cite{G2004}. Specifically, for $t\ge 0$, 
\begin{align} 
\phi_{21}(t)&=\frac{D_{22}}{a_{21}(\omega_+^2-\omega_-^2)}[(r_2^2-\omega_+^2)(a_{22}-\omega_+)\frac{e^{-\omega_+t}}{\omega_+}-(r_2^2-\omega_-^2)(a_{22}-\omega_-)\frac{e^{-\omega_-t}}{\omega_-}] \nonumber\\
\phi_{22}(t)&=\frac{D_{22}}{\omega_+^2-\omega_-^2}[(\omega_+^2-r_2^2)\frac{e^{-\omega_+t}}{\omega_+}-(\omega_-^2-r_2^2)\frac{e^{-\omega_-t}}{\omega_-}]\ ,
\end{align}
where $\omega_{\pm}$ are the eigenvalues of the matrix ${\bf A}$ defined after Eq. (\ref{EqS22}). This suggests to introduce the exponential ansatz $g(t,h)=g(h)e^{-\lambda t}$. Then, by comparing the left-hand and right-hand sides of Eq. (\ref{EqWH}), we readily find  that $\lambda=r_2$ (defined by Eq. (\ref{Eqr2})), and the remaining unknown functions $f(h)$ and $g(h)$ are  obtained by identifying the coefficients of $e^{-\omega_+t}$ and $e^{-\omega_- t}$.
After some  algebra, we get
\begin{align} 
f(h)&=\frac{(r_2-\omega_+)(a_{22}-\omega_+)e^{-\omega_+h}-(r_2-\omega_-)(a_{22}-\omega_-)e^{-\omega_-h}}{a_{21}(\omega_+-\omega_-)}\nonumber\\
g(h)&=\frac{(r_2-\omega_+)(r_2-\omega_-)[(\omega_+-a_{22})e^{-\omega_+h}-(\omega_--a_{22})e^{-\omega_-h}]}{a_{21}(\omega_+-\omega_-)}\ .
\end{align}
In particular, when $h=0$,
\begin{align} 
\label{Eqfg0}
f(0)&=\frac{a_{11}-r_2}{a_{21}}\nonumber\\
g(0)&=\frac{(r_2-\omega_+)(r_2-\omega_-)}{a_{21}}\ .
\end{align}
From the expression of $\langle X_1(t+h)\vert X_2^-(t)\rangle$, we can easily compute the  mean-square  errors $\sigma_{11,2}(h)$ and $\sigma_{12,2}(h)$, and expanding in powers of $h$ finally leads to Eq. (\ref{EqOUT12hatm}) in the main text.

Interestingly, the above calculation has a connection to classical linear filtering and stochastic control theory~\cite{A2006,B2005,AM2008}. To cast the model described by the coupled  Langevin equations  (\ref{EqOU}) as a  linear control problem of the state-space form, we introduce two yet unspecified (but non-zero) parameters $G$ and $K$ such that 
\begin{align} 
\label{Eqnewvar}
GK&=a_{12}\nonumber\\
G-a_{21}K&=a_{22}-a_{11}
\end{align}
and the variable
\begin{equation} 
\label{Eqfilter}
{\widehat X}_1(t)=K X_2(t)\, .
\end{equation}
The Langevin equation for $X_1(t)$ and $X_2(t)$ can then be rewritten as
\begin{subequations}
\label{EqKBfilter:subeqns}
\begin{align} 
\dot X_1(t)&=-a_{11}X_1(t)-G\widehat X_1(t)+\xi_1(t)
\label{EqKBfilter:subeq1}\\
\dot {\widehat X}_1(t)&=-a_{11}\widehat X_1(t)-G\widehat X_1(t)+K[-a_{21}X_1(t)+\xi_2(t)+a_{21}\widehat X_1(t)]\, .
\label{EqKBfilter:subeq2}
\end{align}
\end{subequations}
One realizes that, provided one defines an ``observation variable" $Y(t)\equiv -a_{21}X_1(t)+\xi_2(t)$, the above set of equations can be interpreted as  describing the dynamics of a state variable  $X_1(t)$ (the ``signal process") and of its linear estimator  $\widehat X_1(t)$ given the trajectory $Y^-(t)\equiv \{Y(s)\}_{s\in[-\infty,t]}$ (see also \cite{SDNM2014}). In this representation, $G$ and $K$ denote  the control gain and the observer (or Kalman) gain, respectively~\cite{note80}.  The first two terms in the r.h.s. of Eq. (\ref{EqKBfilter:subeq2}) copy  the model of  the signal process  [including the linear control feedback $-G\widehat X_1(t)$] whereas the third term, which can also be written as $Y(t)+a_{21} \widehat X_1(t)$, corrects the model by a linear feedback that drives $\widehat X_1$ closer to $X_1$ at time $t+dt$. The process $e(t)\equiv Y(t)+a_{21}\widehat X_1(t)$ is  the so-called ``innovation", which is the part of the measurement that provides new information about the state of the system~\cite{KSH2000}. 

In the standard filtering problem one looks for  the linear estimate ${\widehat X}_1^*(t)$ that minimizes the mean-square error $P(t) = \langle [X_1(t)-\widehat X_1(t)]^2\rangle$ at every time $t$. This optimal estimate is known to be ${\widehat X}_1^*(t)=\langle X_1(t)\vert Y^-(t)\rangle$, the orthogonal projection of $X_1(t)$ onto the trajectory $Y^-(t)$. The main insight of the Kalman-Bucy theory is to convert the associated Wiener-Hopf integral equation into a  nonlinear differential equation by exploiting the state-space formulation~\cite{KSH2000}. The specification of the optimal Kalman-Bucy filter thus amounts to computing  the  gain $K^*$ that minimizes the estimation error variance $P(t)$. In the  present stationary case, one simply has (cf. Eqs. (III) and (IV) in section 7 in \cite{KB1961} with $F(t)=-a_{11},H'(t)=-a_{21},G(t)=1,Q(t)=2D_{11},R(t)=2D_{22}$)
\begin{align} 
\label{EqKBfilter1}
K^*=-a_{21}\frac{P^*}{2D_{22}}\ ,
\end{align}
 and $P^*$ is solution of the algebraic Ricatti equation
\begin{align} 
\label{EqKBfilter2}
-2a_{11}P^*-a_{21}^2\frac{(P^*)^2}{2D_{22}}+2D_{11}=0\, .
\end{align}
The physical solution is
\begin{align} 
\label{EqKBfilter3}
K^*=\frac{a_{11} -r_2}{a_{21}}\ ,
\end{align}
which can  be shown to also  be the solution of the Wiener-Kolmogorov causal filter~\cite{W1949,K1992,BS1950} (as the two filters are equivalent in the stationary case). Plugging this expression into the quadratic equation $a_{21}K^2 +(a_{22}-a_{11})K-a_{12}=0$ obtained by eliminating $G$ from Eqs. (\ref{Eqnewvar}), one then finds that the optimal filter is obtained when $r_2$ satisfies Eq. (\ref{EqMcond}) of the main text, {\it i.e.}, $r_2=r_2^*=\omega_{\pm}$. This condition of course requires specific relations between the parameters of the original bi-dimensional Ornstein-Uhlenbeck model. When this occurs, the individual dynamics of  $X_2(t)$ [or, equivalently, $\widehat X_1(t\vert Y^-(t))$]  is then Markovian, which is no coincidence. Indeed, as is well known, the innovation $e(t)$ is a white noise process with the same variance $R$ as the measurement noise when the filter is optimal~\cite{KSH2000}. This fact underscores the property that the Kalman-Bucy filter is so efficient at extracting information from the measurements that one is just left with a white noise afterwards. The filter state $\widehat X_1^*(t\vert Y^-(t))$  is then a sufficient statistic for the conditional distribution of $X_1(t)$. In passing, note that $r_2^*$ and in turn $K^*$ do not depend on the parameter $a_{12}$ that defines the feedback from $2$ to $1$. This illustrates  the separation principle in optimal control theory that  states that the optimal observer gain obtained by minimizing the estimation error is independent of the cost of the control and therefore is independent of the feedback gain $G$.

\vspace{0.25cm}
\textit{3. Entropy production rate}
\\

In the steady state, the entropy production rate in system $1$ is equal to the rate of entropy change in the environment, and is computed from Eq (\ref{Eqsigma12B}) where $\widehat F_{X,t}$,  $D_{XX,t}$, and $J_{XX,t}$ are replaced by $F_1({\bf x})=-a_{11}x_1-a_{12}x_2$, $D_{11}$, and $J_1({\bf x})=F_1({\bf x})P({\bf x}) -D_{11}\partial_{x_1}P({\bf x})-D_{12}\partial_{x_2}P({\bf x})$, respectively. This yields
 \begin{align} 
\label{EqsigmaxOU}
 \sigma_1=\sigma_1^B=\frac{a_{12}}{(a_{11}+a_{22})D_{11}}[a_{12}D_{22}-a_{21}D_{11}+ (a_{11}-a_{22})D_{12}]\ .
\end{align}

\section{Multi-time-step TE rates for the three-component process of Sec. \ref{sec:Kiviet}}
\label{sec:AppKiviet}

\setcounter{figure}{0} \renewcommand{\thefigure}{E.\arabic{figure}} 

\renewcommand{\theequation}{E\arabic{equation}} 

 In this Appendix,  we generalize the  calculations of Sec. \ref{Sec:OU} to the case of a multi-dimensional  OU process. Since the complexity of the calculation  rapidly increases with the number of components,  we only consider the stochastic model studied in section \ref{sec:Kiviet} which remains simple because the third Langevin equation describing the dynamics of the OU colored noise is decoupled from the two other equations  [see Eqs. (\ref{EqLKiviet})]. The matrix ${\bf A}$ in Eq. (\ref{EqOU}) is now a $3\times 3$ matrix with $a_{31}=a_{32}=0$, and $a_{33}>0$. 
 
The main difference with the bi-dimensional Markov case  is that  the (coarse-grained) dynamics of the joint process of interest, ${\bf X}_{1,2}\equiv (X_1,X_2)$, is no longer Markovian because the effective noises affecting the two sub-processes are colored,
\begin{align}
\label{EqEffnoises}
\xi_i^{eff}(t)=\xi_i(t)-a_{i3}X_3(t)=\xi_i(t)-a_{i3}\int_{-\infty}^t ds\: e^{-a_{33}(t-s)}\xi_3(s), \: \: i=1,2\ ,
\end{align}
with covariances
\begin{align}
\label{EqEffcor}
\langle \xi_i^{eff}(t)\xi_i^{eff}(t')\rangle&=2D_{ii}\delta(t-t')-a_{i3}(2D_{i3}-\frac{a_{i3}}{a_{33}}D_{33})e^{-a_{33}\vert t-t'\vert}\nonumber\\
\langle \xi_1^{eff}(t)\xi_2^{eff}(t')\rangle&=2D_{12}\delta(t-t')-2a_{13}D_{23}e^{-a_{33}(t-t')}\Theta(t-t')-2a_{23}D_{13}e^{-a_{33}(t'-t)}\Theta(t'-t)+\frac{a_{13}a_{23}}{a_{33}}D_{33}e^{-a_{33}\vert t-t'\vert}\ .
\end{align}
 Therefore, ${\bf X}_{1,2}^-(t)$ cannot be replaced by ${\bf X}_{1,2}(t)$ in Eqs. (\ref{EqT12ha}) and (\ref{Eqsigmajj}), and Eq. (\ref{EqMMSE}) (or the corresponding equation for $\langle X_1(t+h)\vert {\bf X}_{1,2}^-(t)\rangle$) is no longer valid because the  noises in the time interval $[t,t+h]$ do not average to zero (as they are correlated with the noises  for $s\le t$, which are fixed). To circumvent the problem, we need again to whiten the noises, which amounts to determining four causal functions $\widetilde H_{ij}(t)$ such that 
\begin{align}
\label{EqHtilde}
X_i(t)=\int_{-\infty}^t ds\: [\widetilde H_{i1}(t-s)\widetilde \xi_1(s)+\widetilde H_{i2}(t-s)\widetilde \xi_2(s)], \: \: i=1,2\ ,
\end{align}
where $\widetilde \xi_1(t)$ and $\widetilde \xi_2(t)$ are Gaussian noises with zero-mean and covariances $\langle \widetilde \xi_i(t)\widetilde \xi_j(t')\rangle=2D_{ij}\delta(t-t')$. Then,  Eq. (\ref{Eqsigma22}) is just replaced by 
\begin{align} 
\label{Eqsigma22a}
\sigma_{22}(h)=2\int_0^h dt\:[D_{11}\widetilde H_{21}^2(t)+D_{22}\widetilde H_{22}^2(t)+2D_{12}\widetilde H_{21}(t)\widetilde H_{22}(t)]\ ,
\end{align}
and  $\sigma_{11}(h)$ is given by the symmetric equation. By construction, the matrix  $\widetilde {\bf H}(\omega)=[\widetilde H_{ij}(\omega)]$ must  satisfy  the equation 
\begin{align} 
\label{EqSomega}
{\bf S}^{(2)}(\omega)=\widetilde {\bf H}(\omega)(2{\bf D}) \widetilde {\bf H}^T(-\omega)\, ,
\end{align}
where ${\bf S}^{(2)}(\omega)$ is the $2\times 2$ sub-matrix of ${\bf S}(\omega)$ with elements $S_{ij}(\omega)=\langle X_i(\omega)X_j(-\omega)\rangle$ for $i,j=1,2$.  
Specifically,
\begin{align} 
S_{ij}(\omega)=2D_{ij}\frac{\omega^2+f_{ij}\omega+g_{ij}}{(\omega^2+\omega_+^2)(\omega^2+\omega_-^2)}\ ,
\end{align}
where  $\omega_{\pm}$ are defined  in Appendix \ref{sec:AppOU} after Eq. (\ref{EqS22}), $f_{11}=f_{22}=0$, and  $f_{12}, g_{11},g_{12},g_{22}$ are complicated functions of the model parameters. Note that the numerator and the denominator of  $S_{ij}(\omega)$ are only quadratic polynomials despite the fact that the OU process defined by Eqs. (\ref{EqLKiviet}) is trivariate.  This simplification occurs because $a_{31}=a_{32}=0$ and we also set $\beta_G=\beta_{\mu}$ (as the two timescales $1/\beta_G$ and $1/\beta_{\mu}$ are taken equal  to the measured autocorrelation time of $\mu$ when fitting the model parameters to experiments~\cite{K2014}).

In order to solve Eq. (\ref{EqSomega}),  we seek a matrix $\widetilde {\bf H}(\omega)=[\widetilde H_{ij}(\omega)]$ in the form
\begin{align}
\label{EqHtilde1}
\widetilde {\bf H}(\omega)=\frac{1}{(\omega_+-i\omega)(\omega_--i\omega)}\left(
\begin{array}{cc}
\omega_{11}-i\omega&\omega_{12}\\
\omega_{21}&\omega_{22}-i\omega
\end{array}\right)
\ ,
\end{align}
imposing $\lim_{\omega \to \infty}-i\omega \widetilde {\bf H}(\omega)={\bf 1}$ so that  the response functions  satisfy $\widetilde {\bf H}(t=0^+)={\bf 1}$.  By matching the coefficients of $\omega^n$ in the numerators of $S_{ij}(\omega)$, we  can then determine the  four unknown quantities $\omega_{ij}$. The corresponding non-linear equations are solved numerically and the correct solution is the one that ensures that the elements of the matrix $\widetilde {{\bf H}}^{-1}(\omega)$ have no poles  in the upper-half plane $\mbox {Im}(\omega)>0$ (this is the generalization of  the  minimum-phase condition introduced in Appendix \ref{sec:AppOU}).

It remains  to calculate $\sigma_{22,2}(h)$ from Eq. (\ref{Eqsigma22p}), where $H'_{22}(t)$ is the response function  defined by Eq. (\ref{EqMVA2}) and obtained from the Wiener-Hopf factorization of $S_{22}(\omega)$. This yields
 \begin{align} 
\label{EqH22p}
H'_{22}(\omega)=\frac{\omega'-i\omega}{(\omega_+-i\omega)(\omega_--i\omega)}\ ,
\end{align}
where $\omega'$ is the root of the equation $\omega^2+g_{22}$ with a positive real part.

Finally, expanding $\sigma_{22}(h)$ and $\sigma_{22,2}(h)$  in powers of $h$, using $H'_{22}(t=0^+)=\widetilde H_{22}(t=0^+)=1$, $\dot H'_{22}(t=0^+)=\omega'-(a_{11}+a_{22})$, $\dot {\widetilde H}_{22}(t=0^+)=\omega_{22}-(a_{11}+a_{22})$, and $\dot {\widetilde H}_{21}(t=0^+)=\omega_{21}$, we obtain
\begin{align} 
T_{1 \to 2}(h)&=\frac{1}{2}\ln \frac{2D_{22}h+2D_{22}\dot H'_{22}(t=0^+)h^2+{\cal O}(h^3)}{2D_{22}h+[2D_{22}\dot {\widetilde H}_{22}(t=0^+)+2D_{12}\dot {\widetilde H}_{21}(t=0^+)]h^2+{\cal O}(h^3)}\nonumber\\
&=\frac{1}{2}\ln \frac{D_{22}h+D_{22}(\omega'-a_{11}-a_{22})h^2+{\cal O}(h^3)]}{D_{22}h-[D_{22}(\omega_{22}-a_{11}-a_{22})+D_{12}\omega_{21}]h^2+{\cal O}(h^3)}\ ,
\end{align}
and thus
\begin{align} 
{\cal T}_{1 \to 2}=\frac{1}{2}[\omega'-\omega_{22}-\frac{D_{12}}{D_{22}}\omega_{21}]\ .
\end{align}
${\cal T}_{2 \to 1}$ is given by the symmetric formula.

In order to  compute the backward TE rates, we again have to replace the $a_{ij}$'s by the elements of the matrix ${\bf A}^*={\boldsymbol \Sigma} {\bf A}^T {\boldsymbol \Sigma}^{-1}$. However, this simply amounts to repeating the calculation of the forward TE rates after interchanging the power-spectrum functions $S_{12}(\omega)$ and $S_{21}(\omega)$ [as $S_{12}^\dag(\omega)=S_{21}(\omega)$].

It is useful to  illustrate the above equations numerically to see why the marginal processes $X_1$ and $X_2$ have a quasi-Markovian behavior. Let us for instance consider the case of the low-IPTG experiment. The  three coupled Eqs. (\ref{EqLKiviet})  are then equivalent to the two non-Markovian equations 
 \begin{align} 
\dot X_1&=-0.209 X_1 +0.069X_2 +\xi^{eff}_1\nonumber\\
\dot X_2&=0.084 X_1-0.282X_2+\xi_2^{eff}\ , 
\end{align}
with  
 \begin{align}
\label{EqEffcor1}
\langle \xi_1^{eff}(t)\xi_1^{eff}(t')\rangle&=2D_{11}\:\delta(t-t')-0.003\:e^{-0.33\vert t-t'\vert}\nonumber\\
\langle \xi_2^{eff}(t)\xi_2^{eff}(t')\rangle&=2D_{22}\:\delta(t-t')+0.001\:e^{-0.33\vert t-t'\vert}\nonumber\\
\langle \xi_1^{eff}(t)\xi_2^{eff}(t')\rangle&=2D_{12}\:\delta(t-t')-0.005\:e^{-0.33(t-t')}\Theta(t-t')+0.002\: e^{-0.33\vert t-t'\vert}\ ,
\end{align}
 and $2D_{11}=0.020$, $2D_{22}=0.059$, and $2D_{12}=0.014$.  Although the exponential terms seem to only give small corrections,  these noises cannot be approximated by their white components. Indeed, this would significantly change the values of the TE rates and even reverse the directionality of the information flow (${\cal T}_{1\to 2}\approx 2.2\times 10^{-3},{\cal T}_{2\to 1}\approx 8.6\times 10^{-3}$, to be  compared with the values given in Table \ref{Table1}). On the other hand, by taking the inverse of the response functions $\widetilde H_{ij}(\omega)$, we obtain from Eqs. (\ref{EqHtilde}) the equivalent representation
 \begin{align} 
\label{EqLnew1}
\dot X_1&=-0.069X_1-0.034X_2 -\int_{-\infty}^t ds\: e^{-0.398(t-s)}[0.018X_1(s)-0.002X_2(s)]+\widetilde \xi_1\nonumber\\
\dot X_2&=0.183X_1-0.354X_2-\int_{-\infty}^t ds\: e^{-0.398(t-s)}[0.012X_1(s)-0.002X_2(s)]+\widetilde \xi_2\ , 
\end{align}
with $\langle \widetilde \xi_i(t)\widetilde \xi_j(t')\rangle =2D_{ij}\delta(t-t')$.  Moreover, by using Eq. (\ref{EqMVA2}) and the similar equation for $X_1$, and taking the inverse of the response functions $H'_{ii}(\omega)$, the individual processes $X_1$ and $X_2$ can be represented by
 \begin{align} 
\label{EqLnew2}
\dot X_1&=-0.089X_1 -0.017\int_{-\infty}^t ds\: e^{-0.402(t-s)}X_1(s)+ \xi'_1\nonumber\\
\dot X_2&=-0.286 X_2+0.005\int_{-\infty}^t ds\: e^{-0.205(t-s)}X_2(s)+\xi'_2\ ,
\end{align}
 with $\langle \xi'_i(t) \xi'_j(t')\rangle =2D_{ij}\delta(t-t')$. Let us recall that, by construction, Eqs. (\ref{EqLnew1}) and (\ref{EqLnew2}) yield the same statistical properties of the stationary processes $X_1$ and $X_2$ as the original Eqs. (\ref{EqLKiviet}).  Although these are still non-Markovian equations, neglecting the  contributions of the memory kernels is now  harmless and leads to exactly the same values of the TE rates as the full equations. 
 
 Likewise, for the high-IPTG experiment, the three coupled Eqs. (\ref{EqLKiviet})  are equivalent to the two equations  (similar to Eqs. (\ref{EqLnew1}))
 \begin{align} 
\label{EqLnew3}
\dot X_1&=-1.009X_1+0.232X_2 -\int_{-\infty}^t ds\: e^{-3.254(t-s)}[0.053X_1(s)+0.006X_2(s)]+\widetilde \xi_1\nonumber\\
\dot X_2&=-3.230X_2+\widetilde \xi_2\ .
\end{align}
In this case the white noises $ \widetilde \xi_1$ and $\widetilde \xi_2$ are independent, as $D_{12}=0$.


\begin{thebibliography}{10}
\bibitem{note54} The equivalence stems from the fact that the entropy of a Gaussian distribution is directly proportional to the logarithm of the determinant of its covariance matrix.
\bibitem{note15} When considering the BTE rate in a steady state, it  is implicit that the limit of an infinite trajectory is taken, so that 
${\cal T}^\dag_{X \to Y}$ no longer depends on $T$. Eqs. (\ref{EqDiff2}) and  (\ref{EqDiff5}) then imply Eq. (\ref{EqDiff4}).
\bibitem{note8} One can show that this definition also satisfies Eq. (\ref{EqSigmaBX}), where the  path probability $\widehat P_{Y_0^T}(X_0^T\vert X_0)$  has the {\it same} definition as in the bipartite case. Since the noises are correlated, it may be surprising that $\widehat P_{Y_0^T}(X_0^T\vert X_0)$ is expressed in terms of the same Onsager-Machlup action functional as in the bipartite case  which only involves the element $D_{XX}$ of the diffusion matrix (see \cite{footnote_probdetached}). However, we recall that $\widehat P_{Y_0^T}(X_0^T\vert X_0)$ is the probability of observing the trajectory $X_0^T$ when $Y_0^T$ is considered as an external protocol. It may be viewed as the continuous-time limit of the probability  $\prod_{k=0}^{N-1} P(X_{k+1}\vert {\bf Z}_k)$ where the evolution over the time interval $[0,T]$ is discretized into steps of width $\Delta t=T/N$ with $t_k=k\Delta t$ for $k=0,...N$, $X_k\equiv X(t_k)$ and $Y_k\equiv Y(t_k)$ (see, {\it e.g.},  \cite{SU2012,HS2014}). As can be seen from  Eq. (\ref{EqPropa1}), $P(X_{k+1}\vert {\bf Z}_k)$ only depends on $ D_{XX}$.
\bibitem{note14} Note that we only consider the single-time-step sensory capacities $\overline {C}_i$ because the joint dynamics is non-Markovian and therefore inequality (\ref{EqIneq3}) is not necessarily satisfied. This implies that the sensory capacities $C_i$ defined by Eq. (\ref{EqSensCap}) may be larger than $1$.
\bibitem{note80} The transformation from Eqs. (\ref{EqOU}) to Eqs. (\ref{EqKBfilter:subeqns}) is physically meaningful and Eqs. (\ref{EqOU})  describes the dynamics of a pair signal-observer  at the condition that $K$ and $G$ are real, which requires that $\Delta=(a_{11}-a_{22})^2+4a_{12}a_{21}\ge 0$.
\end{thebibliography}
\end{document}